\documentclass[ 12pt, letterpaper]{article}%
\usepackage{fullpage}
\usepackage{amsmath}
\usepackage{graphicx}%
\usepackage{amsfonts}%
\usepackage{amssymb}
\usepackage{url}
\usepackage{fancyhdr}
\usepackage{isomath}
\usepackage{amsmath}
\usepackage{amsbsy}
\usepackage{amssymb}
\usepackage{amscd}
\usepackage{amsfonts}
\usepackage{graphicx}
\usepackage{verbatim}
\usepackage{euscript}
\usepackage{alltt}
\usepackage{stmaryrd}
\usepackage{float}
\usepackage{appendix}
\usepackage{caption}
\usepackage[english]{babel}

\usepackage{subfigure}
\usepackage{tikz}
\usetikzlibrary{shapes,arrows}
\usepackage{amsmath}
\usepackage{amsbsy}
\usepackage{amssymb}
\usepackage{amscd}
\usepackage{amsfonts}

\newcommand{\bfa}{{\mathbold a}}
\newcommand{\bfb}{{\mathbold b}}

\newcommand{\bfe}{{\mathbold e}}

\newcommand{\bfi}{{\mathbold i}}

\newcommand{\bfm}{{\mathbold m}}
\newcommand{\bfn}{{\mathbold n}}

\newcommand{\bfp}{{\mathbold p}}

\newcommand{\bfu}{{\mathbold u}}
\newcommand{\bfv}{{\mathbold v}}

\newcommand{\bfA}{{\mathbold A}}

\newcommand{\bfD}{{\mathbold D}}

\newcommand{\bfU}{{\mathbold U}}

\newcommand{\bfX}{{\mathbold X}}

\newcommand{\beq}{\begin{equation}}
\newcommand{\eeq}{\end{equation}}
\newcommand{\beqs}{\begin{eqnarray}}
\newcommand{\eeqs}{\end{eqnarray}}
\newcommand{\beql}{\begin{equation} \label}

\newcommand{\bfepsilon}{\mathbold{\epsilon}}
\newcommand{\bfnu}{\mathbold{\nu}}
\newcommand{\bfalpha}{\mathbold{\alpha}}

\newcommand{\bflambda}{\mathbold{\lambda}}

\newcommand{\bfLambda}{\mathbold{\Lambda}}

\newcommand{\grad}{\mathop{\rm grad}\nolimits}
\newcommand{\divergence}{\mathop{\rm div}\nolimits}
\newcommand{\curl}{\mathop{\rm curl}\nolimits}

\usepackage[margin=1in]{geometry}
\newcommand*\circled[1]{\tikz[baseline=(char.base)]{
            \node[shape=circle,draw,inner sep=1pt] (char) {#1};}}

\newcommand{\boxalign}[2][0.97\textwidth]{
 \par\noindent\tikzstyle{mybox} = [draw=black,inner sep=6pt]
 \begin{center}\begin{tikzpicture}
  \node [mybox] (box){%
   \begin{minipage}{#1}{\vspace{-5mm}#2}\end{minipage}
  };
 \end{tikzpicture}\end{center}
}

\usepackage{titlesec}

\begin{document}

\title{A non-traditional view on the modeling of nematic disclination dynamics}
\author{Chiqun Zhang$^1$, Xiaohan Zhang$^1$, Amit Acharya$^1$, \\Dmitry Golovaty$^2$, Noel Walkington$^1$\\\\
$^{1}$Carnegie Mellon University, Pittsburgh, USA \\ $^2$ The University of Akron, Akron, USA\\}

\date{March 1, 2016}
\maketitle

%--------------------%
%Abstract
%--------------------%

\begin{abstract}

\noindent Nonsingular disclination dynamics in a uniaxial nematic liquid crystal is modeled within a mathematical framework where the  kinematics is a direct extension of the classical way of identifying these line defects with singularities of a unit vector field representing the nematic director. It is well known that the universally accepted Oseen-Frank energy is infinite for configurations that contain disclination line defects. We devise a natural augmentation of the Oseen-Frank energy to account for physical situations where, under certain conditions, infinite director gradients have zero associated energy cost, as would be necessary for modeling half-integer strength disclinations within the framework of the director theory. Equilibria and dynamics (in the absence of flow) of line defects are studied within the proposed model.  Using appropriate initial/boundary data, the gradient-flow dynamics of this energy leads to non-singular, line defect equilibrium solutions, including those of half-integer strength. However, we demonstrate that the gradient flow dynamics for this energy is not able to adequately describe defect evolution. Motivated by similarity with dislocation dynamics in solids, a novel 2D-model of disclination dynamics in nematics is proposed. The model is based on the extended Oseen-Frank energy and takes into account thermodynamics and the kinematics of conservation of defect topological charge. We validate this model through computations of disclination equilibria, annihilation, repulsion, and splitting. We show that the energy function we devise, suitably interpreted, can serve as well for the modeling of equilibria and dynamics of dislocation line defects in solids making the conclusions of this paper relevant to mechanics of both solids and liquid crystals.

\end{abstract}

% X
\tikzstyle{decision} = [diamond, draw, fill=blue!20, 
    text width=5em, text badly centered, node distance=1.5cm, inner sep=0pt]
\tikzstyle{block} = [rectangle, draw, fill=blue!20, 
    text width=25em, text centered, rounded corners, minimum height=3em]
\tikzstyle{line} = [draw, -latex']
\tikzstyle{cloud} = [draw, ellipse,fill=red!20, node distance=3cm,
    minimum height=2em]
%% X 

\section{Introduction}

Liquid crystals (LC) are matter in a state whose properties are between liquids and solids. Research on liquid crystals is currently advancing quite rapidly motivated by applications and discoveries in material science as well as in biological systems. There are many types of liquid crystal states, depending on the amount of order in the material. A nematic phase consists of rod like molecules that retain some long-range orientational order. In this work, we are primarily interested in modeling disclinations in a uniaxial nematic liquid crystalline medium, treated by an augmentation of the classical model (cf. \cite{stewart2004static}) where the director order parameter is represented by a unit vector field. 

The classical theories of liquid crystal mechanics like the Oseen-Frank and Ericksen-Leslie models predict unbounded energy in finite bodies with discrete disclinations. Recently, a kinematic augmentation of classical Leslie-Ericksen theory \cite{acharya2013continuum, pourmatin2012fundamental} has been devised that allows alleviating the singularity, with results being demonstrated for the case where the defect field is not allowed to evolve. These works aim to achieve an understanding of the connections between the classical theory of defects, such as solid dislocations and disclinations introduced by Weingarten and Volterra, and the theory of defected liquid crystals, a line of enquiry that began from the work of Kleman  \cite{kleman1973defect}. In \cite{acharya2013continuum}, the model introduces an augmented Oseen-Frank kinematics and involves a director field and an incompatible director distortion field that is not \emph{curl-free}. In \cite{pourmatin2012fundamental}, a finite element based numerical scheme was used to solve for the director fields of prescribed static disclinations and a critical examination presented of the similarities and differences that arise between the modeling of LC disclinations and solid dislocations using the eigendeformation approach \cite{pourmatin2012fundamental}. In this paper, we study this augmented model with natural constitutive modifications to enable the study of equilibria and evolution of LC disclinations, including those of half integer strength. First, a gradient flow dynamics of the augmented energy is utilized and used to calculate equilibrium solutions. However, we find that the gradient flow dynamics for this energy is not suitable for modeling the defect evolution problem, and explain why this must be so. Motivated by the crystal dislocation case, a 2D model based on the augmented energy, thermodynamics, and the kinematics of conservation of defect topological charge is constructed to analyze nematic disclination dynamics. We validate this model through computations for disclination equilibria, annihilation, repulsion, and dissociation.

Non-singular equilibria and dynamics of liquid crystal point and line defects have been studied in the literature, particularly within the Landau de-Gennes (L-dG) framework \cite{de1995physics, sonnet2012dissipative, mottram2014introduction, schopohl1987defect, bauman2012analysis,ravnik2009landau,di2014half,kralj1991nematic,macdonald2012robust,ignat2014uniqueness, ignat2015stability, ignat2013stability,nguyen2010refined,cladis1972non,bethuel1992bifurcation,biscari1997local,canevari2013biaxiality,fatkullin2009vortices,
golovaty2014minimizers,henao2012symmetry,kralj1999biaxial,mkaddem2000fine}. A more limited number of studies have been carried out in the Oseen-Frank and Leslie-Ericksen models as well as Ericksen's model for nematics with variable degree of orientation \cite{frank1958liquid, cladis1972non, bethuel1992bifurcation,   virga1995variational, biscari2003expulsion,biscari2005field, sonnet1997dynamics, gartland2002elastic,hardt1988stable,berlyand2005homogenization,lin1995nonparabolic,
lin2000existence,MR2804649}. The general consensus from the literature is that finite energy line-defects, including those of half-integer strength, can only be predicted by the full L-dG theory among all the models mentioned above.

As a point of departure, Ball and Bedford \cite{ball2014discontinuous} suggest the use of discontinuous order parameter fields, in particular a discontinuous vector order parameter field to represent uniaxial nematics. The exploration there is essentially kinematical and focuses primarily on the appropriate mathematical function spaces to be used, stopping short of demonstrating specific examples of solutions (or approximations thereof) of defect equilibria resulting from the use of energy functions and dynamical models based on their discontinuous kinematics. Our work, in essence, achieves precisely this goal, thus being complementary to \cite{ball2014discontinuous}. While our computational work does not employ discontinuous fields, it is demonstrated and explained why our approach yields, in a sense, the natural practical approximation of such discontinuous limiting director fields.

The work of Gartland \cite{gartland2015scalings} demonstrates how the classical Oseen-Frank energy may be viewed as a constrained form of the Landau-deGennes energy at temperatures below the `supercooling temperature.' Since in this temperature range the bulk Landau-deGennes energy is minimized by $Q$-tensors representing the uniaxial nematic phase, the constrained L-dG energy of nematic configurations that contain line defects is infinite. Our work develops a modification of the Oseen-Frank energy that enables the prediction of finite-energy defect fields, utilizing a core energy regularization that involves a material length scale which may be associated with the `nematic correlation length' $\xi$ as defined in \cite{gartland2015scalings}.

\section{Notation}

The condition that $a$ is defined to be $b$ is indicated by the statement $a := b$. 
The Einstein summation convention is implied unless specified otherwise. The symbol $\bfA \bfb$ denotes the action of a tensor $\bfA$ on
a vector $\bfb$, producing a vector. In the sequel, $\bfa\cdot\bfb$
represents the inner product of two vectors $\bfa$ and $\bfb$; the symbol $\bfA\bfD$
represents tensor multiplication of the second-order tensors
$\bfA$ and $\bfD$.

The symbol $\divergence$ represents the divergence and $\grad$ represents the
gradient. In this paper all tensor or vector indices are written with respect to the basis  $\bfe_i$, $\bfi$=1 to 3, of a rectangular Cartesian coordinate system. The following component-form notation holds: 
\begin{equation*}
 \begin{split}
    \left(\bfA\times\bfv\right)_{im} &= e_{mjk} A_{ij} v_{k}    \\
    \left(\divergence\bfA\right)_{i} &= A_{ij,j}                \\
    \left(\curl\bfA\right)_{im} &= e_{mjk} A_{ik,j}             \\
 \end{split}
\end{equation*}
where $e_{mjk}$ is a component of the alternating tensor $\bfX$.

The following list describes some of the mathematical symbols we use in this work:

$\bfn$: director

$k$: disclination strength

$\theta$: angle of director field

$\bflambda$: layer field

$l$: layer thickness

$\xi$: core width

\section[Augmented Oseen-Frank energy and corresponding gradient flow method] {Augmented Oseen-Frank energy and corresponding gradient flow computations} \label{sec:gradient_theory}

It is generally believed that a theory of nematic line defects cannot be established with a representation of the nematic director by a unit vector field. Indeed, consider a nematic occupying a two-dimensional domain with the director field $\bfn$ taking values in $\mathbb S^1$. Assuming the validity of the universally accepted Oseen-Frank energy density function \cite{frank1958liquid, oseen1933theory} given by 
\begin{gather*}
\mathcal{F}_{OF} = K_1(\divergence \bfn)^2 + K_2 (\bfn\cdot\curl \bfn)^2 + K_3|\bfn\times\curl \bfn|^2 \\+ K_{24} (\divergence \bfn)^2  - \mathrm{tr}(\grad \bfn)^2)
\end{gather*}
where $K_1, K_2, K_3, K_{24}$ are material dependent Frank elastic constants, it can be seen that the planar configuration of a straight, half-integer strength wedge disclination necessarily results in at least one curve $\mathcal C$ in the plane connecting the core of the defect to the external boundary such that the vector field $\bfn$ has to be discontinuous along $\mathcal C$. If the discontinuity were to be approximated by a thin region along $\mathcal C$ characterized by high gradients of the director, the Oseen-Frank energy of the resulting configuration would yield a physically unobserved region of very high energy density. One of our goals in this paper is to propose a model that adequately resolves this problem by augmenting the director model by an additional field. The resulting model is different from the Landau De-Gennes $Q$-tensor model \cite{mottram2014introduction}, and makes close connections to models of line defects in other fields, such as crystal plasticity and phase transitions in solids.

\begin{figure}[H]
\centering
\subfigure[Director field, represented as a vector field, changes direction through a layer in the center of the body, but there is no disclination and the energy is zero.]{
\includegraphics[width=0.45\textwidth]{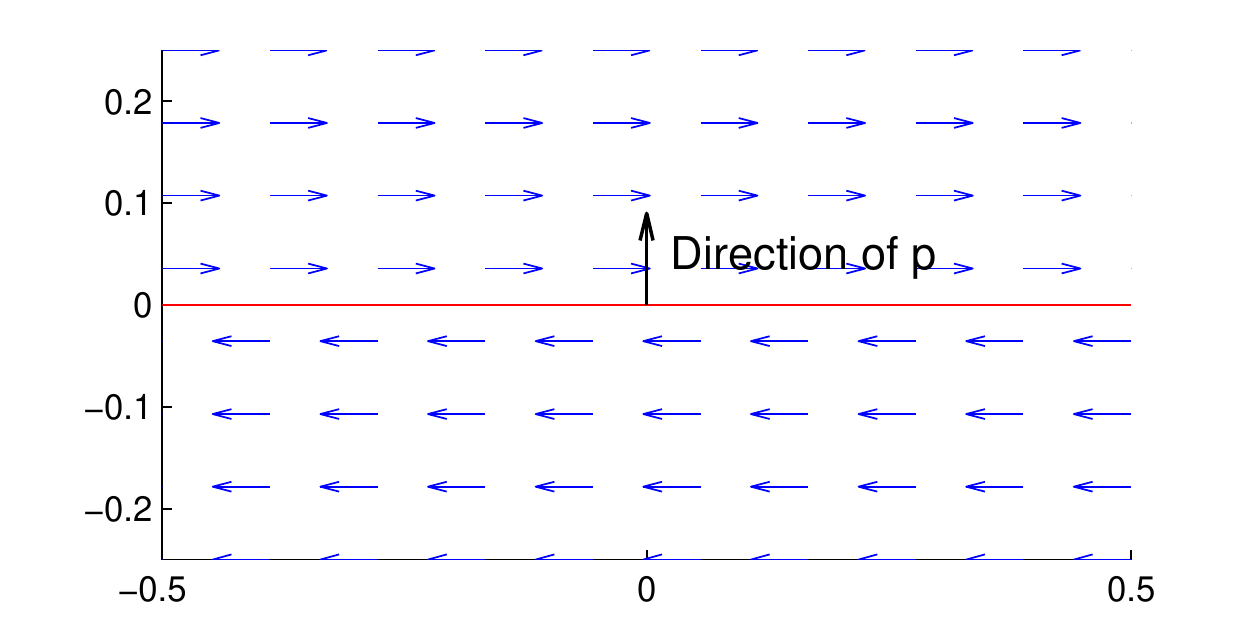}
\label{fig:OF_req_1}}\qquad
\subfigure[The equivalent case with a different vector field.]{
\includegraphics[width=0.45\textwidth]{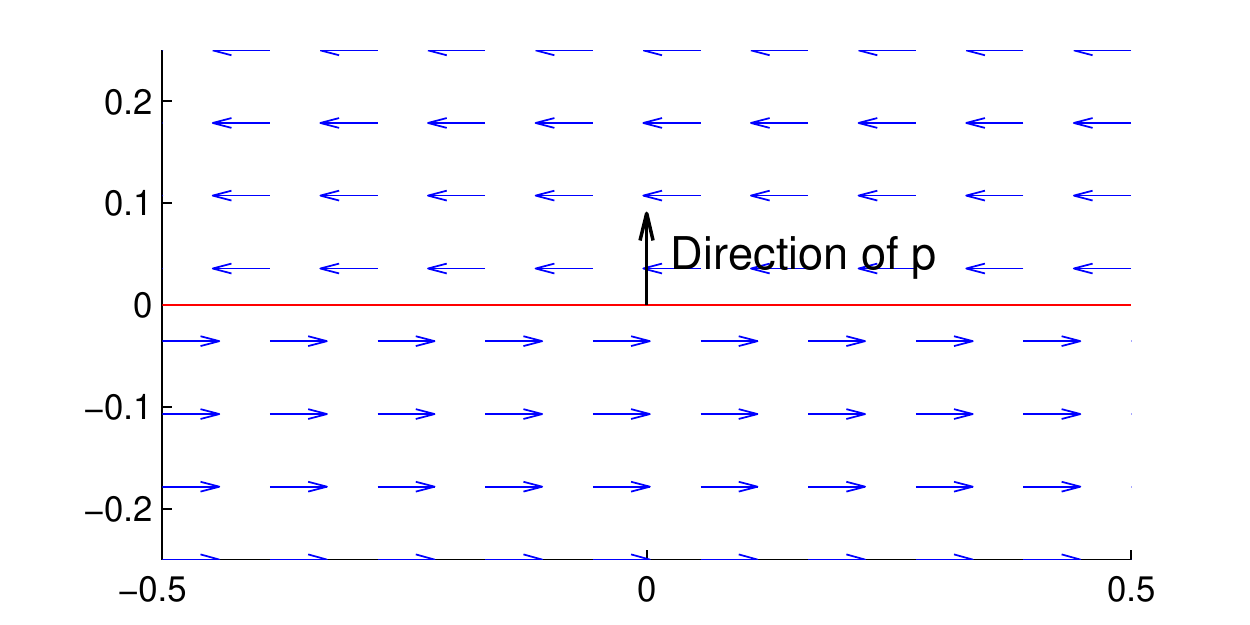}
\label{fig:OF_req_2}}
\subfigure[The director field without artificial arrows.]{
\includegraphics[width=0.5\textwidth]{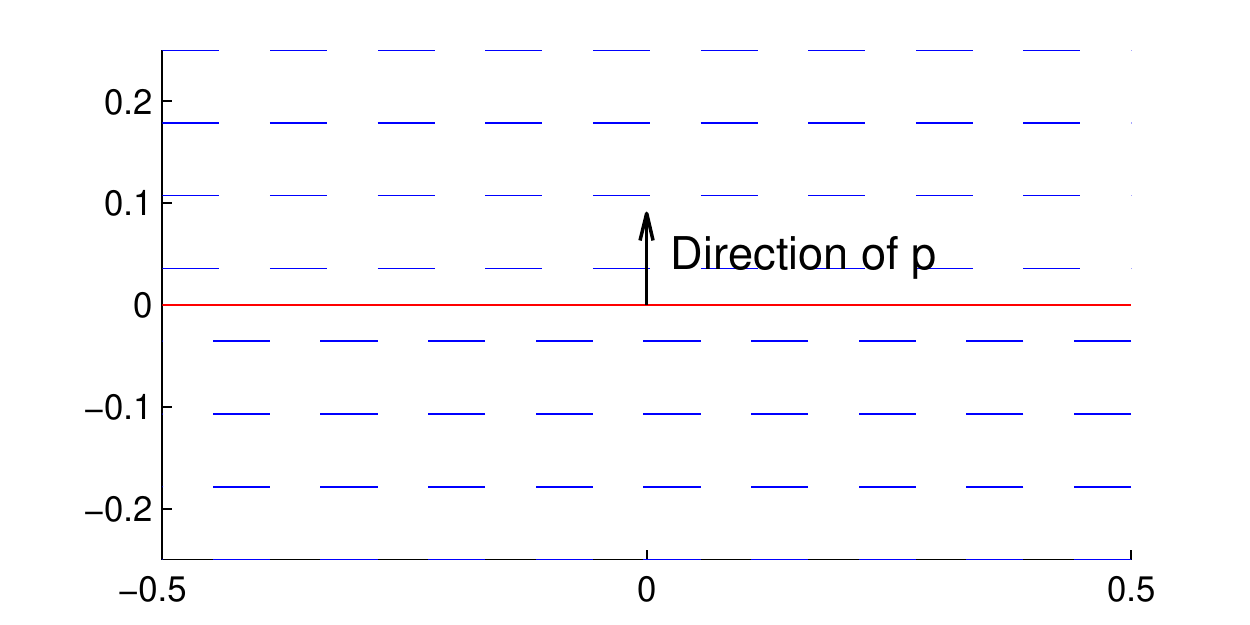}
\label{fig:OF_req_3}}
\caption{Illustration of issues with representation of the director field by a vector field.}
\end{figure}

The Oseen-Frank energy function is a quadratic function in the director field\footnote
{From here onwards, we will use the imprecise short-form of `director field' to refer to the `director vector field.'}
and its gradients. With the half-integer defect as a motivation, it would seem that if director discontinuity associated with the winding of the director by $\pi$ radians were to be assigned a  vanishing energy cost, then progress may be made on modeling line defects. Mathematically, suppose that the director $\bfn$ winds by $\pi$ radians over the distance $l$ in the direction of a unit vector $\bfp$, where $l$ is a parameter with physical dimensions of length. Associating a zero energy cost to the jump of the director by $\pi$ radians across the line perpendicular to $\bfp$ can be stated as a condition
\begin{gather}
\label{eq:e}
0 = \mathcal{F}(\bfn, {\bf 0}) = \mathcal{F}\left(\bfn, \frac{2 \bfn}{l} \otimes \bfp \right) = \mathcal{F}(-\bfn, -\frac{2\bfn}{l}\otimes \bfp) \; \; \forall \,\bfp
\end{gather}
as $l \to 0_+$. This is equivalent to demanding a zero energy cost for a flip of $\pi$ radians over a layer of width $l$ in the limit of vanishing layer width. The second equlity in \eqref{eq:e} stems from the condition
\begin{gather*}
\mathcal{F}_{OF}(\bfn, \grad \bfn) = \mathcal{F}_{OF}(-\bfn, -\grad \bfn)
\end{gather*}
arising from the head-tail symmetry of the nematic molecules\footnote{Of course, the choice of a flip by $\pi$ radians in (\ref{eq:e}) is also intimately connected to head-tail symmetry of the nematic molecules.}. As illustration, these requirements mean there is no energy in the case shown in Figure \ref{fig:OF_req_1} which is physically equivalent to Figure \ref{fig:OF_req_2} and Figure \ref{fig:OF_req_3}. For fixed $\bfn$, this implies a multiple-well structure of the energy density in the director gradient slot of $\mathcal{F}$.

\begin{figure}[H]
\centering
\includegraphics[width=0.5\textwidth]{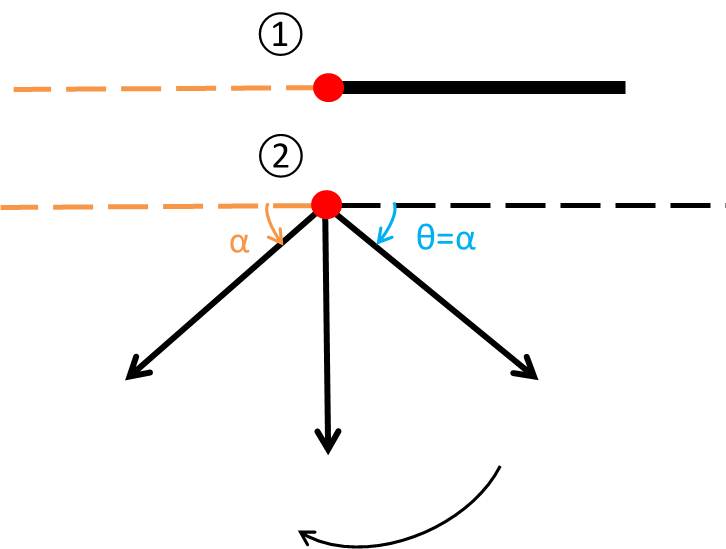}
\caption{Illustration of the reason for non-convexity of the energy density. The angle between the director and its neighbor displays periodicity with change in direction.}
\label{fig:nonconvex_local}
\end{figure}

Figure \ref{fig:nonconvex_local} shows another justification for demanding nonconvexity of the energy-density. It shows two spatial points in the nematic liquid crystal located close to each other, represented as the red dots (point $\circled{1}$ and point $\circled{2}$). The director at point $\circled{1}$ is assumed be in the horizontal direction pointing to the right, shown as the black line. At point $\circled{2}$, the director is considered as a vector originating from this point $\circled{2}$ and rotates clockwise. The angle between the director at point $\circled{1}$ and the director at point $\circled{2}$ is denoted as $\theta$. First, $\theta$ will increase from $0^{\circ}$ to $90^{\circ}$, represented as the blue angle in Figure \ref{fig:nonconvex_local}, and the angle $\alpha$ used to identify the angular separation and gradient for calculation of the the energy density equals $\theta$, which causes the energy density to increase. When the director rotation passes $90^{\circ}$, although the angle $\theta$ between the two directors keeps increasing, the angle $\alpha$ used to calculate the energy density is $\pi-\theta$ (the orange angle in Figure \ref{fig:nonconvex_local}) since physically the director has no direction. Thus, when $\theta$ increases from $90^{\circ}$ to $180^{\circ}$ the energy density decreases. In addition, the energy density will reach its maximum when the angle $\theta$ reaches $90^{\circ}$.

High director gradients, uniform along layers, and with little energy cost from such layers  may be expected in models with the above nonconvexity in the energy density. However, a state with a single disclination is the limit, as the layer width goes to zero, of a continuous, global director configuration that has a high, uniform director gradient in a section of a thin layer, transitioning to gradients of negligible magnitude in the rest of the layer (the parts of the layer $-l/2<y<l/2$ to the right and to the left of $x=0$, respectively, in Figure \ref{fig:gradient_ph_3}. Here $0<l\ll1$.). The transition region in the layer is the core of the disclination. Since the director configuration varies continuously, from its value on the top of the section of the layer with non-negligible gradient to its value at the bottom (of the same section of the layer) along any path going around the core, it is easy to see that such a global configuration has to contain substantial total energy on the whole (see Figure \ref{fig:gradient_ph_3}). Since it is physically reasonable to expect such director configurations to exist without imposed loadings, it is clear that the attainment of such states cannot be a question of global energy minimization, and almost definitely not in a model whose energy density is quadratic in the director gradients (as for example in the 1-constant Oseen-Frank energy density approximation in Leslie-Ericksen theory). It is also believed that disclination cores move under their mutual interaction, even in the absence of applied loads, with speeds unrelated to causes of orientational or positional inertia of the material. Indeed, the Ericksen-Leslie equations governing the director field are most often used without any orientational inertia. Moreover, it seems reasonable to develop models where motion of defects are allowed even in the absence of flow; as justification we quote the following excerpt from Ericksen \cite{ericksen1995remarks}, discussing parallel, straight disclinations:

``Saupe is very familiar with observations of disclinations of this kind, his own and those made by others. Typically, they are observed in specimens contained between a cover plate and a glass slide, in a polarizing microscope. Generally, they do move, but not alway[s] rapidly. There are empirical rules, of a topological nature, for determining the kinds that attract (or repel) each other, such as were discussed by Friedel [12], for example. \emph{As they move, they cause little or no flow; experimentalists tell me that it is hard to detect any so caused, although, they don't doubt that, in principle, there is some.}\footnote{The italicization here is ours.} Dynamical theory does involve viscous contributions, modifying the constitutive equations, etc., associated with the time rate of change of the director as well as the velocity field. From what I know of the theory and observations, I don't believe that one can use equilibrium theory to analyze these phenomena...."

Based on the above observations, it appears to us that accommodating general disclination dynamics, including that of half-integer strength disclinations, within the structure of Leslie-Ericksen theory or Ericksen \cite{ericksen1991liquid} is probably an unattainable goal. Thus, we augment the kinematics (and dynamical structure) of Leslie-Ericksen theory with an additional field that allows for equilibria and motion of non-singular disclinations, as described below. 

We follow the primarily kinematical ideas presented in \cite{pourmatin2012fundamental} and restrict attention to the planar case.  We assume that the director $\bfn$ is a unit vector which therefore can be parametrized with an angle field, i.e., $\bfn = \cos\theta \bfe_1 +\sin\theta \bfe_2$. As in \cite{pourmatin2012fundamental} we introduce a \emph{layer field} $\bflambda$. Furthermore, we assume that the energy $E$, depending on the fields $grad\, \theta$ and $\bflambda$, takes the form

\begin{equation}\label{eqn:gradient_energy}
E = \int_V\left[\frac{K}{2}|\grad\theta - \bflambda|^2+\frac{\epsilon}{2}|\curl\bflambda|^2+\gamma f(\bflambda)\right]dv. 
\end{equation}
Here $K > 0$ is a constant parameter representing 1-constant Oseen-Frank elasticity. The parameter $\epsilon := KCa\xi^2$ depends on the disclination core width $\xi > 0$, a fundamental length scale of the model, a non-dimensional parameter $C$ to control the magnitude of the core energy, and the width of the layer $l = a\xi$, where $a \geq 0$ is a non-dimensional scaling factor. To allow for conventional expectations, we will accommodate the limit $a \rightarrow 0$ and still allow for finite energy disclination solutions (recall that $\xi>0$). The parameter $\gamma$ is defined as $\gamma  := \frac{2PK \hat k}{a\xi^2}$, with $P$ being a non-dimensional penalty parameter, and $\hat k := \frac{1}{2}$. $f$ is a multi-well function with minima of wells at integer multiples of $\frac{2\pi \hat k}{a\xi}$. A typical candidate for the function $f$ that we use in this work is 
\begin{equation} \label{eqn:eta}
f(\bflambda) =1-\cos\left(2\pi \frac{|\bflambda|}{\left( \frac{2\pi \hat k}{a\xi} \right) }\right) = 1-\cos\left( \xi |\bflambda|\left(\frac{\hat k}{a}\right)^{-1} \right).
\end{equation}
Thus  $|\bflambda| = \frac{2\pi k}{a\xi}$ for a strength-$k$ disclination, where $k$ is any integer-multiple of $\hat k = \frac{1}{2}$, minimizes this symmetry related, non-convex energy density term.

%%The parameter $\gamma$ is defined as $\gamma  := \frac{2PK|k|}{ac^2}$, with $P$ %%being a non-dimensional penalty parameter, and $k$ is the disclination strength. $f%%$ is a multi-well function with minima of wells at values of $\frac{2\pi k}{ac}$. %%A typical candidate for the function $f$ that we use in this work is 
%%\begin{equation} \label{eqn:eta}
%%f(\bflambda) =1-cos\left(2\pi \frac{|\bflambda|}{|\frac{2\pi k}{ac}|}\right) = 1-%%cos\left( c |\bflambda|\left(\frac{|k|}{a}\right)^{-1} \right),
%%\end{equation}
%%where $\frac{2\pi k}{ac}$ are the well values corresponding to strength-$k$ %%disclinations. In this work, the disclination strength $k$ is assumed to be any %%integer multiple of $\frac{1}{2}$. 

The intuition behind why the energy (\ref{eqn:gradient_energy}) can serve to represent disclinations is as follows. For a fixed specification of the field $\bflambda$ very similar to as specified in (\ref{eqn:lambda_ic_spec}), it is shown in detail in \cite{pourmatin2012fundamental} that the director and energy density fields of disclination defects are well captured by a model whose static governing equation is the Euler-Lagrange equation of the energy (\ref{eqn:gradient_energy}) for variations only in the field $\theta$. All that then remains to be convinced of is that the configuration of $\bflambda$ specified in (\ref{eqn:lambda_ic_spec}) is close to one that extremizes the energy (\ref{eqn:gradient_energy}), with the associated $\theta$ field being the solution of the Euler-Lagrange equation of (\ref{eqn:gradient_energy}) for $\theta$-variations (i.e. the right-hand-side of (\ref{eqn:grad_dimen})$_1$). This is easy to see as the magnitude of the $\bflambda$ field in (\ref{eqn:lambda_ic_spec}) does lie in the wells of the function $f$. The term penalizing $\curl \bflambda$ in (\ref{eqn:gradient_energy}) smooths out the transition of $\bflambda$ within the layer, as does the elastic energy term (the first term of the integrand in (\ref{eqn:gradient_energy})\footnote{In this special case where $\bflambda$ is expected to have only one non-vanishing component.}). This transition layer in $\curl \bflambda$ within the layer signifies the core region of a disclination, and the parameter $\epsilon$ characterizes the core energy of the defect, with $\sqrt{\epsilon}$ roughly setting the core-width in the equilibrium solution. It is to be noted that the core energy (i.e the second term in the integrand in (\ref{eqn:gradient_energy})) does not penalize the vertical gradients of the $\bflambda$ field in (\ref{eqn:lambda_ic_spec}) across the horizontal boundaries of the layer.

The various parameters of the model have the following physical dimensions: $[E] = Force \times Length$, $[K] =  Force$, $[\bflambda] = Length^{-1}$, $[\epsilon] =  Force \times Length^2$, $[\xi] = Length$, $[\gamma] = Force \times Length^{-2}$.

To obtain the gradient flow equations, the first variation of the energy $E$ is,
\begin{equation*}
\delta E = \int_V \left\{K (\theta_{,i}-\lambda_i)\delta\theta_{,i}-K(\theta_{,i}-\lambda_i)\delta\lambda_i+\epsilon e_{ijk}e_{irs}\lambda_{s,r}\delta\lambda_{k,j}+{\gamma}\frac{\partial f}{\partial \lambda_i}\delta\lambda_i\right\}dv.
\end{equation*}
Integrate by parts and assume boundary terms to vanish. Then we obtain
\begin{equation*}
\delta E = \int_V \left\{K (-\theta_{,ii}+\lambda_{i,i})\delta\theta_{i}+\left({\gamma}\frac{\partial f}{\partial \lambda_k}-K(\theta_{,k}-\lambda_k)-\epsilon e_{ijk}e_{irs}\lambda_{s,rj}\right)\delta\lambda_k\right\} dv.
\end{equation*}
Extracting terms for $\theta$ and $\lambda$ respectively, we obtain the evolution equations
\begin{eqnarray} \label{eqn:grad}
\begin{aligned}
\frac{\partial\theta}{\partial t} &= M_1 K(\theta_{,ii}-\lambda_{i,i})  \\
\frac{\partial \lambda_k}{\partial t} &= M_2\left(-{\gamma}\frac{\partial f}{\partial \lambda_k}+K(\theta_{,k}-\lambda_k)+\epsilon e_{ijk}e_{irs}\lambda_{s,rj}\right)
\end{aligned}
\end{eqnarray}
\textit{Here $M_1$ and $M_2$ represent mobility coefficients. Their physical dimensions are $[M_1] = Velocity \times Length \times Force^{-1}$ and $[M_2] = Velocity \times Force^{-1} \times Length^{-1}$.}

To non-dimensionalize the above equations, we introduce the following dimensionless variables,
\begin{eqnarray*}
\tilde{x_i} = \frac{1}{\xi}x_i;\quad \tilde{s} = K M_2 t; \quad \tilde{\gamma} = \frac{\xi^2}{K}\gamma = \frac{2P \hat k}{a}; \quad \tilde{\bflambda} = \xi \bflambda; \quad    \tilde{\epsilon} = \frac{1}{K \xi^2} \epsilon= Ca
\end{eqnarray*}
Also, we assume $M_1=M_2\xi^2$; this is justified by the fact that we view the gradient flow equation for $\theta$ as simply a device to achieve equilibrium in $\theta$ with $\bflambda$ fixed. Indeed, in all gradient-flow results presented in the following, we have checked our results to ensure that they are invariant to solving directly for the equilibrium of $\theta$ for fixed $\bflambda$. Then the non-dimensionalized version of (\ref{eqn:grad}) reads as:
\begin{eqnarray*}
\begin{aligned}
\frac{\partial\theta}{\partial \tilde{s}} &= (\theta_{,ii}-\tilde{\lambda}_{i,i})  \\
	\frac{\partial \tilde{\lambda}_k}{\partial \tilde{s}} &= -{\tilde{\gamma}}\frac{\partial f}{\partial \tilde{\bflambda}_k}+(\theta_{,k}-\tilde{\lambda}_k)+\tilde{\epsilon} e_{ijk}e_{irs}\tilde{\lambda}_{s,rj}  \\
\text{where} \quad f &=1-\cos\left( |\tilde{\bflambda}|\left(\frac{\hat k}{a}\right)^{-1}\right).
\end{aligned}
\end{eqnarray*}
After substituting the expressions for $\tilde{\gamma}$ and $\tilde{\epsilon}$, the nondimensional evolution equations are
\boxalign{
\begin{eqnarray} \label{eqn:grad_dimen}
\left.
\begin{aligned}
\frac{\partial\theta}{\partial {s}} &= (\theta_{,ii}-{\lambda}_{i,i})  \\
\frac{\partial {\lambda}_k}{\partial s} &= -2P\sin\left(|\bflambda|\left(\frac{\hat k}{a}\right)^{-1}\right)\hat{\lambda}_k + (\theta_{,k}-{\lambda}_k)+Ca e_{ijk}e_{irs}{\lambda}_{s,rj} 
\end{aligned}
\right\} \text{in the body $B$}
\end{eqnarray}}
where $\hat{\bflambda}$ is the unit vector in the direction of $\bflambda$, and \emph{we have removed all tildes for convenience}. For the purposes of Section \ref{sec:gradient_flow}, all symbols henceforth represent non-dimensional quantities.

In all that follows, we think of our computational solutions employing $a > 0$ as approximations of the limiting case $a = 0$ which assigns no physical significance to the layer. In Section \ref{sec:energy_indep} we show that our equilibrium disclination solutions show a trend to finite total energy even in that limit. \emph{Thus, the nondimensionalized model effectively has two non-dimensional constants, $C, P$.}

\section{Static results from gradient flow}\label{sec:gradient_flow}

We evaluate the gradient flow model by presenting results for straight wedge disclinations. All calculations are done on a square domain of non-dimensional extent $L\times L$ with $L=50$. Unless otherwise specified, we assume $a=1$, $C=1$ and $P=20$. 

We compute results for four cases in this section, namely strength half disclinations ($k= \pm 0.5$) and strength one disclinations ($k=\pm 1$). The initial condition for the layer field for calculations in this section is defined as 
\begin{equation}\label{eqn:lambda_ic_spec}
\bflambda = \begin{cases}
\frac{-2k\pi}{a}\bfe_2, & \text{ if $|x_2|<{\frac{a}{2}}$  and $x_1 > 0$} \\
0, & otherwise.
\end{cases}
\end{equation}

The initial condition on the $\theta$ field is based on Frank's solution \cite{frank1958liquid}, 
\begin{equation}\label{eqn:gradient_ini}
\theta = k\ \tan^{-1} \left(\frac{x_1}{x_2}\right) + Q 
\end{equation}
\textit{where $Q$ is a constant. Here, $Q$ is set to be $-\frac{\pi}{4}$ and the range of the $arctan$ function is assumed to be $[-\frac{3\pi}{2}, \frac{\pi}{2}]$}.

A zero-moment boundary condition is imposed to solve for the $\theta$ field, for each given $\bflambda$. In the following calculations, $\theta$ at the boundary point ($x_1=25, x_2=-\frac{a}{2}$) is fixed to be $0$.

\subsection{Strength $+\frac{1}{2}$ disclination} \label{sec:stat_1/2}

\begin{figure}[H]
\centering
\subfigure[Plot of $\lambda_2$ of initialization. $\bflambda$ is non-zero only inside the layer, with $\lambda_2$ as only non-zero component.]{
\includegraphics[width=0.4\linewidth]{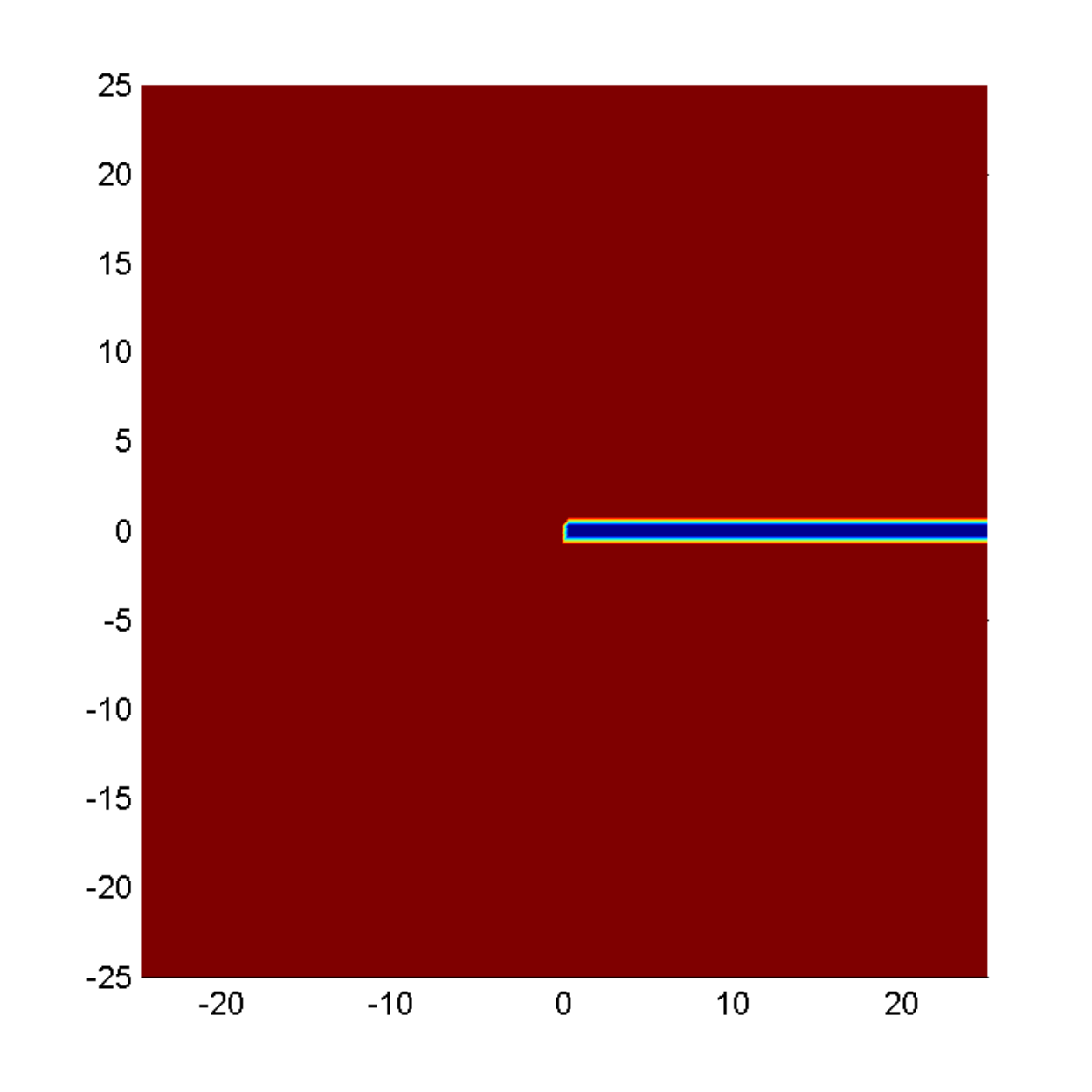}
\label{fig:gradient_ph_1}}\qquad
\subfigure[The energy density plot for this positive half disclination.]{
\includegraphics[width=0.4\linewidth]{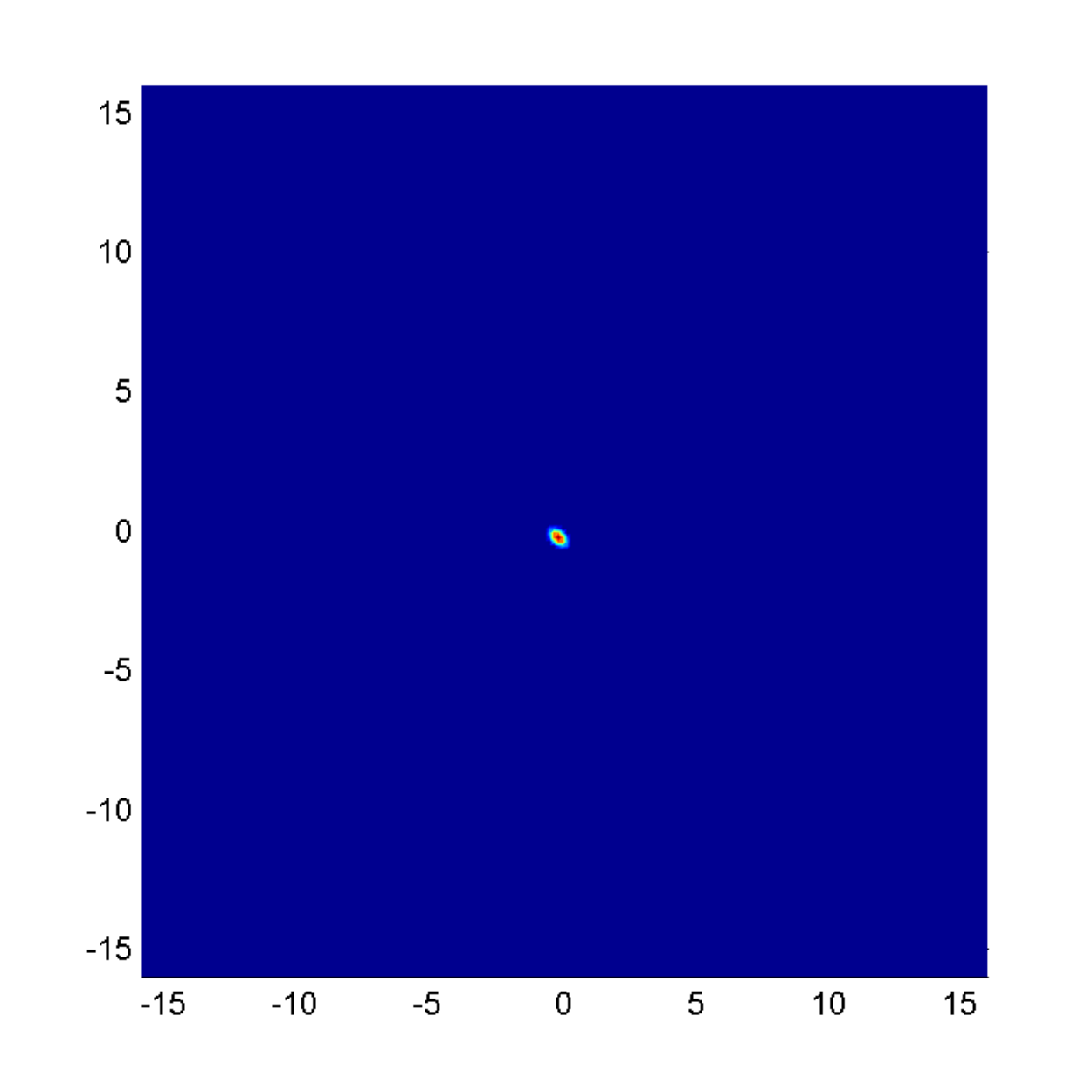}
\label{fig:gradient_ph_4}}
\subfigure[Director field $\theta$ on the whole body at $l/L=0.02$.]{
\includegraphics[width=0.38\textwidth]{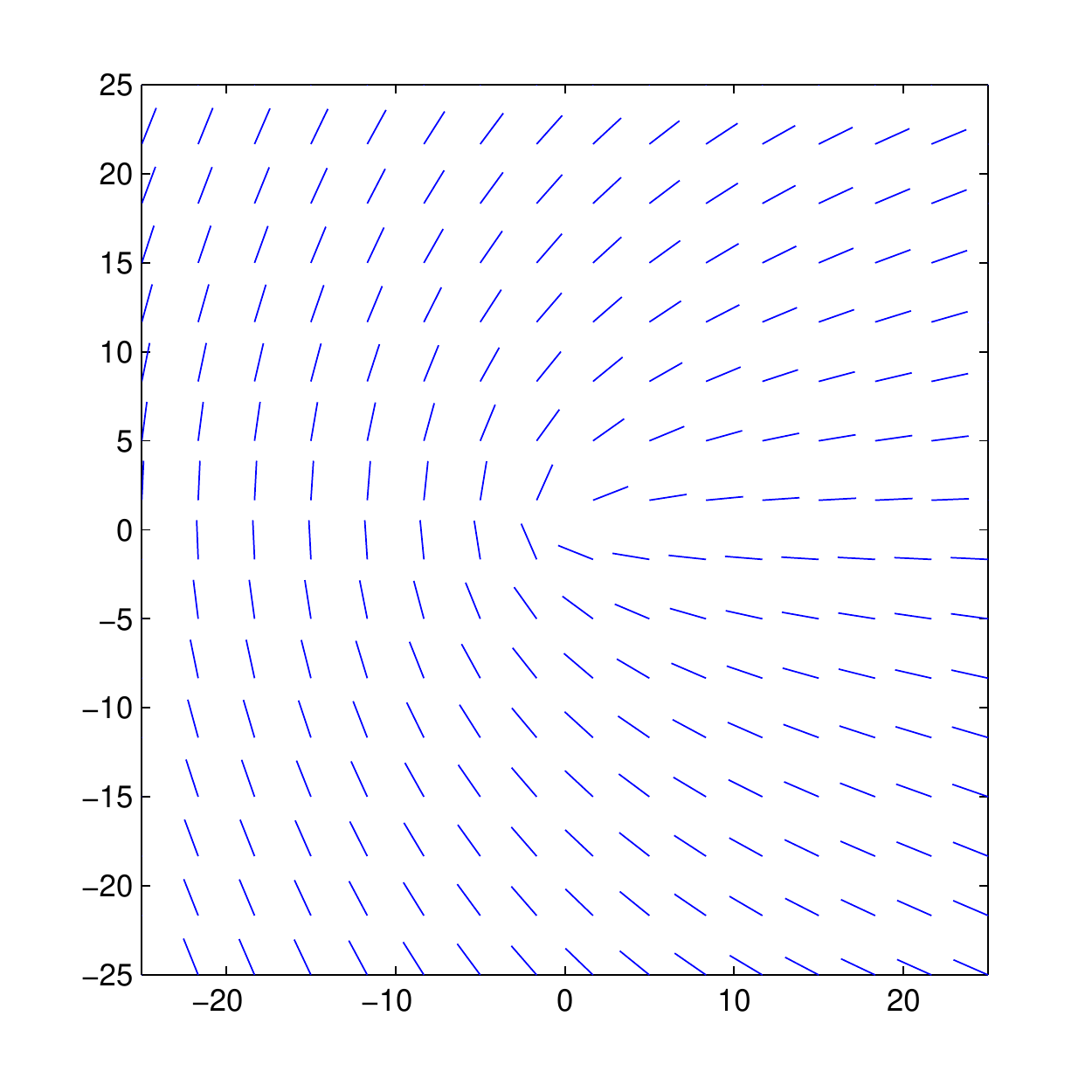}
\label{fig:gradient_ph_2}}\qquad
\subfigure[Director field $\theta$ on the whole body at $l/L=0.005$. ]{
\includegraphics[width=0.38\textwidth]{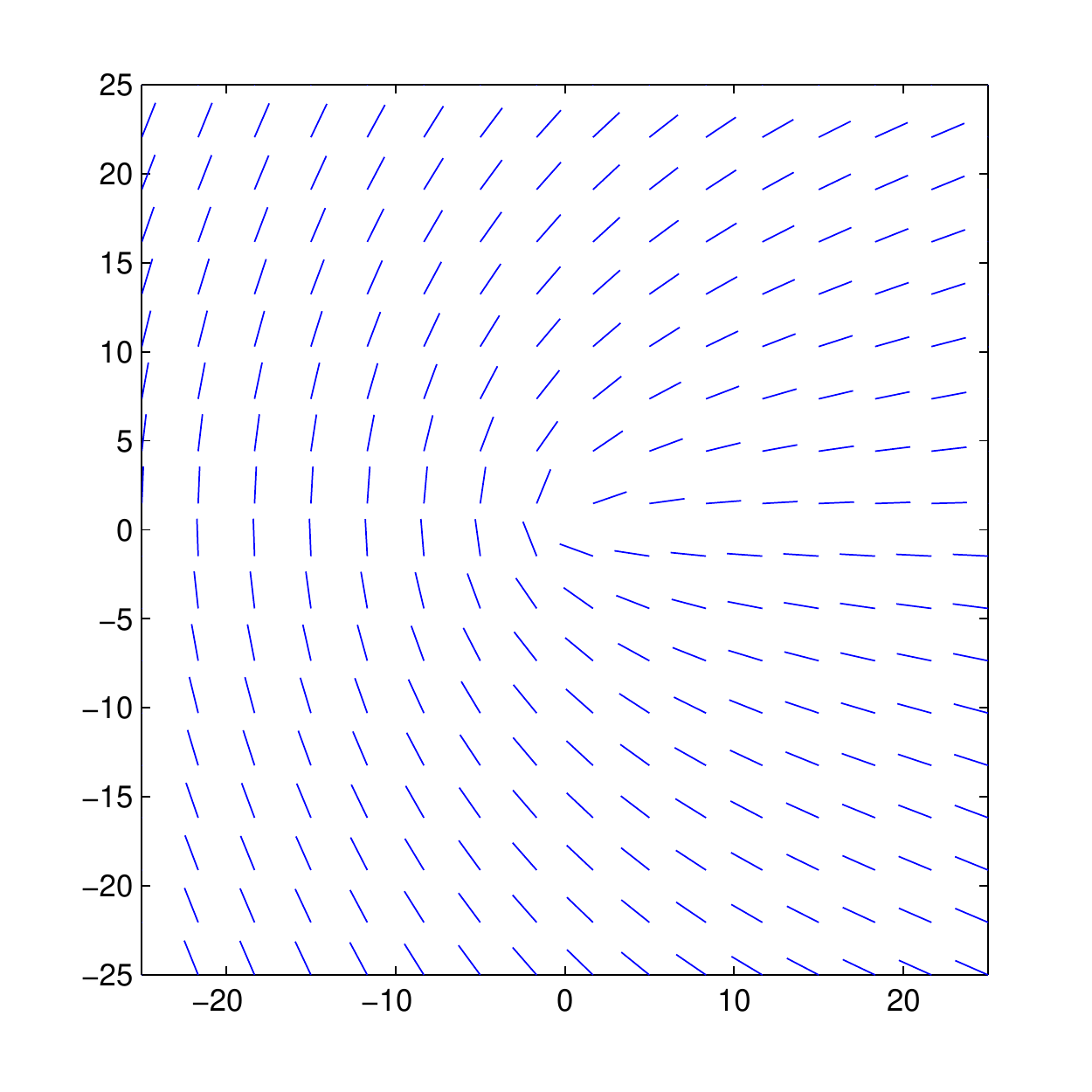}
\label{fig:gradient_ph_2}}\qquad
\subfigure[Magnified view of the director field at $l/L=0.005$ near the core.]{
\includegraphics[width=0.55\textwidth]{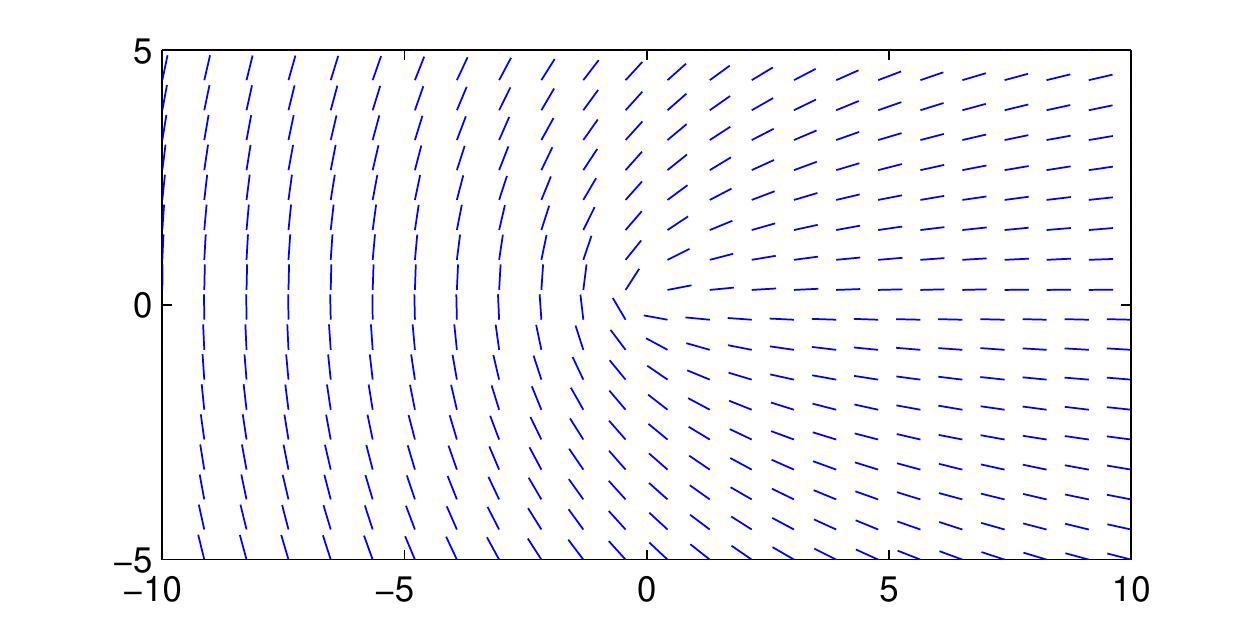}
\label{fig:gradient_ph_3}}
\caption{Results for strength $+\frac{1}{2}$ disclination.}\label{figure_3}
\end {figure}

For a positive half disclination, $k=0.5$, the director rotates $\pi$ radians clockwise while traversing a loop clockwise from the bottom of the layer to the top, starting from an orientation of $\theta = \pi$ with respect to the positive x-axis at the bottom of the layer. The initial prescription of the $\bflambda$ field is shown in Figure \ref{fig:gradient_ph_1}. $\lambda_2$ is the only non-zero component inside the layer and thus the director distortion field is not $curl$-free at the disclination core where the layer terminates. Figure \ref{fig:gradient_ph_4}, \ref{fig:gradient_ph_2} and \ref{fig:gradient_ph_3} are computational equilibria obtained from the gradient flow evolution from the initial conditions described in (\ref{eqn:gradient_ini}). Equilibrium is considered achieved if the magnitudes of the `rates' of evolution become less than $10^{-4}$ for both $\theta$ and $\bflambda$ on the entire domain. The director field over the whole body is represented with dashed line field in Figure \ref{fig:gradient_ph_2}. A magnified view of the core area is shown in Figure \ref{fig:gradient_ph_3}.\emph{In this paper, the spacing of the dashed curves do not represent spacing of the computational mesh.} Figure \ref{fig:gradient_ph_4} shows the energy density distribution for this case. The energy is concentrated in the core and the location of the layer is energetically `invisible'. 

\begin{figure}[H]
\centering
\subfigure[Magnified view of the equilibrated director field near the layer for $+\frac{1}{2}$ disclination. The director turns in the layer but the corresponding energy is as it should be.]{
\includegraphics[width=0.8\linewidth]{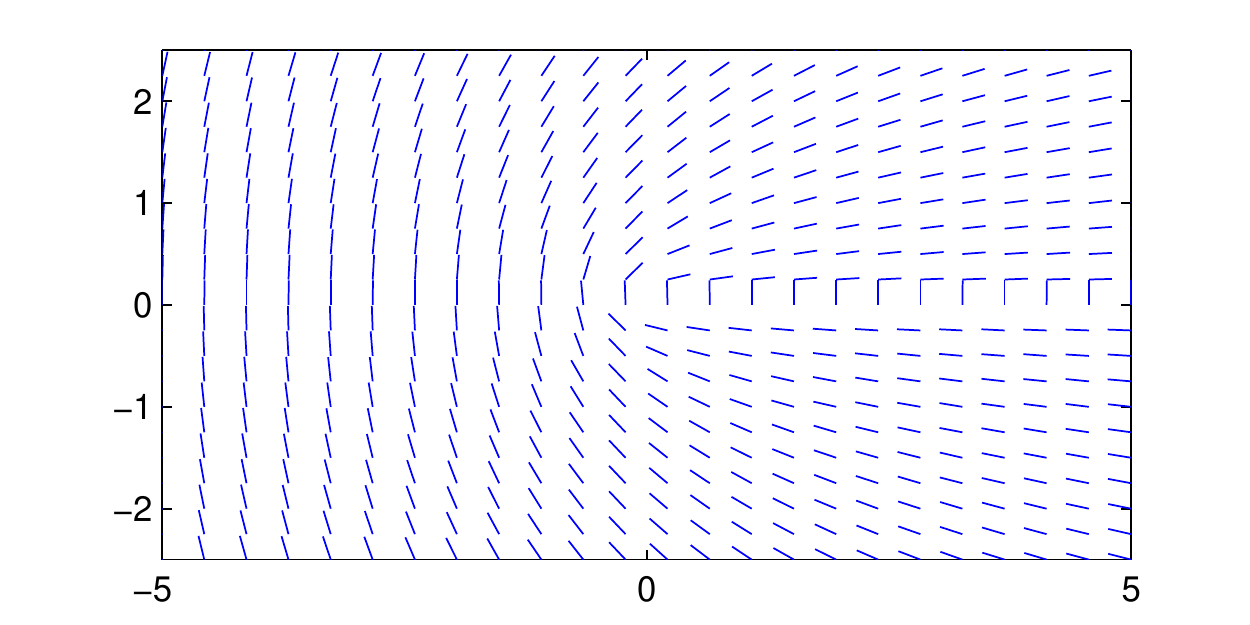}
\label{fig:gradient_ph_5}}
\subfigure[Magnified view of the energy density on the same scale as \ref{fig:gradient_ph_5}.]{
\includegraphics[width=0.8\linewidth]{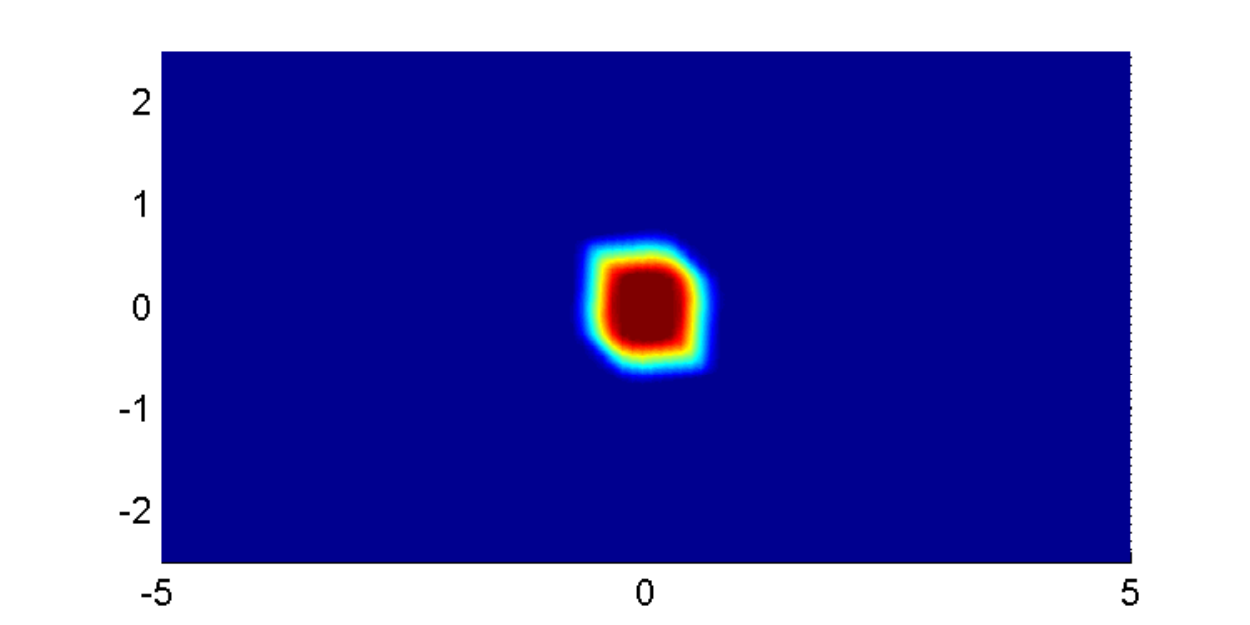}
\label{fig:gradient_ph_6}}
\caption{Magnified view of director field and energy density field near the layer for a $+\frac{1}{2}$ disclination at $l / L=0.005$.}
\end {figure}

Figure \ref{fig:gradient_ph_5} shows the director field within the layer  at $l/L=0.005$. As shown in Figure \ref{fig:gradient_ph_5}, the director field actually rotates within the layer \emph{but with no energy cost.} In the limit $a \rightarrow 0$ this `rotation' of the director field in the layer becomes `invisible', portraying a discontinuity without energy cost, except at the core which is physically realistic.

\subsection{Strength $-\frac{1}{2}$ disclination} \label{sec:stat_-1/2}

\begin{figure}[H]
\centering
\subfigure[Plot of $\lambda_2$ of initialization. $\bflambda$ is non-zero only inside the layer where $\lambda_2$ is the only non-zero component. Compared to the positive half disclination \ref{fig:gradient_ph_1}, $\bflambda$ in this case has the same magnitude but opposite sign.]{
\includegraphics[width=0.4\linewidth]{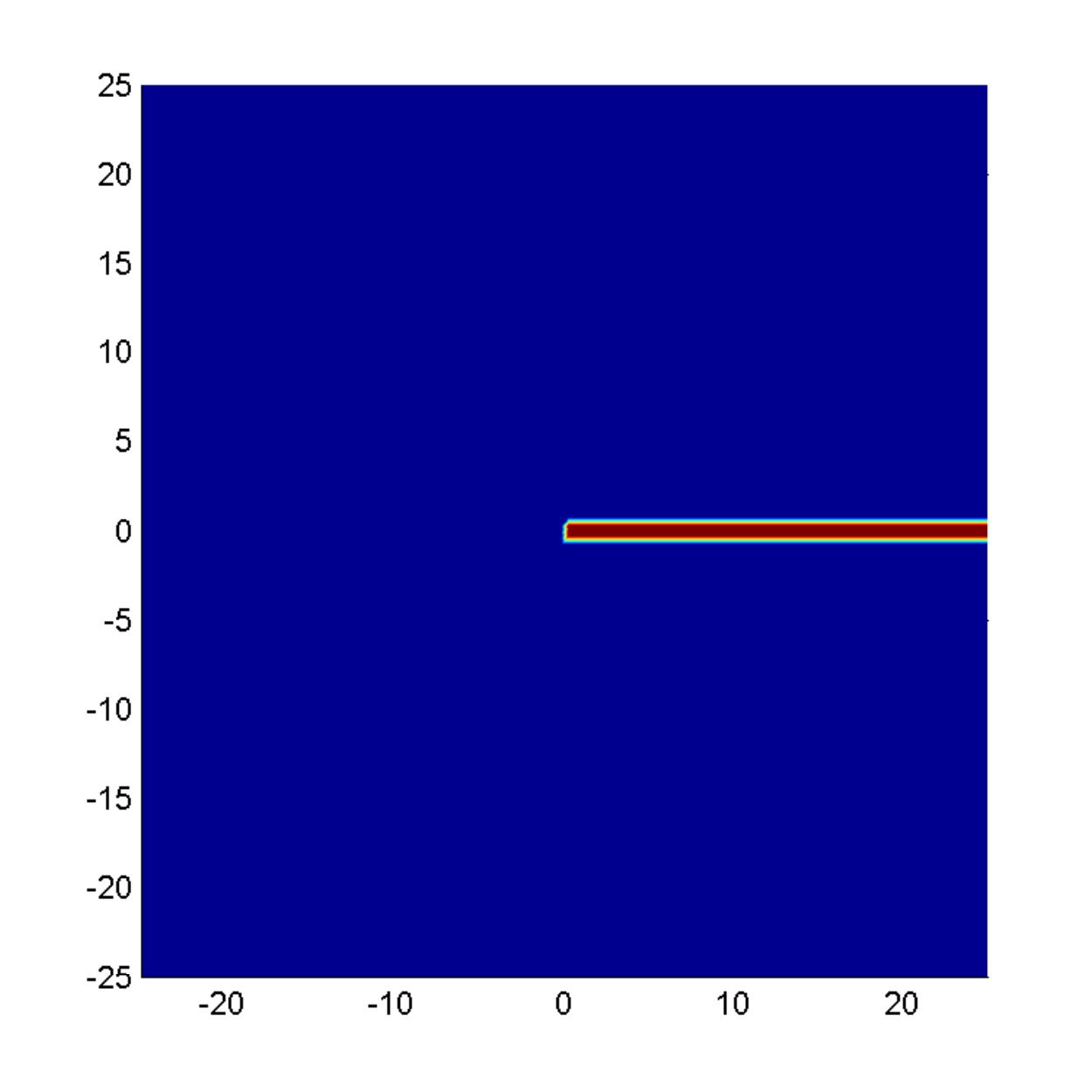}
\label{fig:gradient_nh_1}}\qquad
\subfigure[The energy density plot for this strength $-\frac{1}{2}$ disclination.]{
\includegraphics[width=0.4\linewidth]{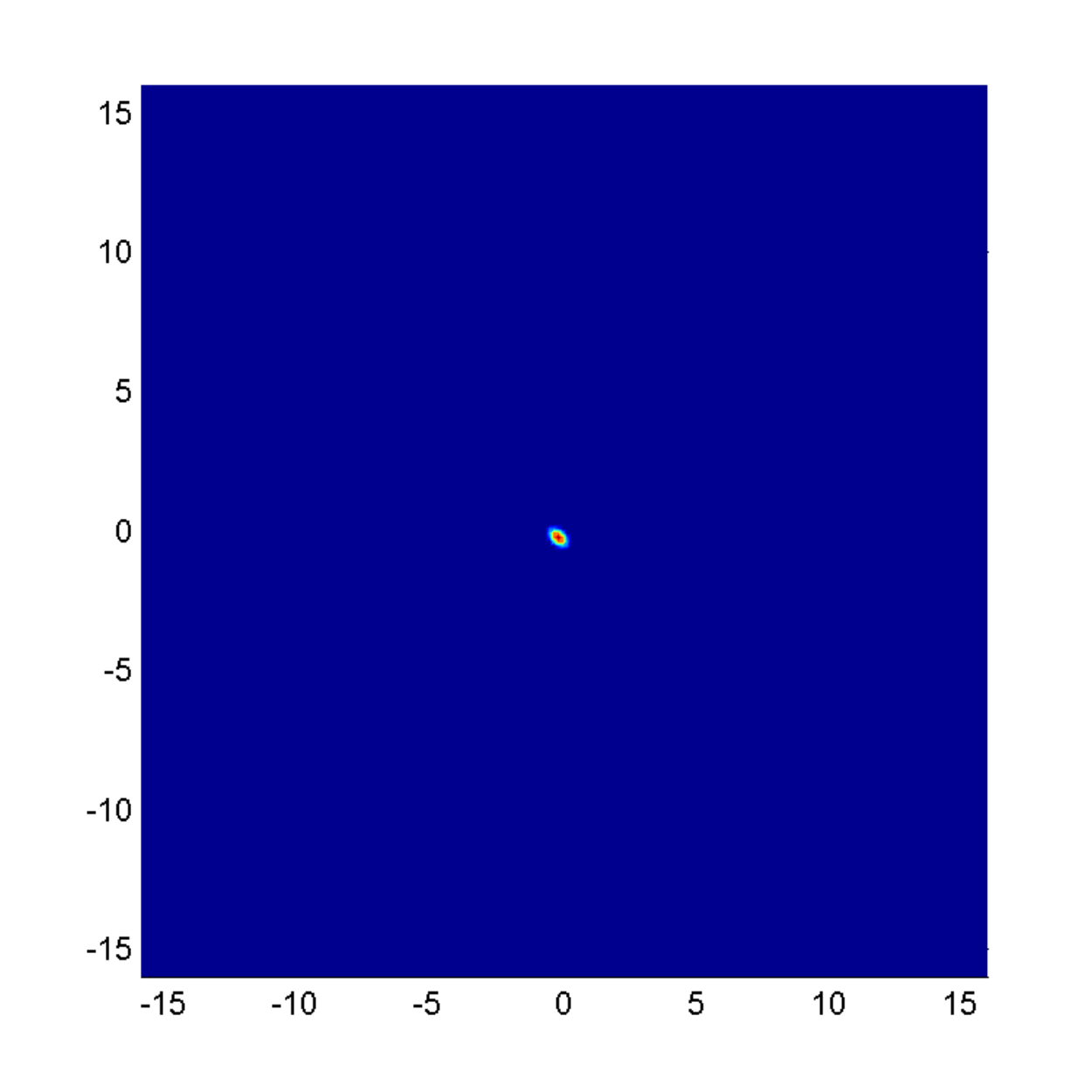}
\label{fig:gradient_nh_4}}
\subfigure[Director field $\theta$ on the whole body at $l/L=0.02$.]{
\includegraphics[width=0.38\textwidth]{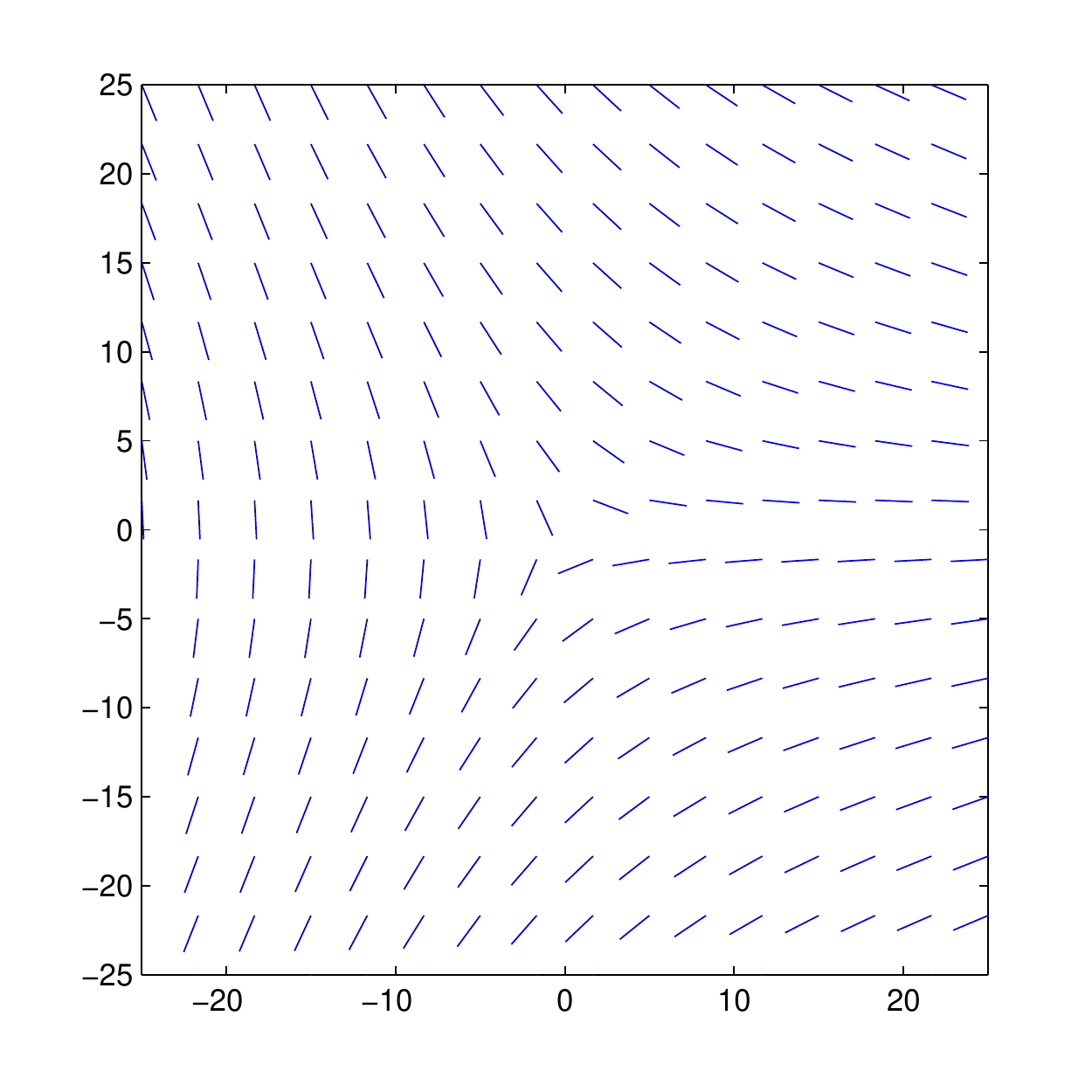}
\label{fig:gradient_nh_2}} \qquad
\subfigure[Director field $\theta$ on the whole body at $l/L=0.005$. ]{
\includegraphics[width=0.38\textwidth]{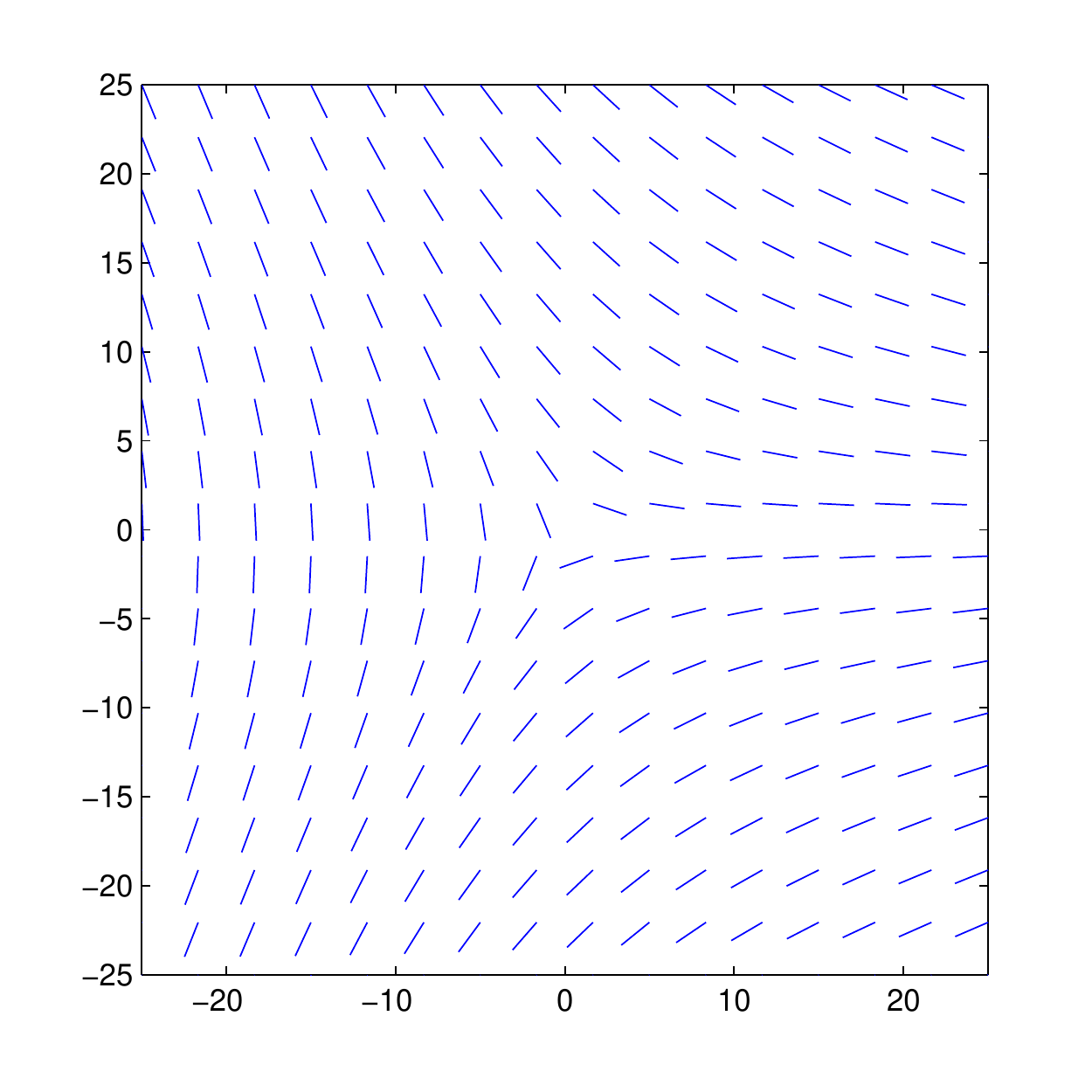}
\label{fig:gradient_nh_2}} \qquad
\subfigure[Magnified view of the director field at $l/L=0.005$ near the core.]{
\includegraphics[width=0.55\textwidth]{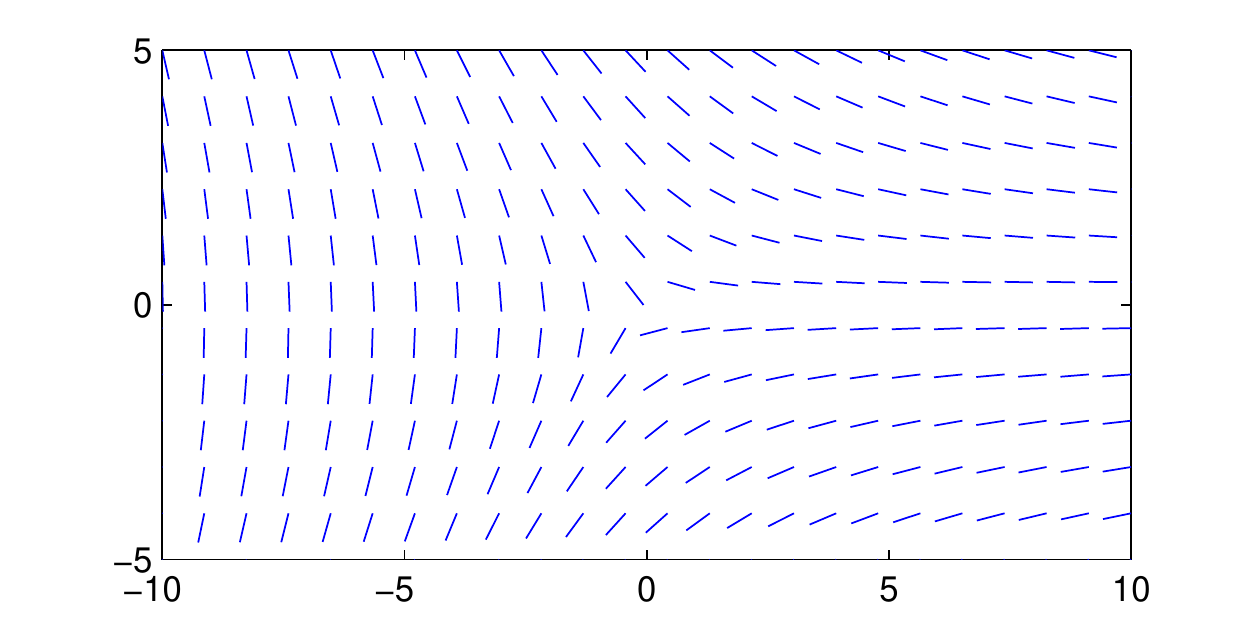}
\label{fig:gradient_nh_3}}
\caption{Results for strength $-\frac{1}{2}$ disclination.}
\end {figure}

For the negative half disclination $k=-0.5$, the director rotates $\pi$ radians anticlockwise while traversing a loop clockwise from the bottom of the layer to the top, starting from a $\theta = \pi$ orientation with respect to the positive $x$-axis at the bottom of the layer.  Figure \ref{fig:gradient_nh_1} shows the initial condition on the $\bflambda$ field for this case. The prescribed value of $\bflambda$ inside the layer has the same magnitude as for the positive half disclination, but with opposite sign. Figure \ref{fig:gradient_nh_2} shows the equilibrated director field over the whole body. A magnified view of the core is shown in Figure \ref{fig:gradient_nh_3}. Figure \ref{fig:gradient_nh_4} shows the energy density distribution for the equilibrium of this case. 

\subsection{Strength $\pm 1$ disclination} \label{sec:stat_1}

\begin{figure}[H]
\centering
\subfigure[Director field $\theta$ for a $+1$ disclination.]{
\centering
\includegraphics[width=0.4\textwidth]{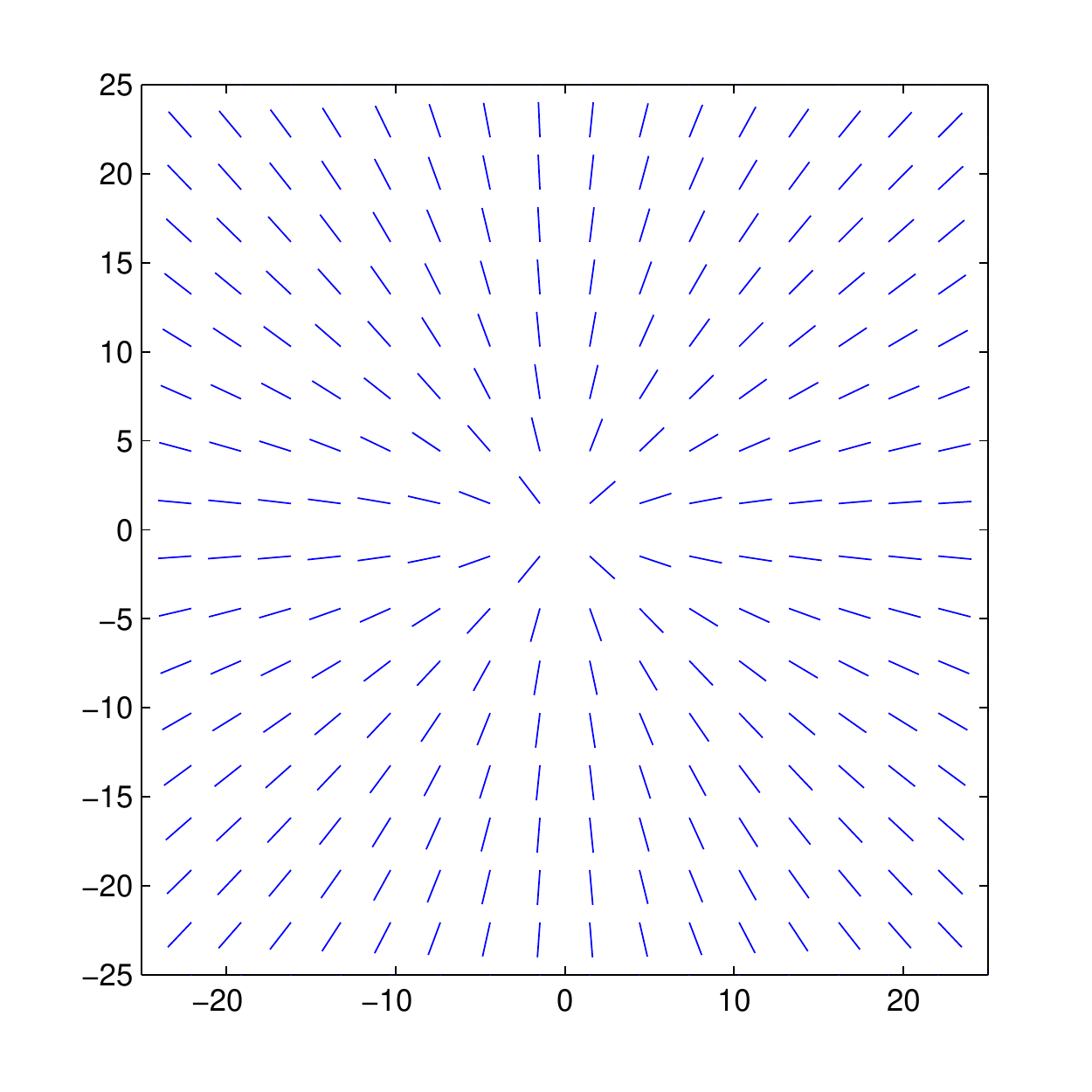}
}\qquad
\subfigure[Director field $\theta$ for a $-1$ disclination.]{
\includegraphics[width=0.4\textwidth]{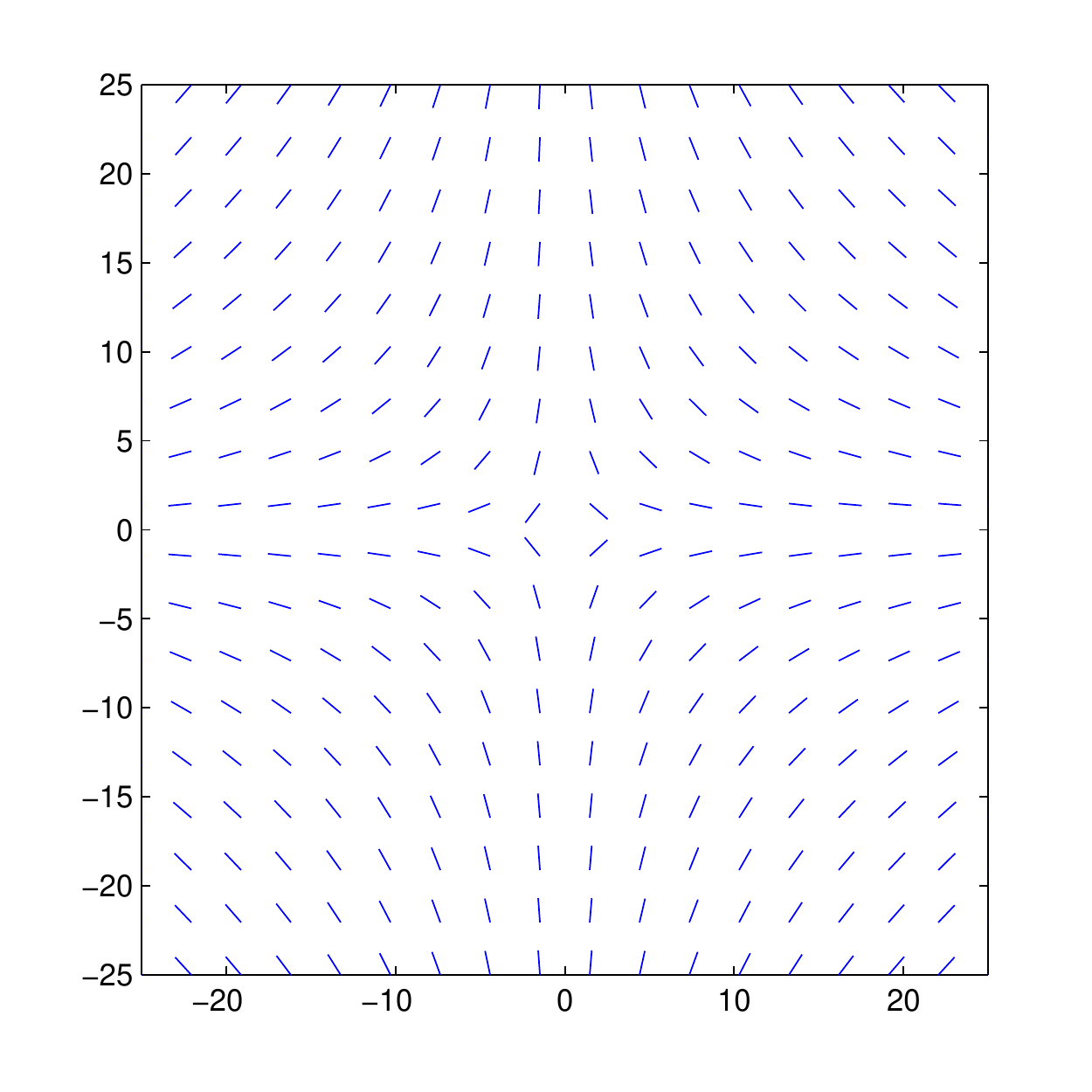}
}
\caption{The equilibrated director results for $\pm 1$ disclinations.}\label{fig:gradient_one}
\end {figure}

Now consider $k=\pm 1$, which implies a director rotation of $2\pi$ radians across the layer. Following the definition of $\bflambda$, we can prescribe $\bflambda$ fields for one disclination as well. Figure \ref{fig:gradient_one} presents the equilibrated director results of $\pm 1$ disclinations. Since strength $\pm1$ disclinations contain higher energy than the sum of the total energies of two half disclinations, strength $\pm1$ disclinations are not stable and tend to dissociate into two strength $\pm \frac{1}{2}$ disclinations. The capability of our model in representing this physical process will be discussed in Section \ref{sec:one_split}.

\subsection{Comparisons with Frank's analytical solution}

\begin{figure}[H]
  \centering
\subfigure[Overall energy density comparison along $x$ axis for strength $+\frac{1}{2}$ disclination.]{
\includegraphics[width=0.4\textwidth]{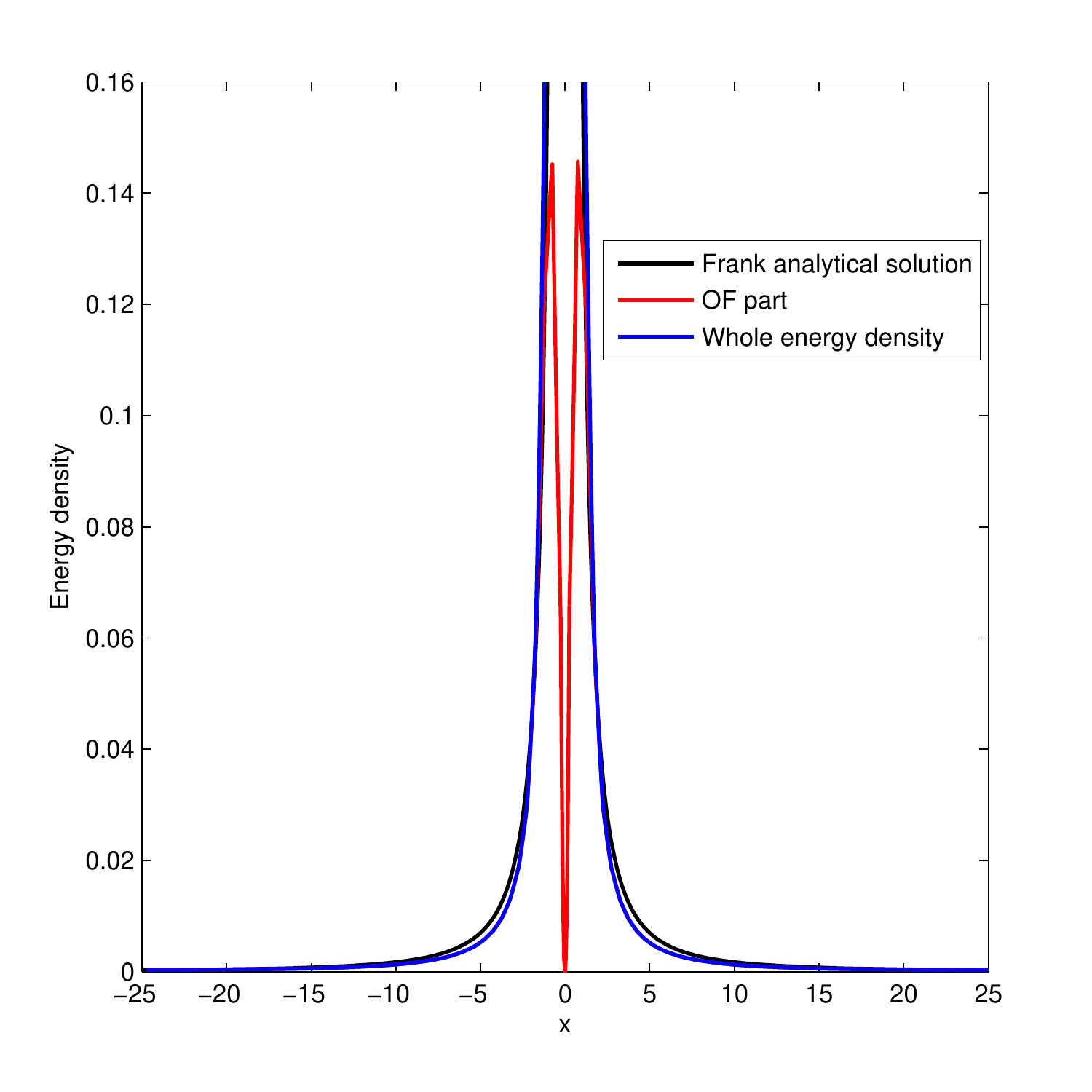}
}\qquad
\subfigure[Energy density comparison along $x$ axis near strength $+\frac{1}{2}$ disclination core.]{
\includegraphics[width=0.4\textwidth]{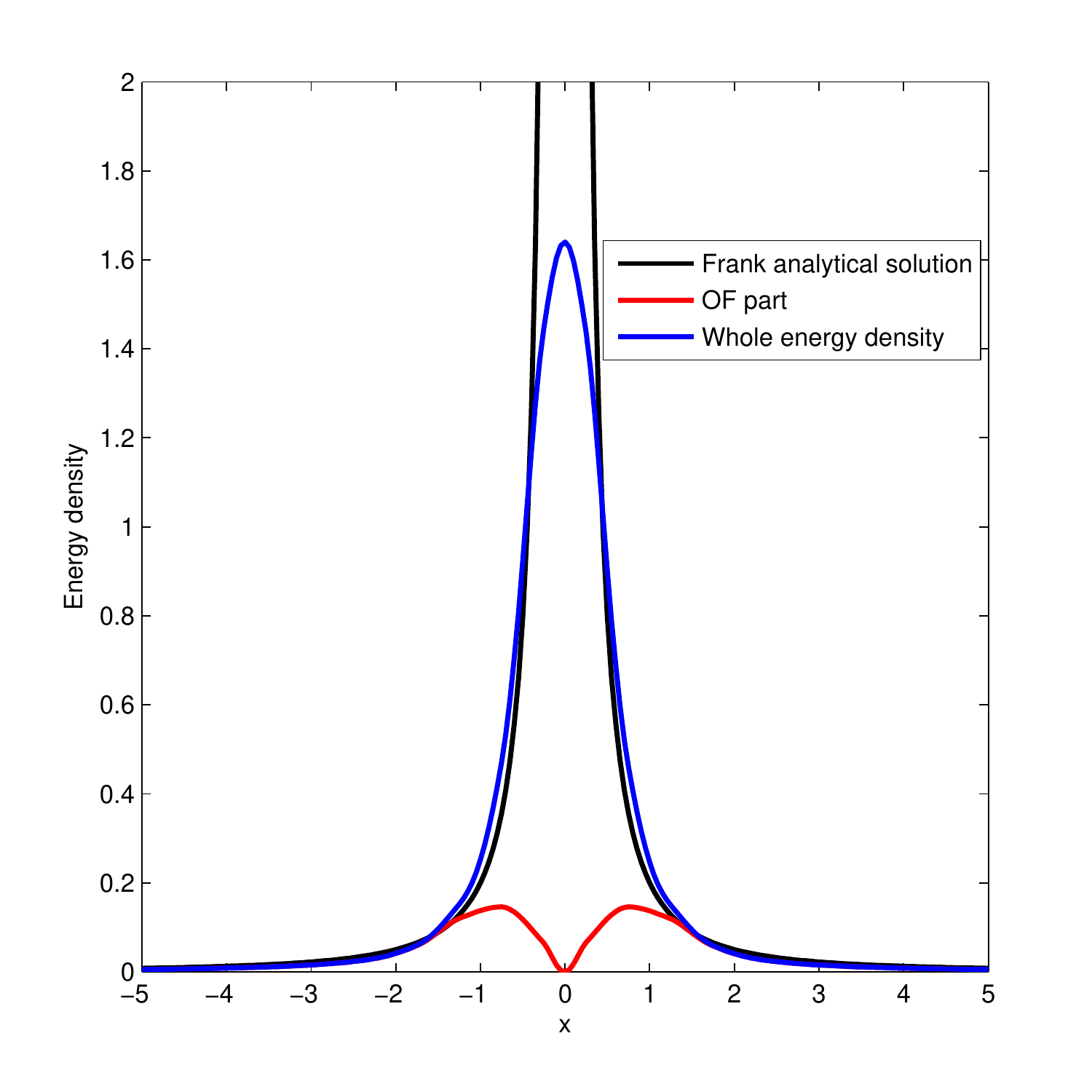}
}
\subfigure[Overall energy density comparison for strength $+1$ disclination.]{
\includegraphics[width=0.4\textwidth]{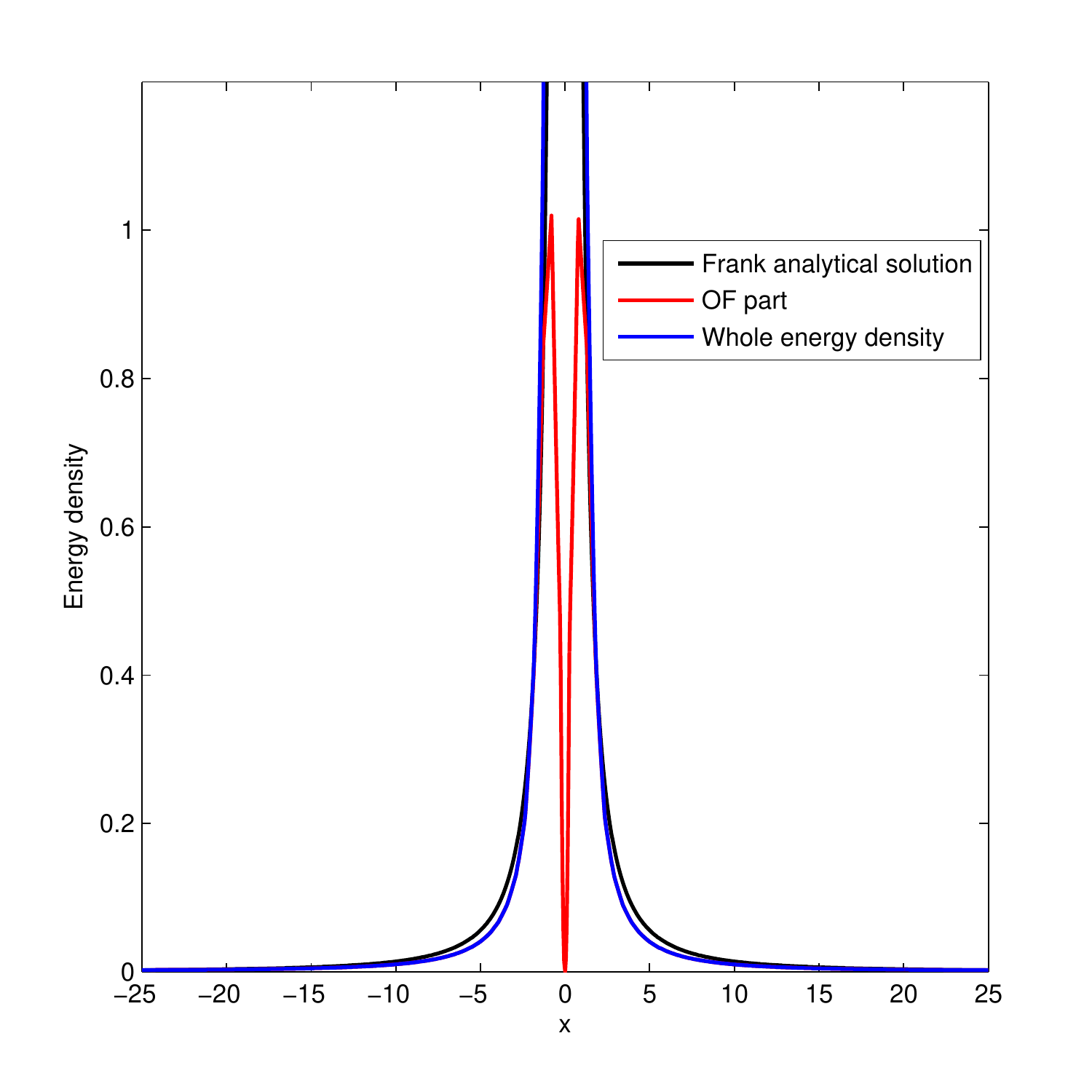}
}\qquad
\subfigure[Energy density comparison near strength $+1$ disclination core.]{
\includegraphics[width=0.4\textwidth]{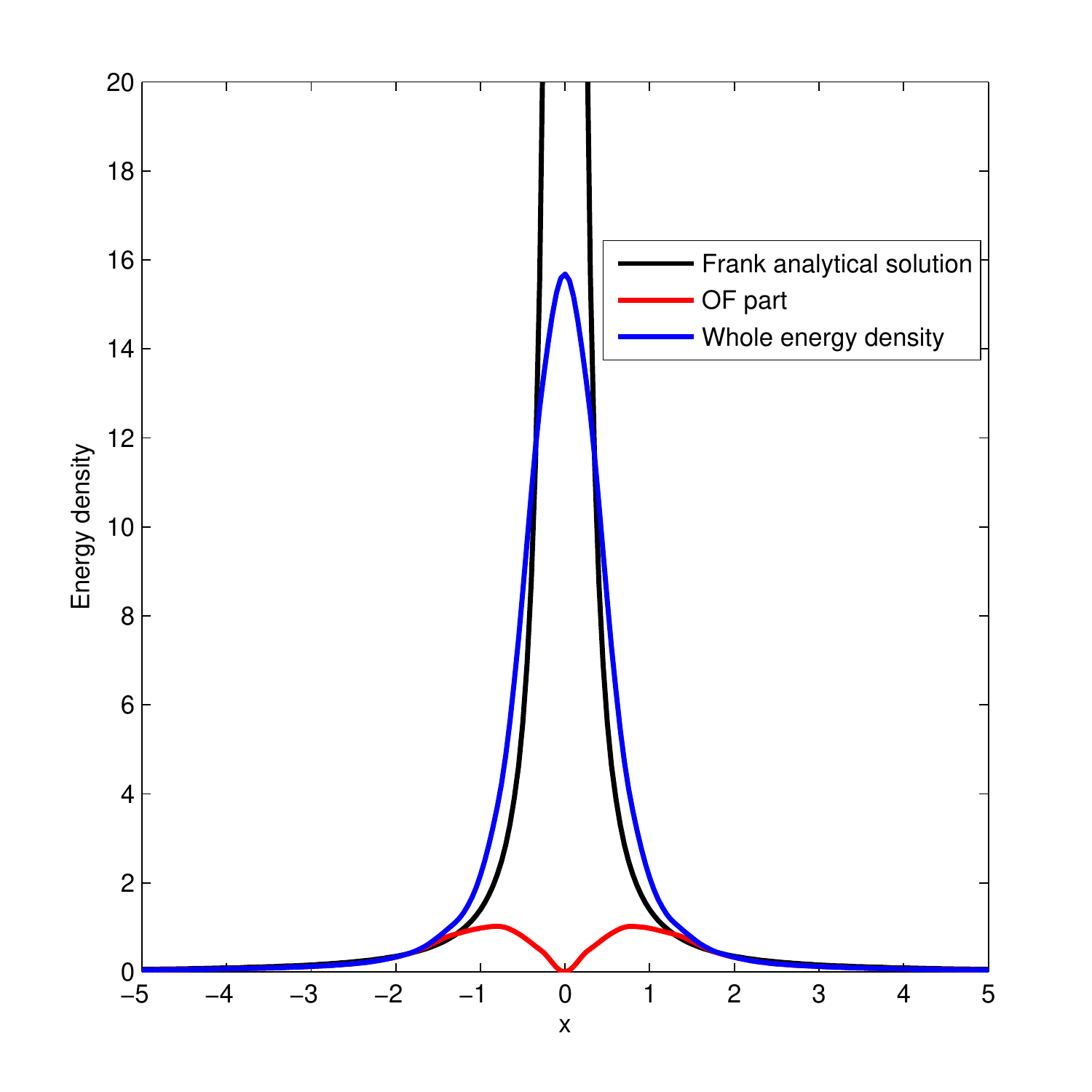}
}
\caption{Energy density comparisons between Frank analytical results and our results along $x$ axis, in both overall domain and near-core area, indicating a good agreement.}
\label{fig:comparison_energy_density}
\end{figure}

The angle of the director field with the $x_1$ axis in Frank's solution \cite{frank1958liquid} is 
\begin{equation*}
\theta = K \tan^{-1} \left(\frac{x_1}{x_2}\right) + q
\end{equation*}
where $q$ is a constant. For the purpose of evaluating the energy for the domain involved, it suffices to consider $\grad \theta$ given as
\begin{equation*}
\frac{K}{r^2}(-x_1\bfe_1+x_2\bfe_2).
\end{equation*}
Thus the energy density variation along the $x_1$ axis of the domain for this solution is 
\begin{equation*}
\psi = \frac{1}{2} |\grad\theta|^2 = \frac{K^2}{2} \left(\frac{1}{x_1}\right)^2.
\end{equation*}
Figure \ref{fig:comparison_energy_density} shows the various contributions for the energy density in our model, as well as a comparison of the energy density field with that of the Frank analytical solution.

\begin{figure}[H]
  \centering
\subfigure[Overall energy density comparison along $y$ axis for strength $+\frac{1}{2}$ disclination.]{
\includegraphics[width=0.4\textwidth]{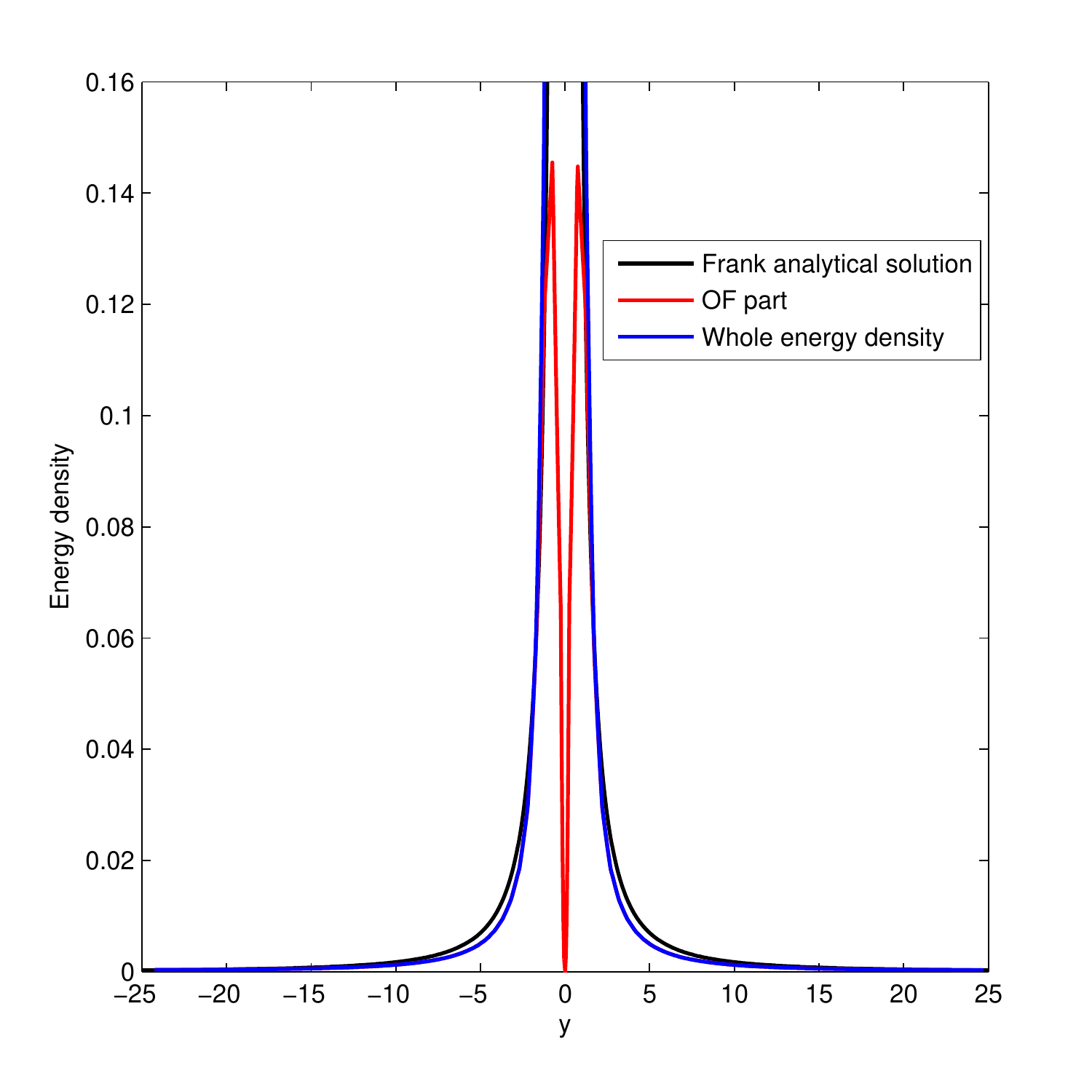}
}\qquad
\subfigure[Energy density comparison along $y$ axis near strength $+\frac{1}{2}$ disclination core.]{
\includegraphics[width=0.4\textwidth]{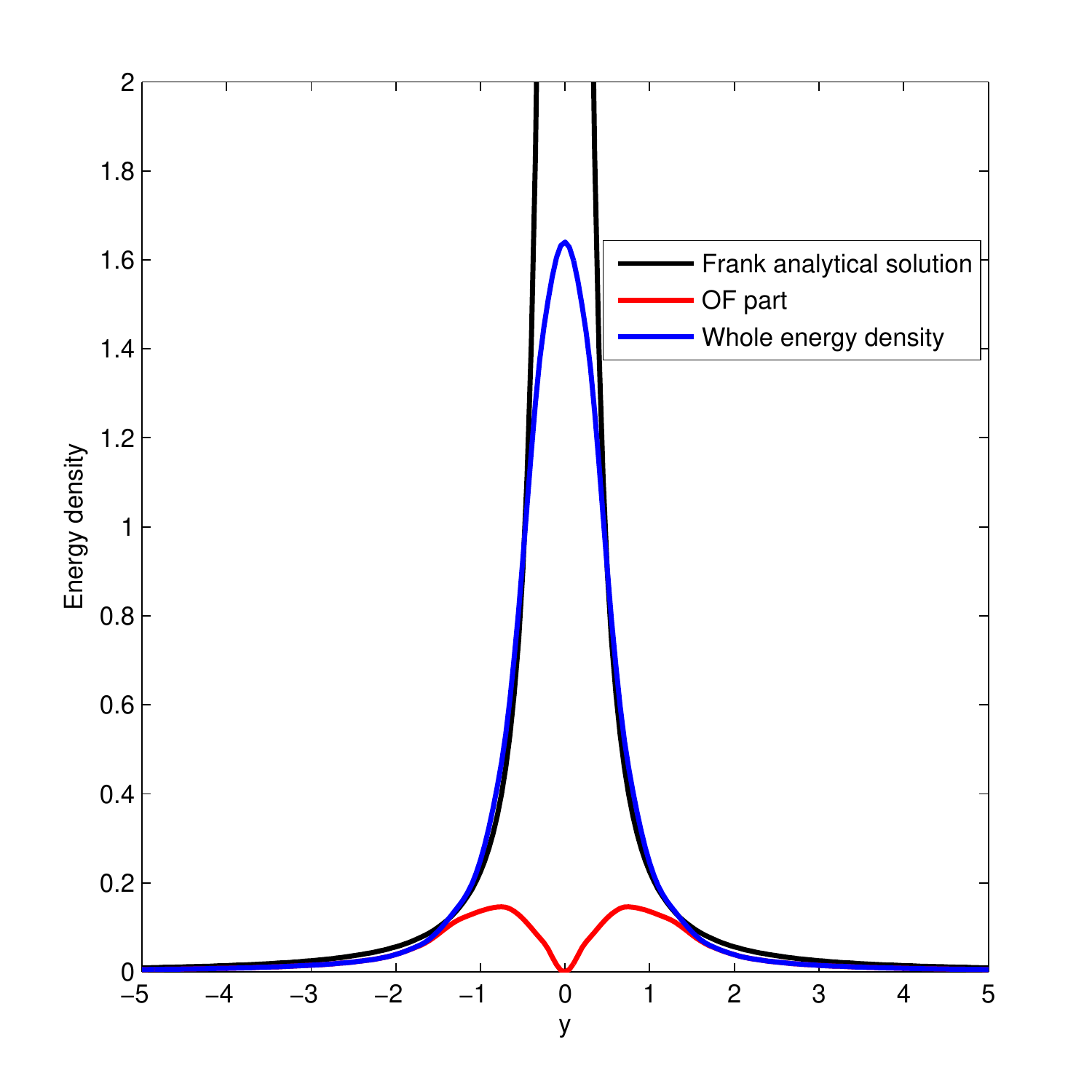}
}
\caption{Energy density comparisons between Frank analytical results and our results of $+1/2$ disclination along $y$ axis, for both the overall domain and near-core area, indicating a good agreement.}
\label{fig:comparison_energy_density_y}
\end{figure}

The energy density should decay as $1/r^2$ when moving away from core where $r$ is the distance from core. In Figure \ref{fig:comparison_energy_density}, the black line is the energy density along the horizontal axis from the Frank analytical solution, labeled as $Frank$ $analytical$ $solution$; the red line is the contribution of the energy density from the Oseen-Frank part $\frac{K}{2}(\grad\theta-\bflambda)^{2}$ in our model, labeled as $OF$ $part$; and the blue line is the whole energy density from our model, labeled as $Whole$ $energy$ $density$. The overall comparisons as well as the comparisons near the core area for both $+1/2$ and $+1$ disclinations are presented in Figure \ref{fig:comparison_energy_density}. These comparisons show good agreement between the energy density and that of the Frank analytical result outside the core. Inside core, our results are nonsingular while the Frank analytical results blow up. Figure \ref{fig:comparison_energy_density_y} shows the energy density comparisons for strength $+1/2$ disclination along the $y$ axis. The energy densities are symmetrically distributed along both the $x$ and $y$ axes and they show good agreement with the Frank solution. The profiles for strength $-\frac{1}{2}$ and strength $-1$ disclinations also follow the correct trends.

\begin{figure}[H]
\centering
\includegraphics[width=0.7\textwidth]{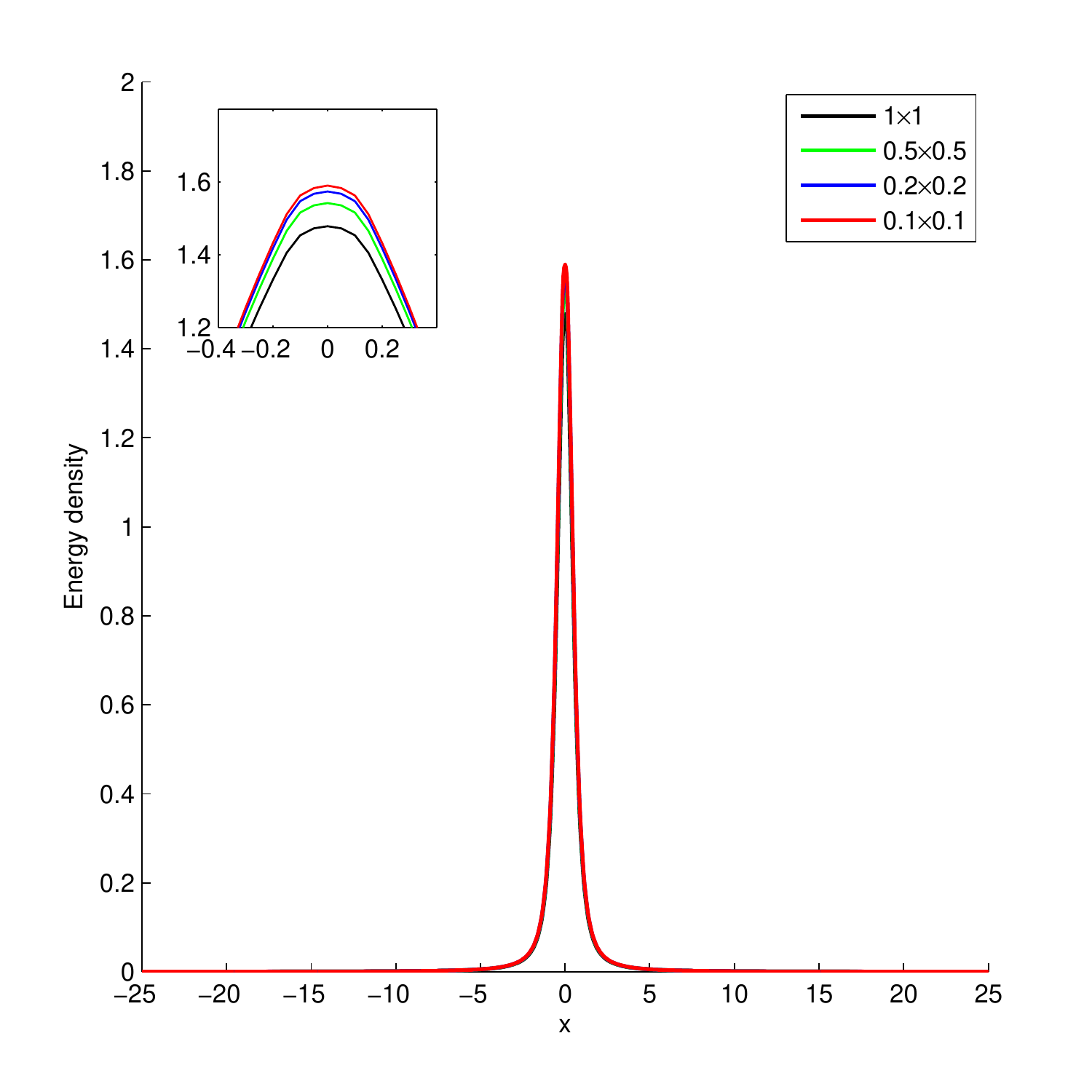}
\caption{Energy density plots along layer direction with different meshing. In the legend, the form $a\times b$ represents the element size, where $a$ is the element size in the $x$ direction and $b$ is the element size in the $y$ direction. The domain size is $50\times 50$. The inset plot is a magnified view at the center of the core. The energy density results converge with mesh refinement.}
\label{fig:energy_density_convergence}
\end{figure}

Figure \ref{fig:energy_density_convergence} shows a convergence study of our approximate solutions for the energy density along the $x$-axis for the $+1/2$ disclination. In Figure \ref{fig:energy_density_convergence}, the lines of different color represent mesh sizes from $1$ to $0.1$. For a fixed problem defined in Section \ref{sec:stat_1/2}, the energy density results converge with mesh refinement.

\subsection{Variation of total energy as a function of layer thickness}\label{sec:energy_indep}

For nematic disclinations, a layer where the director vector `unwinds' is to be considered as an approximation to the physical case of a sharp  discontinuity in the director vector field. Thus it is necessary to demonstrate, at least approximately, that in the limit $a \rightarrow 0$ the total energy of the body with a disclination remains non-zero but finite.

\begin{figure}[H]
\centering
\includegraphics[width=0.7\textwidth]{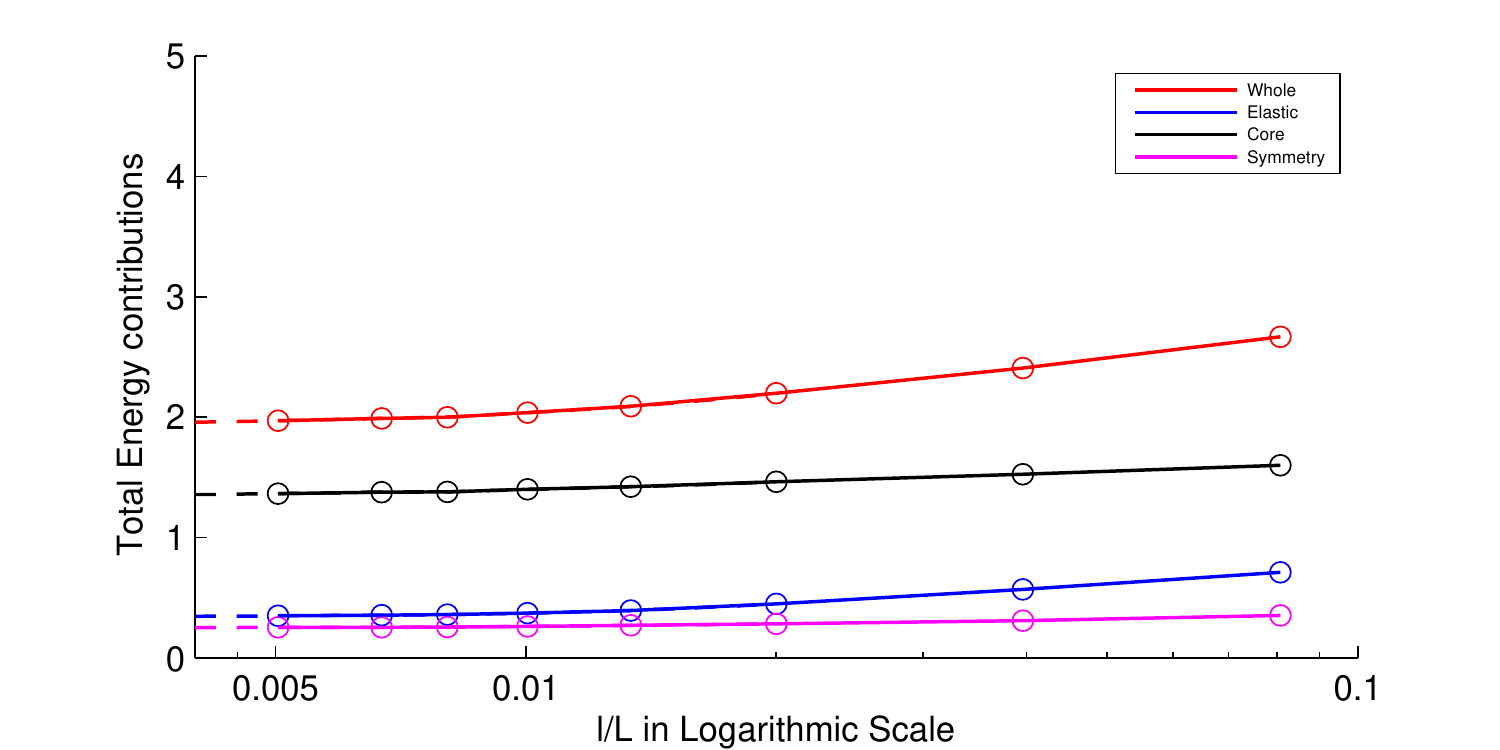}
\caption{Trends of different parts of the total energy as the layer width tends to zero. The total energy as well as the individual contributions converge as $a$ tends to zero.}
\label{fig:energy_con}
\end{figure}

Recall the nondimensionalized energy in this work takes the form 
\begin{equation*}
E = \int_V\left[\frac{1}{2}|\grad\theta - \bflambda|^2+\frac{Ca}{2}|\curl\bflambda|^2+\frac{2P|k|}{a} f(\bflambda)\right]dv. 
\end{equation*}
Figure \ref{fig:energy_con} is a the plot of total non-dimensional value for a $+\frac{1}{2}$ disclination as $a$ tends to zero. The red line, labeled as $Whole$, is the value of total non-dimensional energy $E$; the blue line, labeled as $Elastic$, is the contribution from $\frac{1}{2}|\grad \theta - \bflambda|^2$; the black line, labeled as $Core$, is the contribution from $\frac{Ca}{2}|\curl\bflambda|^2$;  and the green line, labeled as $Symmetry$, is the contribution from $\frac{2P|k|}{a} f$. This plot shows that the total energy as well as the individual contributions converge as $a$ tends to zero. The circles represent values obtained from the calculations at different $l/L$ ratios. The total non-dimensional energy shows a trend of converging to a finite value of $1.915$; the Frank elastic contribution part converges to $17.5\%$ of the total energy; the contribution from the disclination core converges to $69.3\%$ of the total energy; and the contribution from the symmetry-related component converges to $13.2\%$ of the total energy.

\subsection{Shortcoming of the gradient flow dynamics for this energy function} \label{sec:shortcoming}

\begin{figure}[H]
\centering
\subfigure[Initial prescription for $|\bflambda|$ field.]{
\includegraphics[width=0.4\linewidth]{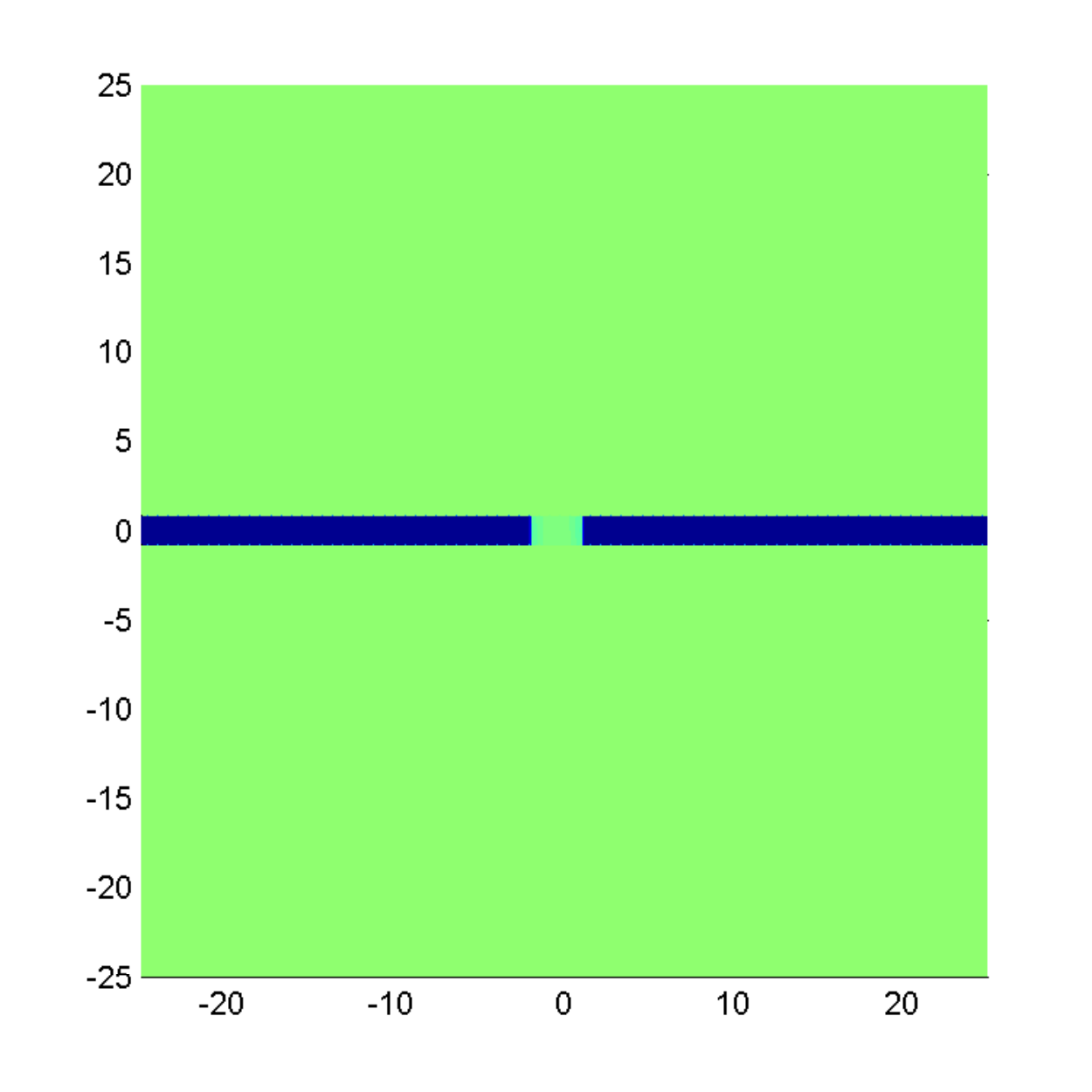}
\label{fig:gradient_anni_ini}}\qquad
\subfigure[Initial prescription for $\theta$ field.]{
\includegraphics[width=0.4\linewidth]{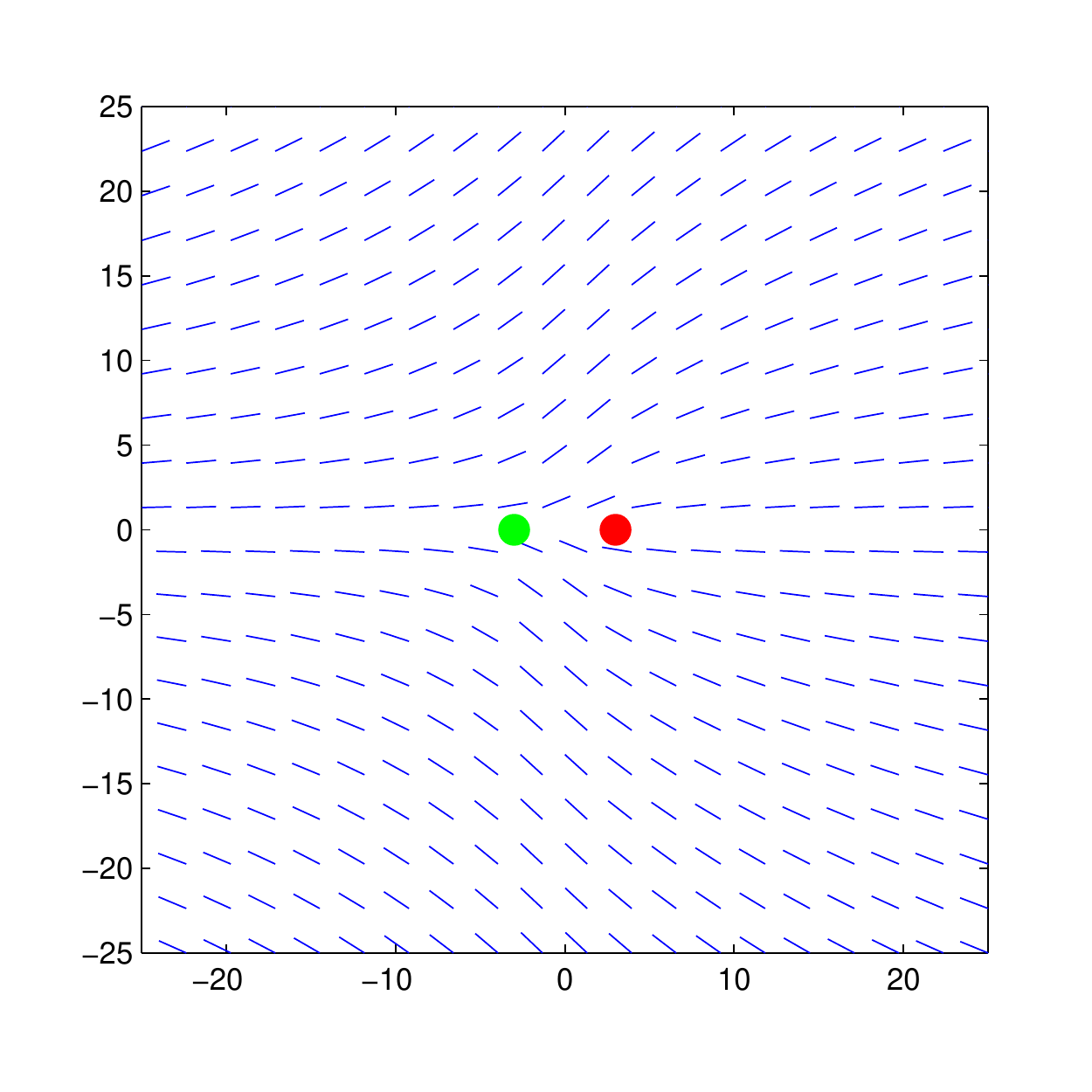}
\label{fig:gradient_anni_ini_theta}}
\end{figure}

In spite of the fact that the gradient flow method for this energy works very well in the computation of defect equilibria as demonstrated in Sections \ref{sec:stat_1/2} - \ref{sec:stat_1}, it is not able to predict the motion of disclinations. To illustrate this point, we consider disclination annihilation as an example. Figure \ref{fig:gradient_anni_ini} shows the corresponding initial $|\bflambda|$ field, i.e., a half disclination dipole is prescribed within the layer as initial condition. Figure \ref{fig:gradient_anni_ini_theta} shows the initialization of the $\theta$ field, where the red dot represents a strength $+1/2$ disclination and the green dot represents a strength $-1/2$ disclination core. 

\begin{figure}[H]
\centering
\subfigure[Director field $\theta$ for the disclination annihilation problem using the gradient flow method.]{
\includegraphics[width=0.4\textwidth]{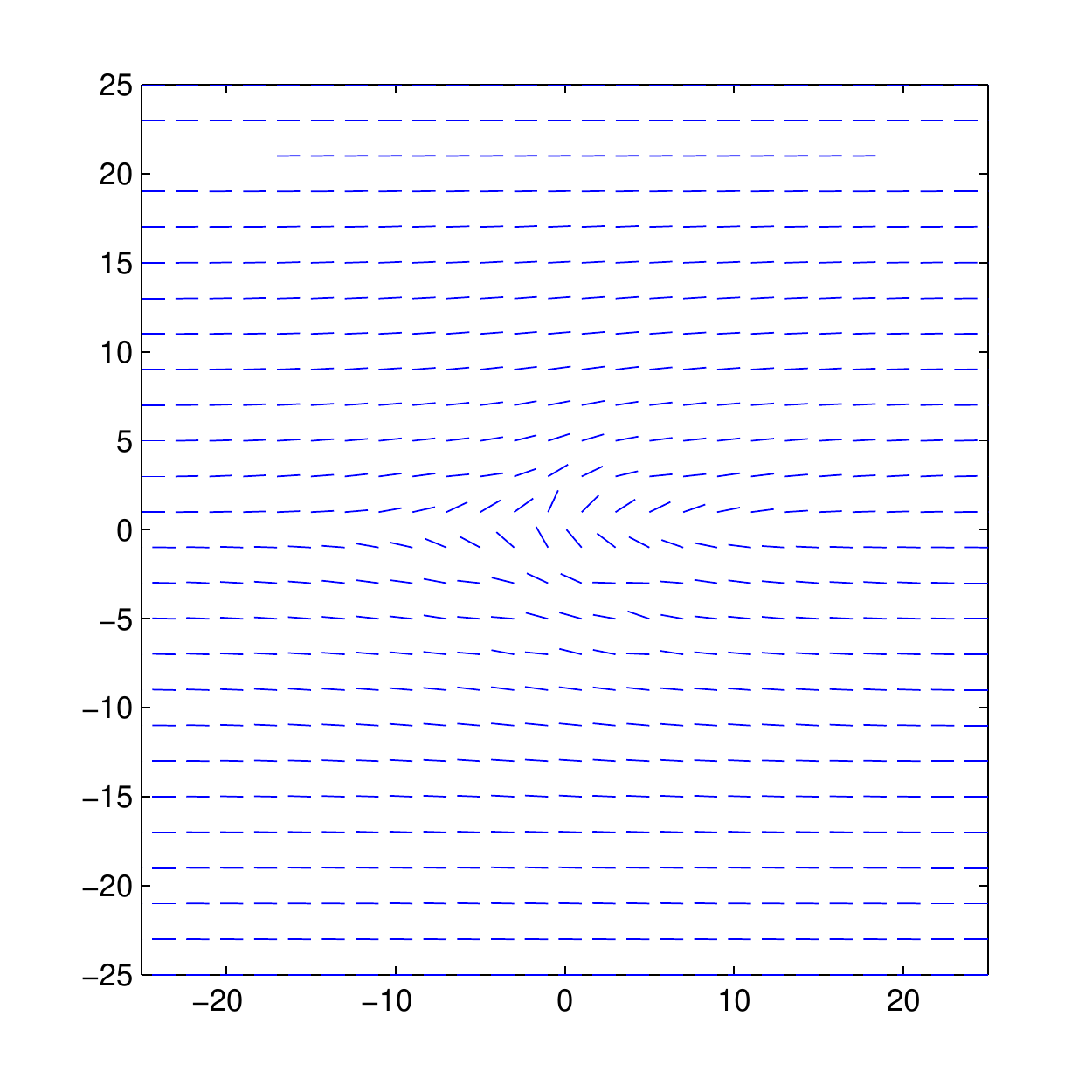}
\label{fig:shortcoming_director}
}\qquad
\subfigure[Energy density in relaxed state achieved by the gradient flow calculation.]{
\includegraphics[width=0.4\textwidth]{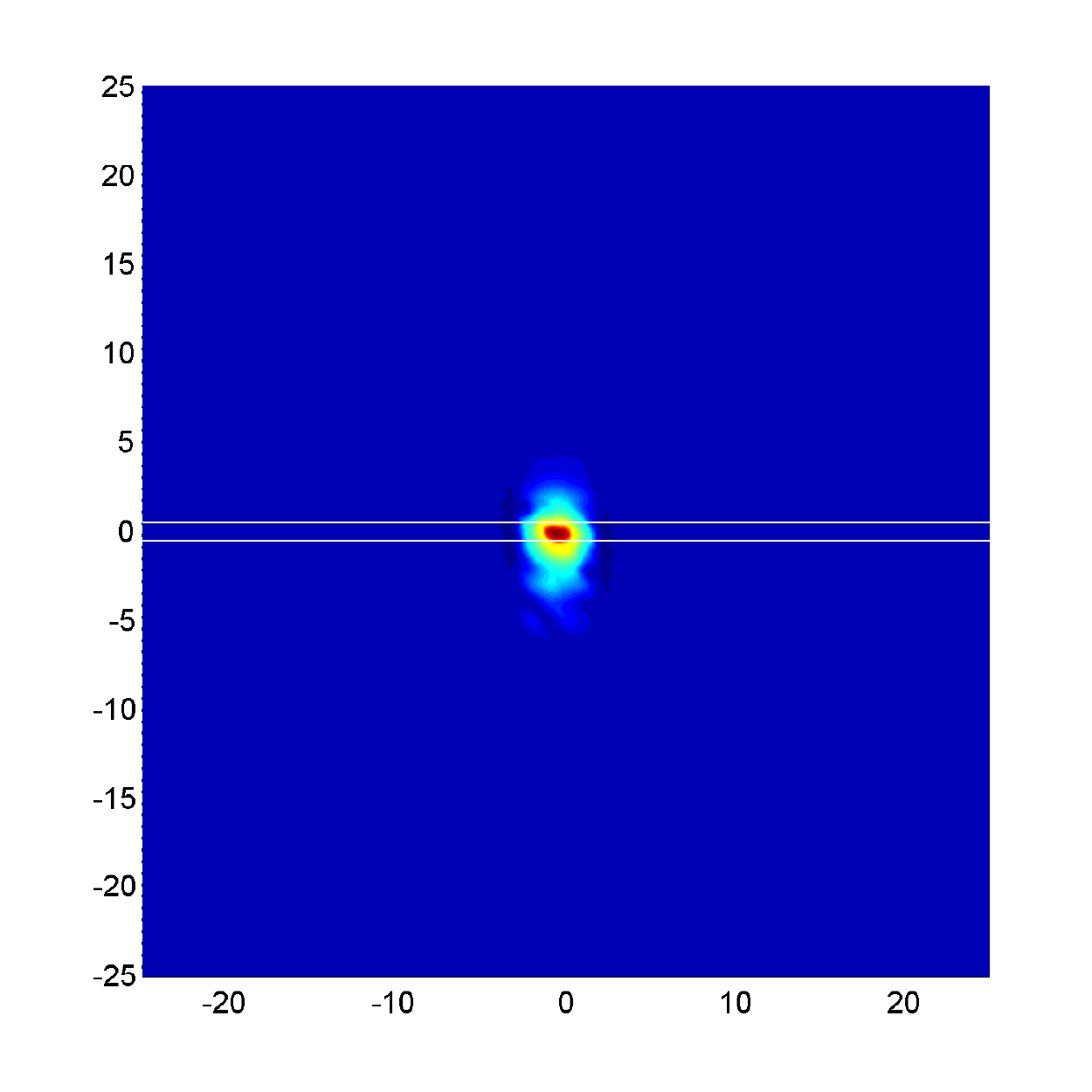}
\label{fig:shortcoming_energy}
}
\caption{Director field and energy density plot for disclination annihilation using the gradient flow method. The two white lines are artificially inserted to display the top and bottom layer boundaries. The results from the gradient flow calculation do not match physical expectation.}\label{fig:gradient_anni_result}
\end{figure}

The physical expectation is that on evolution those two disclinations merge with each other and annihilate, leaving no energy in the end. Recall that for the equilibrium solutions, $P = 20$. With this relatively high penalty on the non-convex term, we find that while the two oppositely charged disclinations evolve to their equilibrium configurations, they simply do not evolve from their equilibrium positions and annihilate, contrary to physical expectation. This can be understood as follows: invoking a dynamical process for the evolution requires continuous evolution in time of the fields at any spatial point. For a disclination to move,  the value of $|\bflambda|$ at a spatial point ahead of the core has to rise continuously from 0 to 2$\pi k/a$ (for a prescribed value of $k$) over a finite time interval. However, for the intervening states in this path, states that are not minima of the wells of the function $f$ have to be sampled, and this leads to a large energy barrier - for large $P$ - that has to be overcome by the driving forces arising from director gradients ($\grad \theta$). What occurs in the calculations is that large restoring energetic forces arise from the multiwell term that forces the spatial point (just ahead of the core) to stay at the minima of the \textit{0-well} of the function $f$. Hence the disclination cannot move. 

A natural remedy then is to think of reducing the penalty on the non-convex term - giving it the flavor of a physical component of the total energy function rather than an artificial mathematical device to represent a constraint limiting $|\bflambda|$ values to discrete states. To this end, we set $P = 2$. This raises another problem. The results from the gradient flow clearly do not match our expectation; there is a clear energy pattern near the core area, as observed in Figure \ref{fig:shortcoming_energy}, where the two white lines are artificially inserted to display the top and bottom layer boundaries. Even worse, there is a large area outside the layer where the corresponding director profile is inhomogeneous, as shown in Figure \ref{fig:shortcoming_director}. Clearly, the physical expectation is that the disclinations should annihilate moving in a straight line leaving behind a homogeneous director field with horizontal orientation everywhere except the layer, and zero energy everywhere (including the layer). This does not happen because with a lower penalty, $\bflambda$ can evolve from $\textbf{0}$, not only along the layer but elsewhere as well wherever there is a driving force, and, indeed, since there are director gradients outside the layer where $|\bflambda| = 0$, there is no impediment to growth of $\bflambda$ at such points, since a steady state of (\ref{eqn:grad_dimen})$_1$ is given by $\theta_{,i} = \lambda_i$, up to constraints posed by Dirichlet boundary conditions as well as the incompatibility of the field $\bflambda$.

\section{A dynamic model for nematic disclinations in 2D} \label{sec:theory_layer}

We seek an alternative to the gradient flow dynamics of the energy (\ref{eqn:grad_dimen}) to model energetically driven disclination dynamics. We follow the ideas in \cite{acharya2013continuum} motivated from the field of dislocation dynamics in solids to derive an appropriate model for the dynamics of straight wedge disclinations (a 2d model,) based on the statements of balance of mass, linear and angular momentum, the second law of thermodynamics, and a conservation statement for topological charge of these lines. We first show the derivation of the general 2D theory, and then derive a simple layer model from the theory as a particular example. In this section, $\bflambda$ and $\theta$ have the same meanings as in Sections \ref{sec:gradient_theory}.

\subsection{Derivation for general 2D case}

As before, we assume that the energy $E$ is given in the form of 
\begin{eqnarray*}
 E&=& \int_V [\psi(\grad\theta - \bflambda, \curl(\bflambda)) + \gamma f(\bflambda)]dv,
\end{eqnarray*}
where $\gamma= \frac{2PK \hat{k}}{a\xi^2}$ with the same definition as in Section \ref{sec:gradient_theory}, $f$ is a multi-well function with wells at $ \frac{2\pi \hat{k}}{a\xi}$, with $a \rightarrow 0$. For the sake of numerical approximation, we shall choose $a$ as a positive scalar that allows us to approximate director discontinuities of infinite magnitude. $k=\frac{n}{2}$ ($n$ can be any integer) is the disclination strength.
To be concise in the following derivations, we denote
\begin{eqnarray*}
\bfe &:=& \grad\theta - \bflambda \\
\bfb &:=& \curl(\bfe).
\end{eqnarray*}
$\bfb = \curl(\grad \theta - \bflambda)$ represents the departure of the director distortion from being the director gradient. In the absence of defects,  $\bfe = \grad \theta$ and hence $\bfb = \curl(\grad \theta) = \bf0$. Thus, $\bfb$ is considered as the defect field. 

Balancing the content of topological charge carried by defect lines within arbitrary area patches, a conservation law for the defect field \cite{acharya2013continuum} emerges in the form
\begin{eqnarray*}
\frac{ \partial \bfb}{\partial  t} &=& -\curl(\bfb \times \bfv) \\
-\curl \left(\frac{\partial \bflambda}{\partial t}\right) &=& -\curl(\bfb \times \bfv) \\
\frac{\partial \bflambda}{\partial t} &=& \bfb \times \bfv.
\end{eqnarray*}
The mechanical dissipation is the conversion of mechanical energy into heat, namely the difference between external power supplied to the body and the sum of the total rate of change of kinetic energy and the rate of change of free energy. In this case, the dissipation reads as (we ignore kinetic energy and flow here for simplicity)
\begin{equation*}
D = \int_{\partial V} \dot{\theta}\,\bfm\,\bfnu da - \int_{V} \dot{\psi} dv - \int_{V} \gamma \dot{f} dv \geq 0.
\end{equation*}
where $\bfnu$ is the normal vector on the boundary $\partial V$ and $\bfm$ is the moment given by $\bfLambda^\intercal \bfe_3$ with $\bfLambda$ is the couple stress tensor. In the following, superposed dots are meant to represent material time derivatives (in the language of continuum mechanics), but since we are ignoring flow, they are identical to spatial time derivatives. Apply the divergence theorem to the dissipation and require the second law of thermodynamics to be in effect to obtain
\begin{eqnarray*}
D &=& \int \{ (\dot{\theta} m_i)_{,i} - \dot{\psi} - \gamma \dot{f}\} dv \geq 0 \\
\Rightarrow \quad D &=& \int \left(m_i - \frac{\partial \psi}{\partial e_i}\right) \dot{\theta_{,i}} - \left(-\frac{\partial \psi}{\partial  e_i}\dot{\lambda_i} + \frac{\partial \psi}{\partial b_i}\dot{b_i} + \gamma \frac{\partial f}{\partial \lambda_i}\dot{\lambda_i}\right) dv \geq 0.
\end{eqnarray*} 
Since nematic elasticity has to be recovered by the model, $(m_i - \frac{\partial \psi}{\partial e_i}) \dot{\theta_{,i}} = 0$ is necessary for every possible $\dot{\theta_{,i}}$ when dissipative mechanisms are inoperative (i.e. $\dot{ \bflambda} = 0 \Rightarrow \dot{\bfb} = 0$). Thus, to satisfy this requirement, we choose $m_i = \frac{\partial \psi}{\partial e_i}$, and perform the following manipulations:
\begin{gather*}
\int -\left[ -\frac{\partial \psi}{\partial e_i} \dot{\lambda_i} + \frac{\partial \psi}{\partial b_i}\dot{b_i} + \gamma \frac{\partial f}{\partial \lambda_i}\dot{\lambda_i}\right] dv \geq 0 \\
\int -\left[ -\frac{\partial \psi}{\partial e_i} (\bfb \times \bfv)_i + \frac{\partial \psi}{\partial b_i}(-e_{ijk}(\bfb \times \bfv)_{k,j}) + \gamma \frac{\partial f}{\partial \lambda_i}(\bfb \times \bfv)_i \right] dv \geq 0 \\
\int -\left[ -\frac{\partial \psi}{\partial e_i} (\bfb \times \bfv)_i + \left(\frac{\partial \psi}{\partial b_i}\right)_{,j}(e_{ijk}(\bfb \times \bfv)_{k}) + \gamma \frac{\partial f}{\partial \lambda_i}(\bfb \times \bfv)_i \right] dv \geq 0 \\
\int \left[\frac{\partial \psi}{\partial e_k}e_{krs}b_r v_s + \left(\curl \frac{\partial \psi}{\partial \bfb}\right)_k e_{krs} b_r v_s - \gamma \frac{\partial f}{\partial \lambda_k}e_{krs}b_r v_s\right] dv \geq 0 \\
\int \left\{ e_{krs}\left[\frac{\partial \psi}{\partial e_k} + \left(\curl \frac{\partial \psi}{\partial \bfb}\right)_k - \gamma \frac{\partial f}{\partial  \lambda_k}\right]b_r\right\}v_s dv \geq 0.
\end{gather*}
Based on the second law of thermodynamics, we need to ensure a non-negative dissipation as stated in the above inequality. To fulfill this requirement, the simplest and most natural choice is to require
\begin{equation*}
\bfv \quad \text{parallel to} \quad \left[\bfm + \curl \frac{\partial \psi}{\partial \bfb} - \gamma \frac{\partial f}{\partial \bflambda}\right]\times \bfb.
\end{equation*}
It is characterized in the most simple of circumstances by choosing $\bfv$ of the form
\begin{equation*}
\bfv = \frac{1}{B_m |\bfb|^m}\left[\left(\bfm + \curl\left(\frac{\partial \psi}{\partial \bfb}\right)- \gamma \frac{\partial f}{\partial \bflambda}\right) \times \bfb\right]
\end{equation*}
with $m = 0$ and $B_m$ is a material constant required on dimensional grounds. The parameter $m$ can probe different types of behaviors. With this choice of $\bfv$, we can verify that the dissipation is larger or equal to zero globally, which means the second law of thermodynamics is satisfied by our model.

Recall that 
\begin{eqnarray}\label{eqn:subs}
\begin{aligned}
\frac{ \partial \bfb}{\partial  t} &= -\curl(\bfb \times \bfv) \\
\frac{\partial \bflambda}{\partial t} &= \bfb \times \bfv.
\end{aligned}
\end{eqnarray}
After substituting $\bfv$ in (\ref{eqn:subs}), the evolution equations for $\bfb$ and $\bflambda$ can be written as
\begin{eqnarray} \label{eqn:general_eqns_lc}
\begin{aligned}
\frac{\partial \bfb}{\partial t} &= -\curl \left[\bfb \times \frac{1}{B_m |\bfb|^m} \left\{\left(\bfm + \curl\left(\frac{\partial \psi}{\partial \bfb}\right)- \gamma \frac{\partial f}{\partial \bflambda}\right) \times \bfb\right\}\right] \\
\frac{\partial \bflambda}{\partial t} &= \frac{1}{B_m |\curl \bflambda|^m} \curl\bflambda \times \left[\left(\bfm + \curl\left(\frac{\partial \psi}{\partial (\curl \bflambda)}\right)- \gamma \frac{\partial f}{\partial \bflambda}\right) \times \curl \bflambda\right].
\end{aligned}
\end{eqnarray}
We have ignored flow, and assume that balance of linear momentum and mass are trivially satisfied. Balance of angular momentum, assuming no director momentum is given by 
\begin{equation*}
\divergence (\bfm) = 0.
\end{equation*}
This reduces to the governing equation
\begin{equation}
\divergence (\grad \theta - \bflambda) = 0.
\end{equation}
The utility of (\ref{eqn:general_eqns_lc}) over the gradient flow dynamics (\ref{eqn:grad_dimen}) is the presence of a non-vanishing $\curl \bflambda$ in the evolution of the $\bflambda$ field in (\ref{eqn:general_eqns_lc})$_2$. At spatial points where $\bflambda$ is zero and $\curl \bflambda = 0$, $\bflambda$ cannot evolve (regardless of the value of the penalty parameter $P$); however, at the boundaries of the core region where one might expect $\bflambda = \bf0$ but $\curl\bflambda \neq \bf0$, evolution is possible allowing motion of the core.

\subsection{A `layer' model}\label{sec:layer_model}
\subsubsection{Model description}
Based on the above formalism for the general 2-d case, we build a simple layer model to explore several physically fundamental behaviors of disclination defects. The model is directly adapted from \cite{acharya2014dislocation} that was developed for dislocation dynamics in solids, with a translation for symbols representing the different fields in the two models.

\emph{In the following, we will interchangeably refer to the coordinates $x_1$ as $x$ and $x_2$ as $y$. A subscript $x$, $y$, or $t$, even when not following  a subscript comma, will refer to partial differentiation with respect to those independent variables.}

\begin{figure}[H]
\centering
\includegraphics[width=0.7\linewidth]{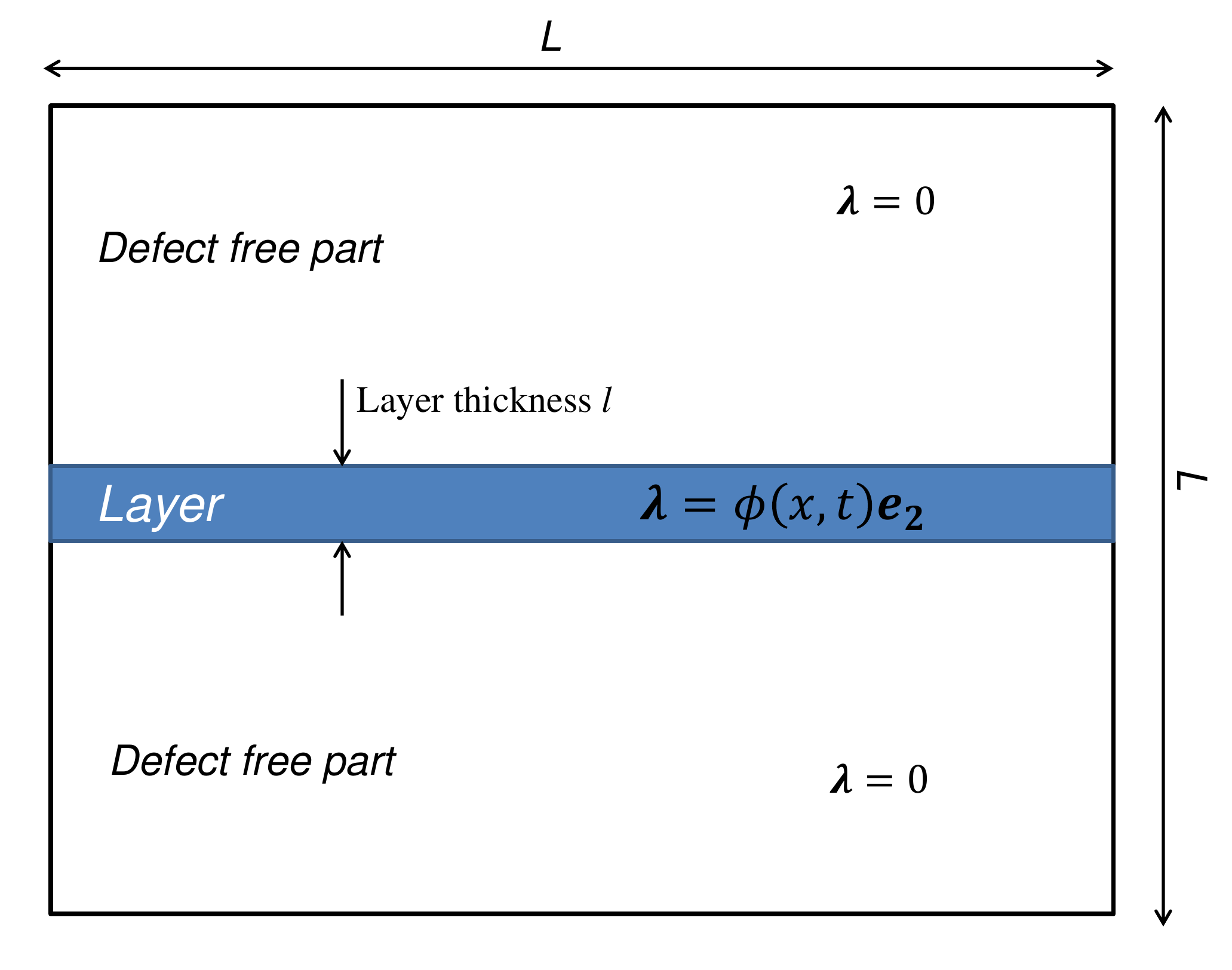}
\caption{Geometry for layer problem. $\bflambda$ has only non-zero component $\lambda_2$ inside the layer.}
\label{fig:dynamic_body}
\end {figure}

The fundamental assumption is that disclinations are allowed to move in a horizontal line, regularized here to a thin layer (with the correct scaling properties so that total energy remains finite even in the limit $a \rightarrow 0$). Consider a square geometry with a layer $\mathcal{L}$ of thickness $l = a \xi $, as shown in Figure \ref{fig:dynamic_body},
\begin{gather*}
\mathcal{V} = \{ (x,y):(x,y)\in [-L/2, +L/2] \times [-L/2, +L/2]\} \\
\mathcal{L} = \{ (x,y):(x,y)\in [-L/2, +L/2] \times [-l/2, +l/2]\} \\
0 \leq l < L, \quad L > 0.
\end{gather*}

The stored energy density function takes the same form as (\ref{eqn:gradient_energy}). Then the dissipation can be written as
\begin{gather*}
D = \int_\mathcal{L} \left(K(\grad \theta - \bflambda) - \gamma\frac{\partial f}{\partial \bflambda}\right) : \dot{\bflambda} dv + \int_\mathcal{L} \frac{\partial \psi}{\partial \bfb} : \curl(\bfb \times \bfv) dv \\
D = \int_\mathcal{L} \left(K(\grad \theta - \bflambda) - \gamma\frac{\partial f}{\partial \bflambda}\right) : (\bfb \times \bfv) dv + \int_\mathcal{L} \curl \left(\frac{\partial \psi}{\partial \bfb}\right) : (\bfb \times \bfv) dv + \int_{\partial \mathcal{L}} \frac{\partial \psi}{\partial \bfb} : (\bfb \times \bfv) \times \bfnu da
\end{gather*}
where $ \dot{\bflambda} = \bfb \times \bfv$, $ \dot{\bfb} = -\curl ( \bfb \times \bfv)$, $\bfb = -\curl \bflambda$, and $\bfnu$ is the unit normal vector of the layer boundary.

In this model, we assume $\bflambda$ takes the form
\begin{equation*}
\bflambda(x, y, t) = \begin{cases}
\phi(x,t)\,\bfe_2, & \text{in the layer ($|x_2| < \frac{l}{2}$)} \\
0, & \text{otherwise}.
\end{cases}
\end{equation*}
Therefore $\bfb$ is also non-zero only in the layer, with component form

\begin{eqnarray*}
&&\bfb = -\curl \bflambda = -e_{ijk} \lambda_{k,j} \bfe_i = -e_{321} \lambda_{2,1} \bfe_3 = -\phi_x\bfe_3 \\
&&\curl \bfb = e_{ijk} b_{k,j} \bfe_i = e_{213} b_{3,1} \bfe_2 = \phi_{11} \bfe_2.
\end{eqnarray*}
We assume $\bfv$ to be of the form, 
\begin{gather*}
 \bfv  = v_1(x,y,t)\bfe_1 =: v(x,t) \bfe_1.
\end{gather*}
Substitute $\bflambda$ in $f$ (\ref{eqn:eta}), 
\begin{equation*}
f = 1-\cos\left( \xi |\phi|\left(\frac{|k|}{a}\right)^{-1}\right).
 \end{equation*}

We assume boundary condition $\phi_x(\pm \frac{L}{2}, t) = 0$.

From $\dot{\bfb} = -\curl (\bfb \times \bfv)$, we have 
\[
\phi_t(x,t) = -\phi_x(x,t)v(x,t).
\]

Since $\bfb \times \bfv$ points in the direction of $\bfe_2$, the same direction of $\bfnu$, then $(\bfb \times \bfv)\times \bfnu= \bf0$. Thus, only the layer is relevant for the dissipation and this becomes
\begin{gather*}
D =\int_\mathcal{L} v(x,t) \left[ K(\theta_y - \phi) - \gamma\frac{\partial f}{\partial \phi} + \epsilon \phi_{xx}\right] (- \phi_x) dv
\end{gather*}
We note that all terms in the above equation depend only on the $x$ coordinate except for $\theta_y$ which also depends on the $y$ coordinate. To build the simplest possible model consistent with thermodynamics, it is essential to average $( \theta_y - \phi)$ over the layer\cite{acharya2014dislocation}. For any feasible $v(x,t)$, the dissipation can be rewritten as
\begin{gather*}
D=\int_\mathcal{L} v(x, t) \left[\tau(x,t) - \gamma \frac{\partial f(\phi(x,t))}{\partial \phi(x,t)} + \epsilon \phi_{xx}(x,t)\right](-\phi_1(x,t))dv +R
\end{gather*}
where
\begin{gather*}
R = \int_\mathcal{L} v(x,t)[\theta_2(x,y,t) -\phi(x,t)- \tau(x,t)](-\phi_x(x,t))dv.
\end{gather*}
If we make the choice
\begin{gather*}
\tau = \frac{K}{a \xi} \int_{-\frac{a \xi}{2}}^{\frac{a \xi}{2}} \left(\theta_{y}(x,y,t) - \phi(x,t) \right) dy,
\end{gather*}
it is immediate that $R=0$ due to the definition of $\tau$. We make the constitutive assumption for the velocity as
\begin{gather*}
v(x,t) = \frac{-1}{B_m |\phi_x|^m} \{ \phi_x [\tau - \tau^b + \epsilon \phi_{xx}] \} \\
\frac{\partial \phi}{\partial{t}} =\frac{|\phi_x|^{2-m}}{B_m}\left(\tau -\tau^b + \epsilon \phi_{xx}\right) \\
\text{where} \quad \tau = \frac{K}{a \xi} \int_{-\frac{a \xi}{2}}^{\frac{a \xi}{2}}(\theta_{y} - \phi) dy; \quad \tau^b = \gamma \frac{\partial f}{\partial \phi}; \quad f = 1-\cos\left( \xi |\phi|\left(\frac{|k|}{a}\right)^{-1}\right).
\end{gather*}
Here, $B_m$ is a non-negative coefficient characterizing energy dissipation with physical dimensions depending on $m$. The parameter $m$ can be chosen to probe different types of behavior. Especially, the model for $m=2$ is the analog of the gradient flow case (\ref{eqn:grad_dimen}) with layer restriction. $m = 0$ has been shown to demonstrate possible pinning of defects in computational experiments \cite{zhang2015single}.

By choosing the following dimensionless variables 
\begin{gather*}
\tilde{x} = \frac{1}{\xi}x; \quad \tilde{y} = \frac{1}{\xi}y; \quad \tilde{\epsilon} = \frac{1}{K \xi^2}\epsilon=Ca; \quad
\tilde{\tau} = \frac{\xi}{K}\tau; \quad \tilde{\tau}^b = \frac{ \xi }{K}\tau^b; \\
\quad  \tilde{s}=\frac{K}{\xi^{4-2m} B_m}t; \quad \tilde{\phi} = \xi \phi
\end{gather*}
we arrive at the dimensionless governing equations as described below. 
\begin{equation*}
\left\{
\begin{aligned}
& \theta_{\tilde{x} \tilde{x}} + \theta_{\tilde{y} \tilde{y}} - \tilde{\phi}_{\tilde{y}} =0 \quad \text{in }  \mathcal{V}\\
&\frac{\partial \tilde{\phi}}{\partial{\tilde{s}}} ={|\tilde{\phi}_{\tilde{x}}|^{2-m}}\left(\tilde{\tau} - \tilde{\tau}^b + \tilde{\epsilon} \tilde{\phi}_{\tilde{x} \tilde{x}}\right) \quad \text{in } \mathcal{L}.
\end{aligned}
\right.
\end{equation*}
\emph{After removing tildes for simplicity}, the dimensionless system that governs the problem reads as 
\boxalign{
\begin{eqnarray}
\label{eq:sum_gov_eq}
\left\{
\begin{aligned}
& \theta_{xx} + \theta_{yy} - \phi_{y} =0 \quad \text{in } \mathcal{V}\\
&\frac{\partial \phi}{\partial{s}} ={|\phi_x|^{2-m}}\left(\tau - \tau^b + Ca \phi_{xx}\right) \quad \text{in } \mathcal{L}
\end{aligned}
\right.
\end{eqnarray}
where 
\begin{equation*}
\begin{aligned}
\tau =  \frac{1}{a}\int^{a/2}_{-a/2} \left( \theta_{y} - \phi \right) dy, \quad
\tau^b = {2 P} sin\left(\phi\left(\frac{|k|}{a}\right)^{-1}\right).
\end{aligned}
\end{equation*}}

The corresponding numerical scheme for the above dimensionless system is developed in Appendix \ref{App:1}.

\section{Disclination annihilation, repulsion, and dissociation} \label{sec:app_layer}

We explore several disclination dynamic cases (in the absence of flow) within the 2D layer model. The domain is shown in Figure \ref{fig:dynamic_body} with geometry $50\times50$. The parameter $a = 1$ is assumed the same as in the gradient flow simulations. The layer field $\bflambda$ is prescribed and restricted within a thin layer whose thickness is $a$, so that the disclination can only move along the $x$ direction. The penalty parameter $P$ is set to $1$ (recall that in the gradient flow simulations $P=20$, and $P = 2$ was unsuccessful in recovering physically expected equilibria). In the following, we will demonstrate and discuss results on disclination annihilation, repulsion, and dissociation.

\emph{In this section, all cases are calculated from $\phi$ evolution equations with $m=0$, unless otherwise mentioned.}

\subsection{Disclination annihilation}

\begin{figure}[H]
\centering
\subfigure[Initialization of $\phi$ for disclination annihilation. A $\phi$ field corresponding to a strength $+\frac{1}{2}$ and a strength $-\frac{1}{2}$ disclination is prescribed.]{
\includegraphics[width=0.45\textwidth]{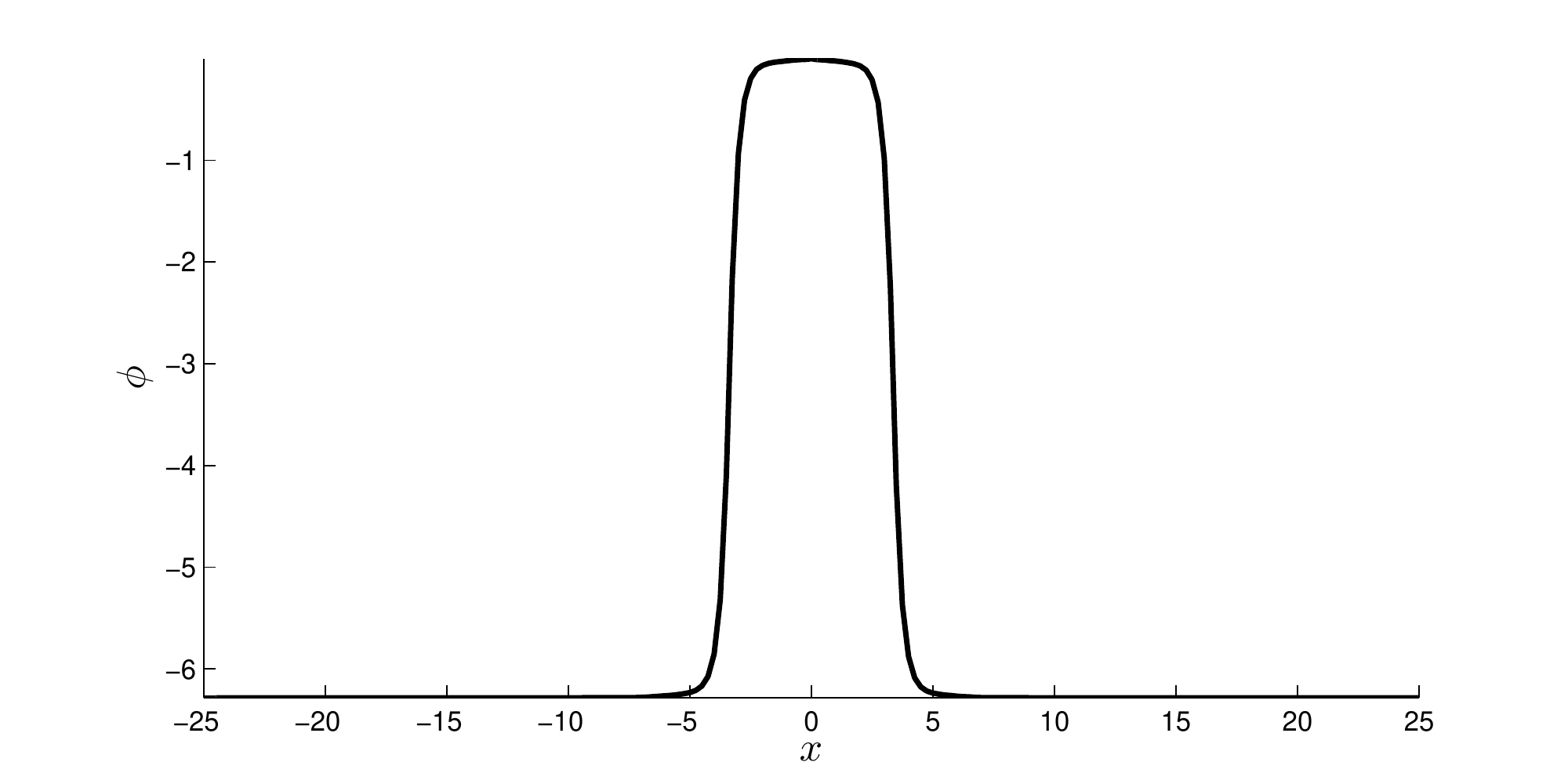}
\label{fig:dynamic_annih_1a}} \qquad
\subfigure[Director field corresponding to the initialized $\phi$. ]{
\includegraphics[width=0.45\textwidth]{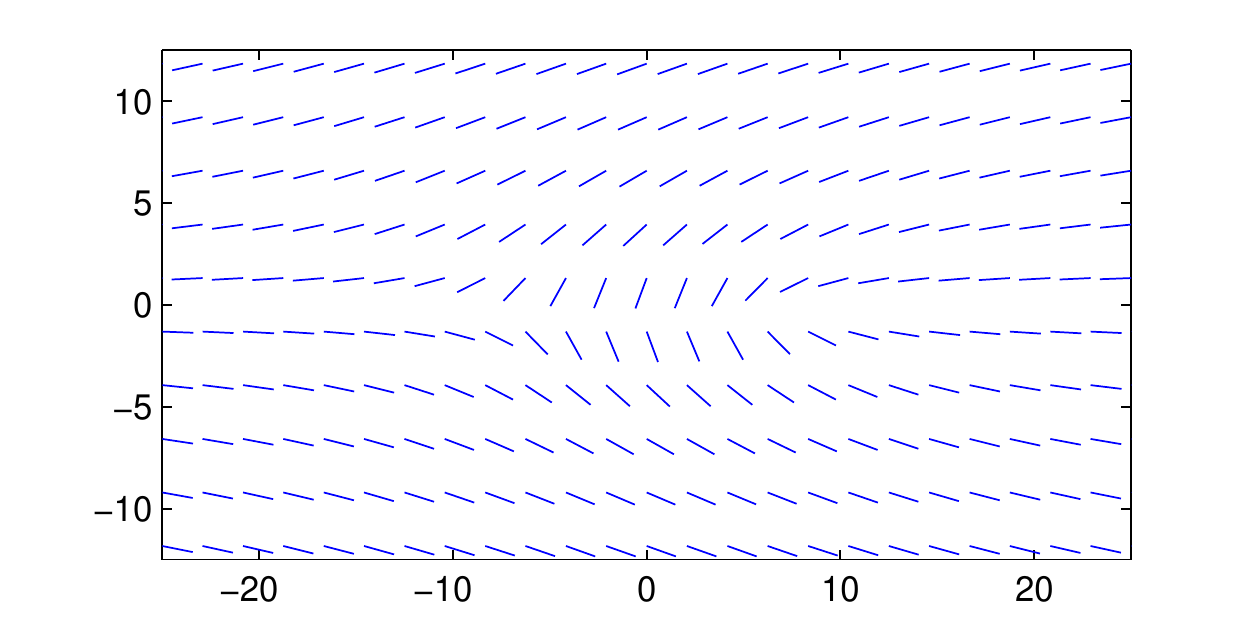}
\label{fig:dynamic_annih_1b}}
\caption{Initialization for the disclination annihilation problem.}\label{fig:dynamic_annih}
\end {figure}

\begin{figure}[H]
\centering
\includegraphics[width=0.7\linewidth]{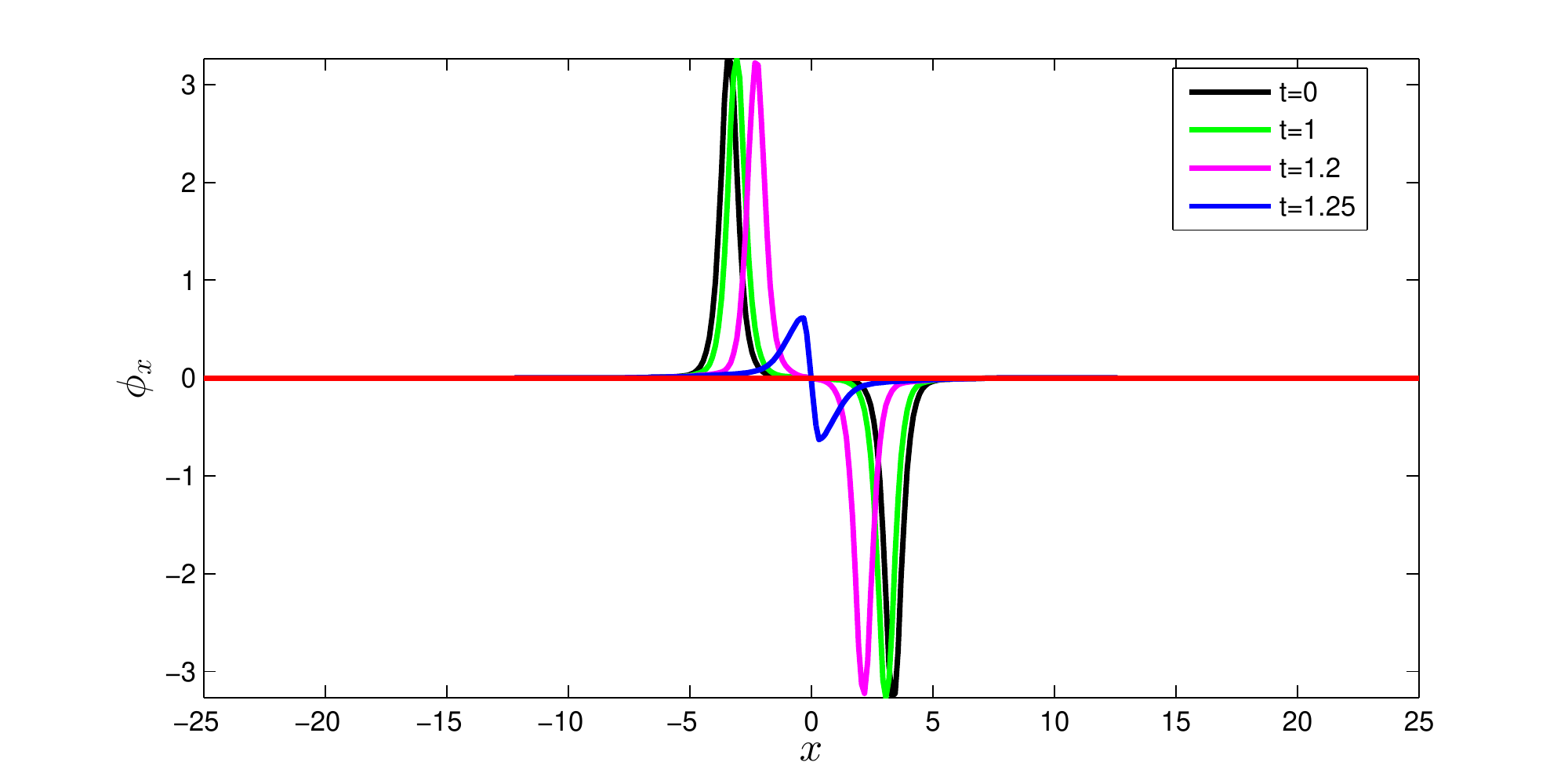}
\caption{$\phi_x$ snapshots at different time steps. The bumps represent disclination cores. The disclination dipoles eventually annihilates.}
\label{fig:dynamic_annih_2}
\end {figure}

We start with disclination annihilation, which the gradient flow approach failed to predict in Section \ref{sec:shortcoming}. Even in this dynamic `layer problem', if $P = 20$, and $m = 2$ (i.e. the analog of the gradient flow in the layer case), we find that, as expected, the oppositely charged disclinations do not annihilate. Within the layer ansatz and now setting $P = 1$, initially, a disclination dipole, i.e., two disclinations with opposite signs of $+\frac{1}{2}$ and $-\frac{1}{2}$, is prescribed as shown in Figure \ref{fig:dynamic_annih_1a}. The horizontal axis in Figure \ref{fig:dynamic_annih} represents the $x$ axis along the layer and the vertical axis shows the magnitude of $\phi$. Figure \ref{fig:dynamic_annih_1b} shows the director field corresponding to the initialized $\phi$ prescription obtained by solving (\ref{eq:sum_gov_eq})$_1$. Figure \ref{fig:dynamic_annih_2} shows the snapshots of defect movement during the simulation. The vertical axis shows the gradient of $\phi$ along the layer ($\phi_x$), representing the location of the core. Different colors represent the results at different times and at each time two opposite bumps are interpreted as a disclination dipole because $\curl \bflambda = -\bfb = \phi_x \bfe_3$. As time evolves, these two cores move toward each other, and finally merge. In the final result, the disclination dipole annihilates and no disclination exists in the body. 

\begin{figure}[H]
\centering
\subfigure[Director snapshot at t=0.]{
\includegraphics[width=0.45\textwidth]{figure/annihilation/1.pdf}
}
\subfigure[Energy density plot at t=0.]{
\includegraphics[width=0.45\textwidth]{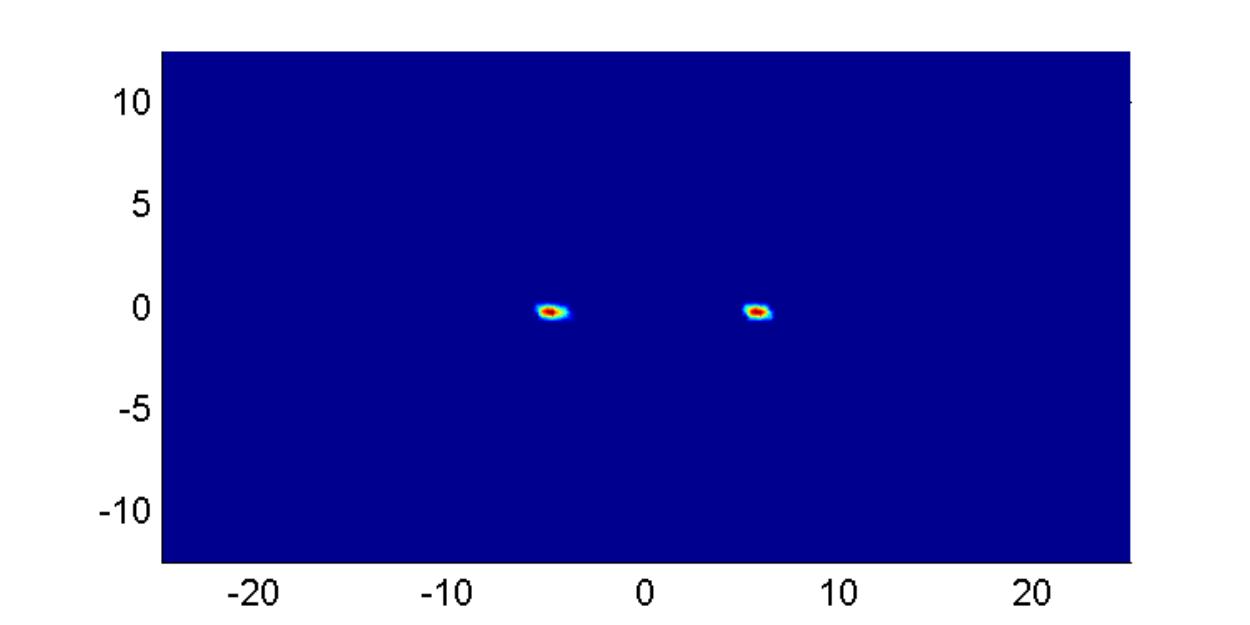}
}
\subfigure[Director snapshot at t=1.]{
\includegraphics[width=0.45\textwidth]{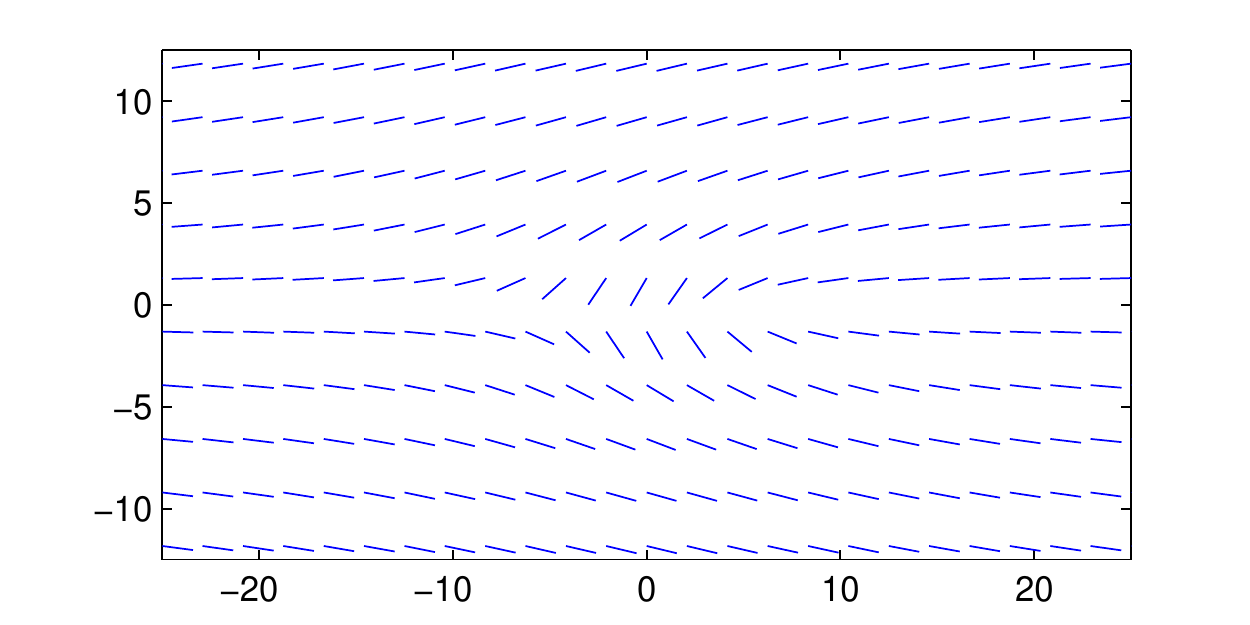}
}
\subfigure[Energy density plot at t=1.]{
\includegraphics[width=0.45\textwidth]{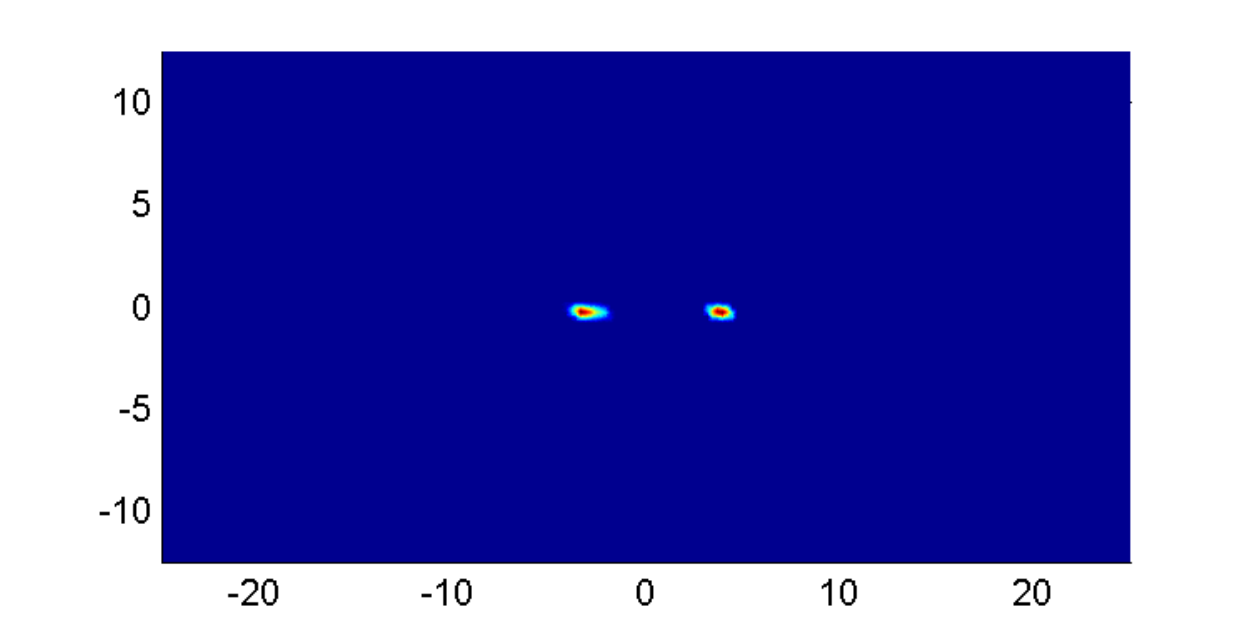}
}
\subfigure[Director snapshot at t=1.2.]{
\includegraphics[width=0.45\textwidth]{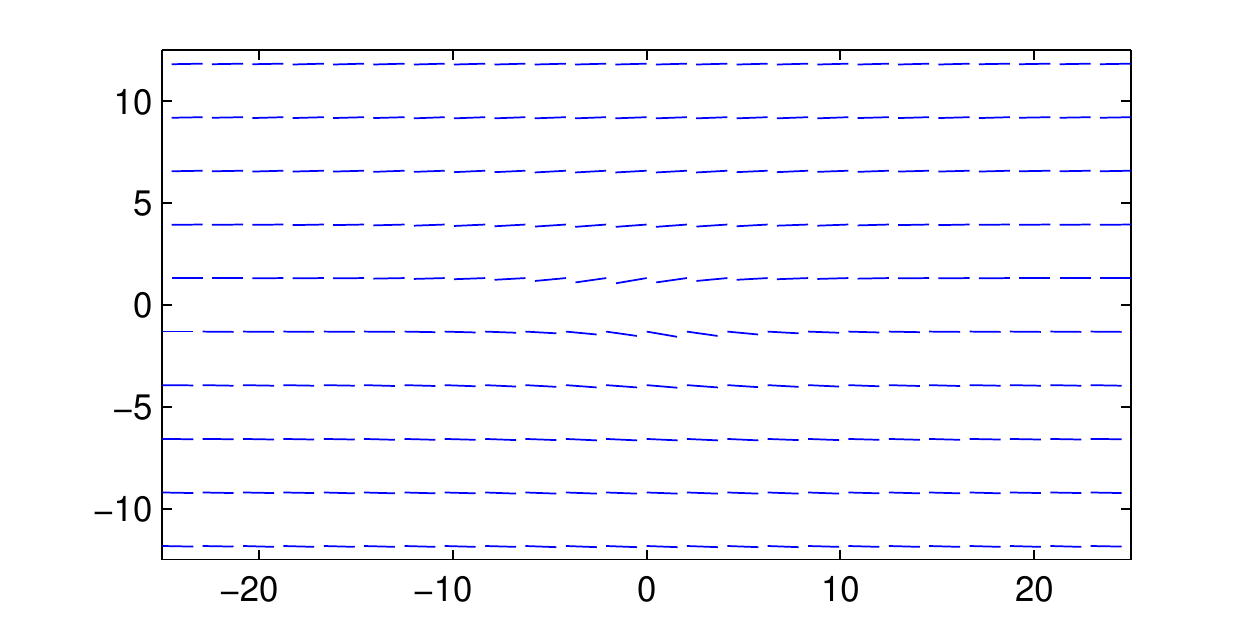}
}
\subfigure[Energy density plot at t=1.2.]{
\includegraphics[width=0.45\textwidth]{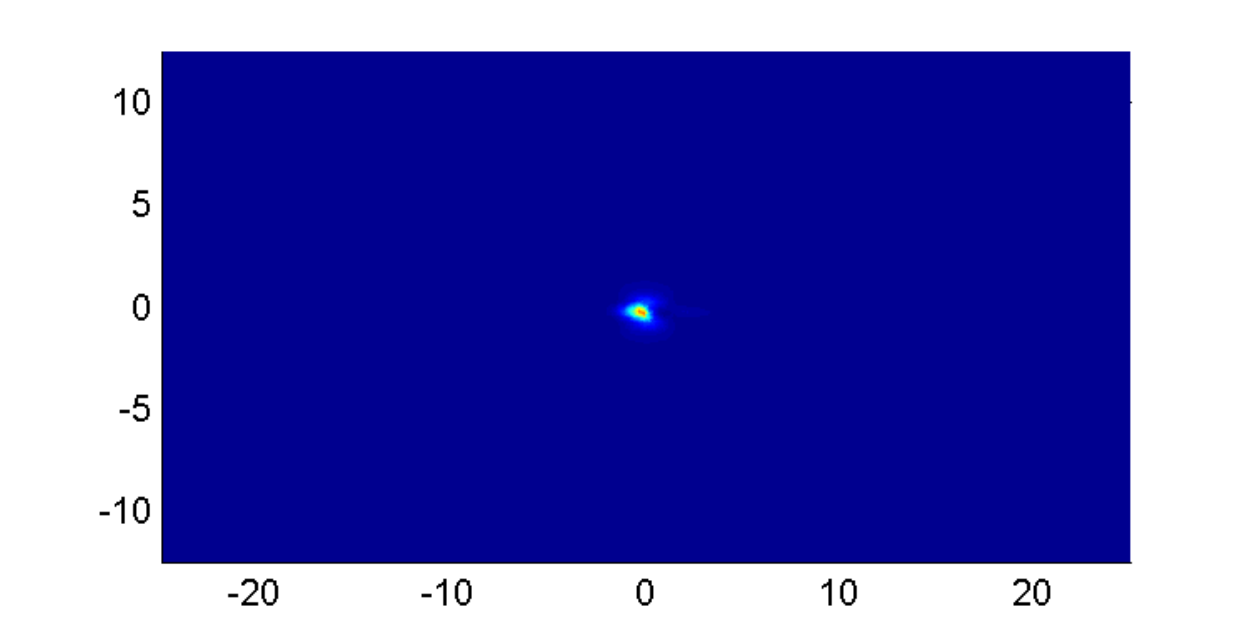}
}
\subfigure[Director snapshot at t=1.25.]{
\includegraphics[width=0.45\textwidth]{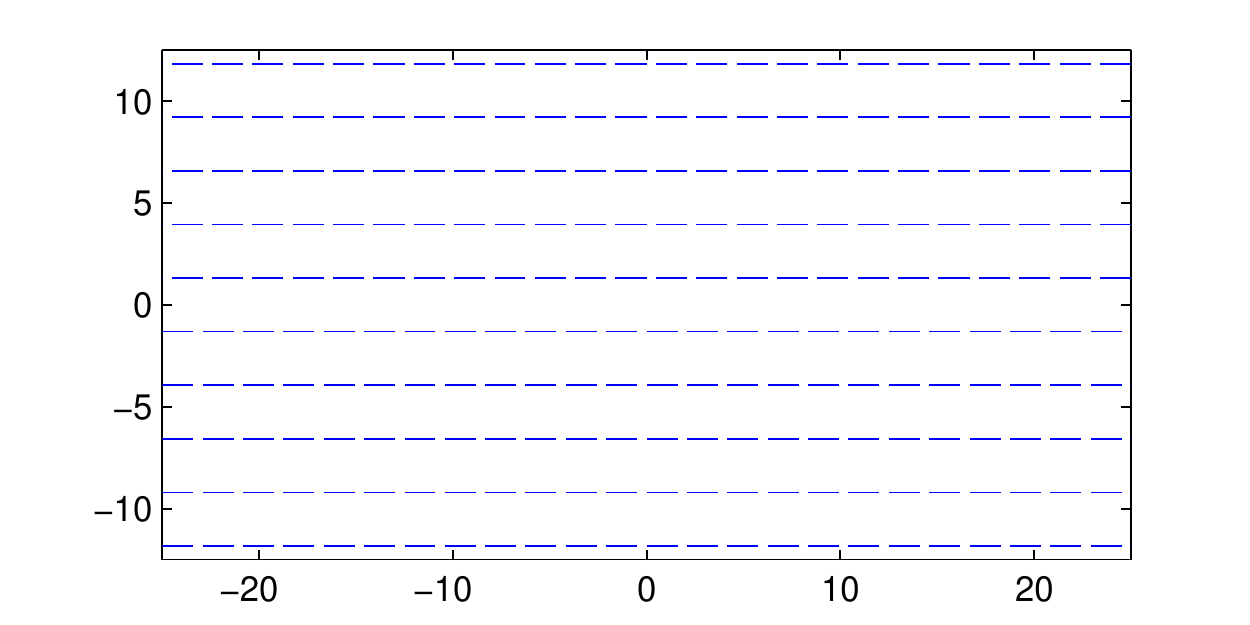}
}
\subfigure[Energy density plot at t=1.25.]{
\includegraphics[width=0.45\textwidth]{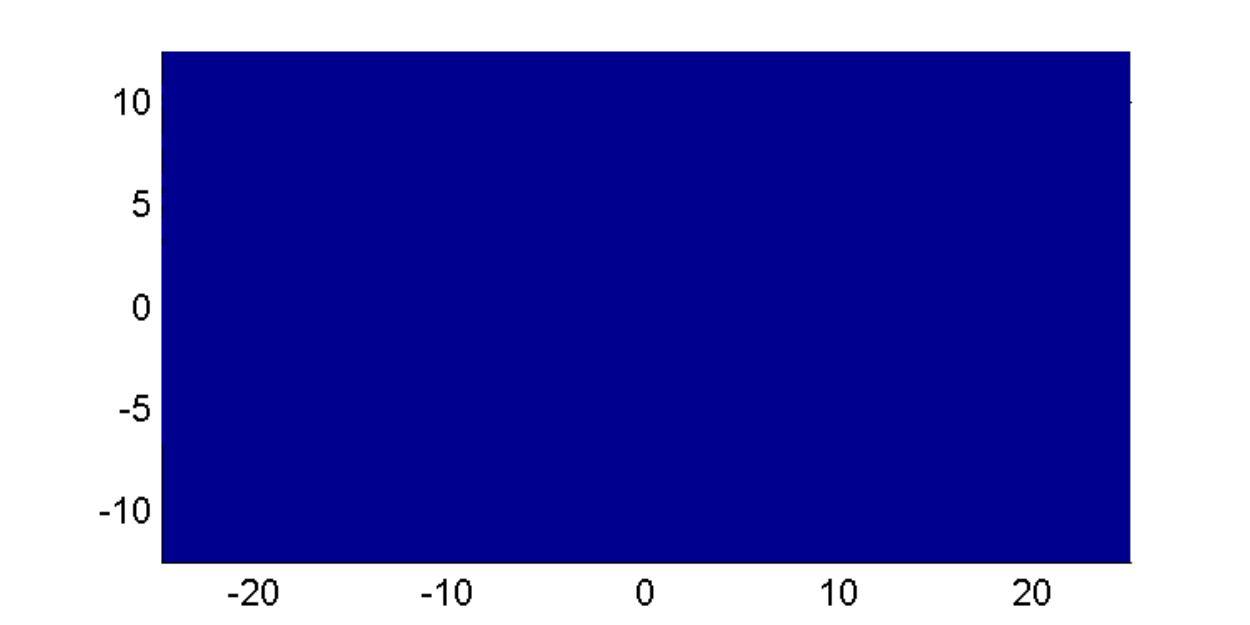}
}
\caption{Snapshots for the director field and energy density at different time steps. The disclination dipole merges and annihilates.}\label{fig:dynamic_annih_3}
\end {figure}

 Figure \ref{fig:dynamic_annih_3} shows the snapshots of the director field at different times and their corresponding energy density fields at that time. Both the director snapshots and energy density plots show that the disclination dipole annihilates in the end, which leads to zero energy. 
 
\subsection{Disclination repulsion}\label{sec:disclination_rep}

\begin{figure}[H]
\centering
\subfigure[Initialization of $\phi$ for disclination repulsion. A $\phi$ fields corresponding to a pair of strength $-\frac{1}{2}$ disclinations is prescribed.]{
\includegraphics[width=0.45\textwidth]{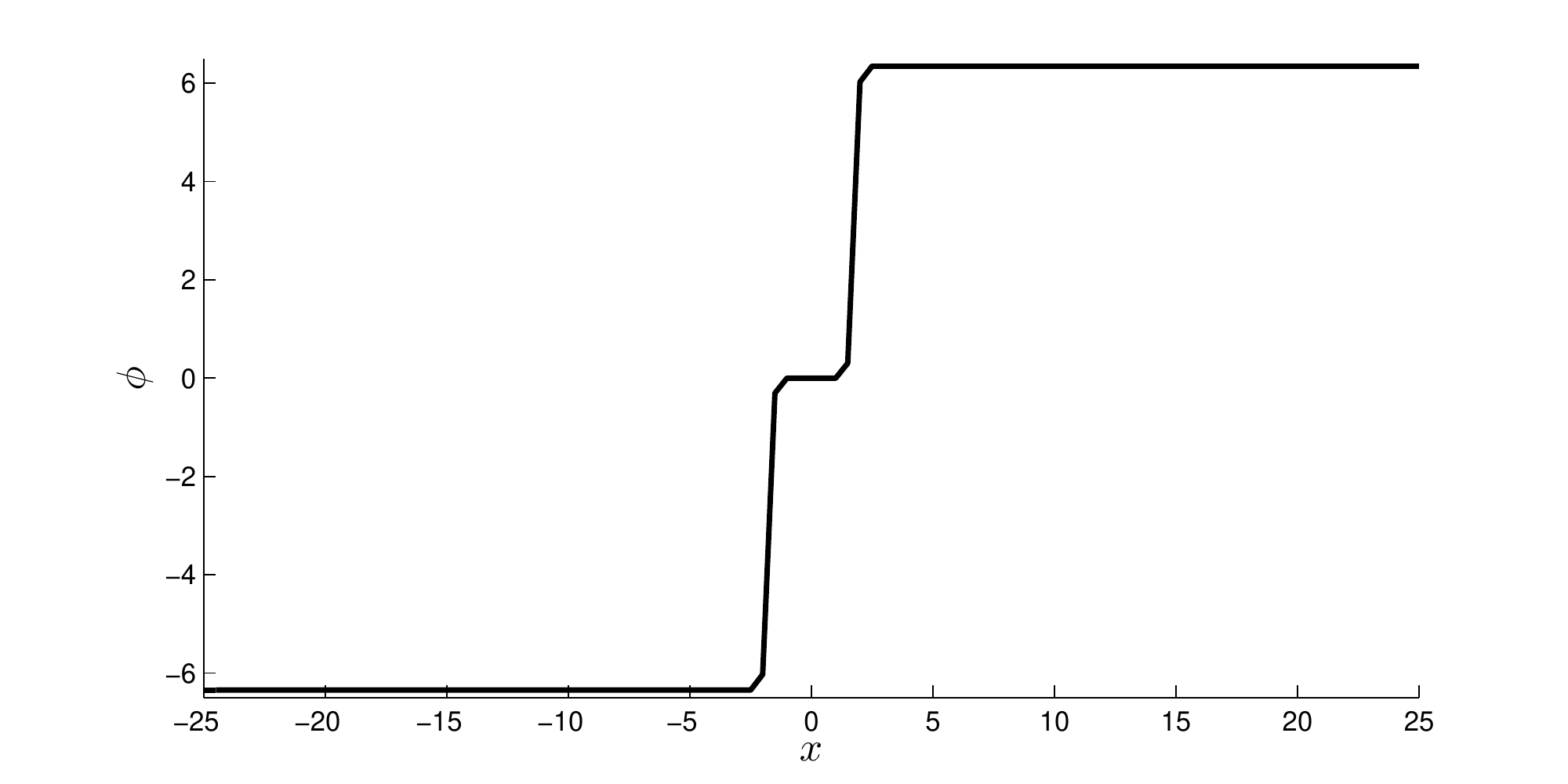}
\label{fig:dynamic_rep_1a}}\qquad
\subfigure[The director field corresponding to the initialized $\phi$. ]{
\includegraphics[width=0.45\textwidth]{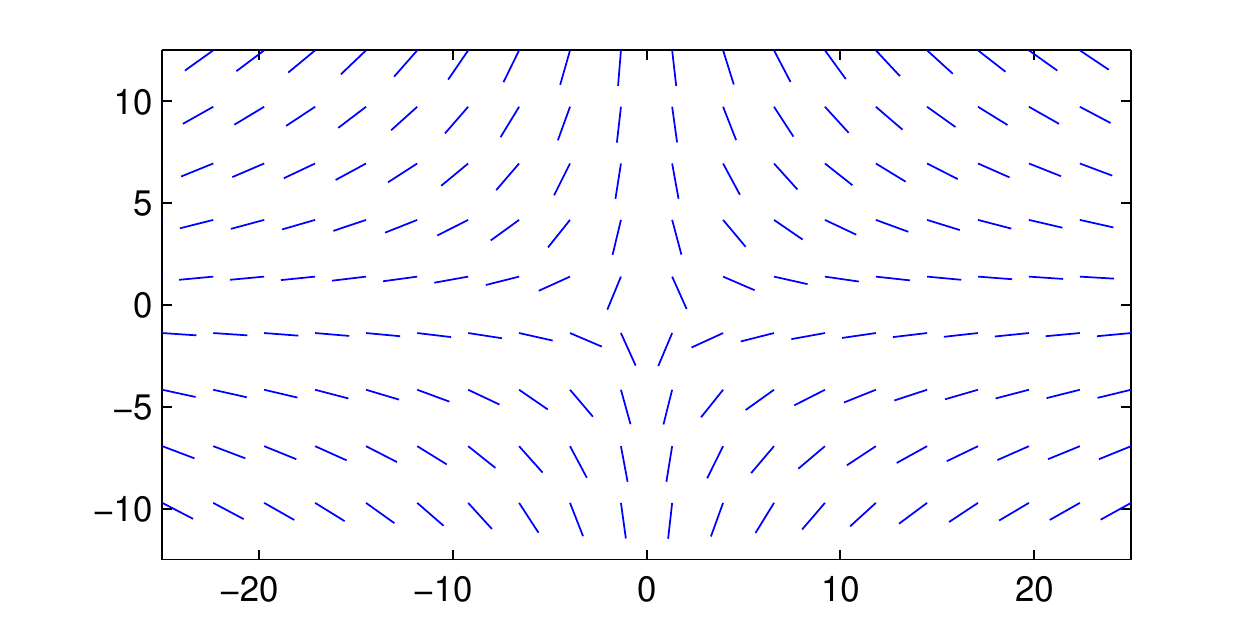}
\label{fig:dynamic_rep_1b}}
\caption{Initialization for disclination repulsion.}
\end {figure}

\begin{figure}[H]
\centering
\includegraphics[width=0.7\linewidth]{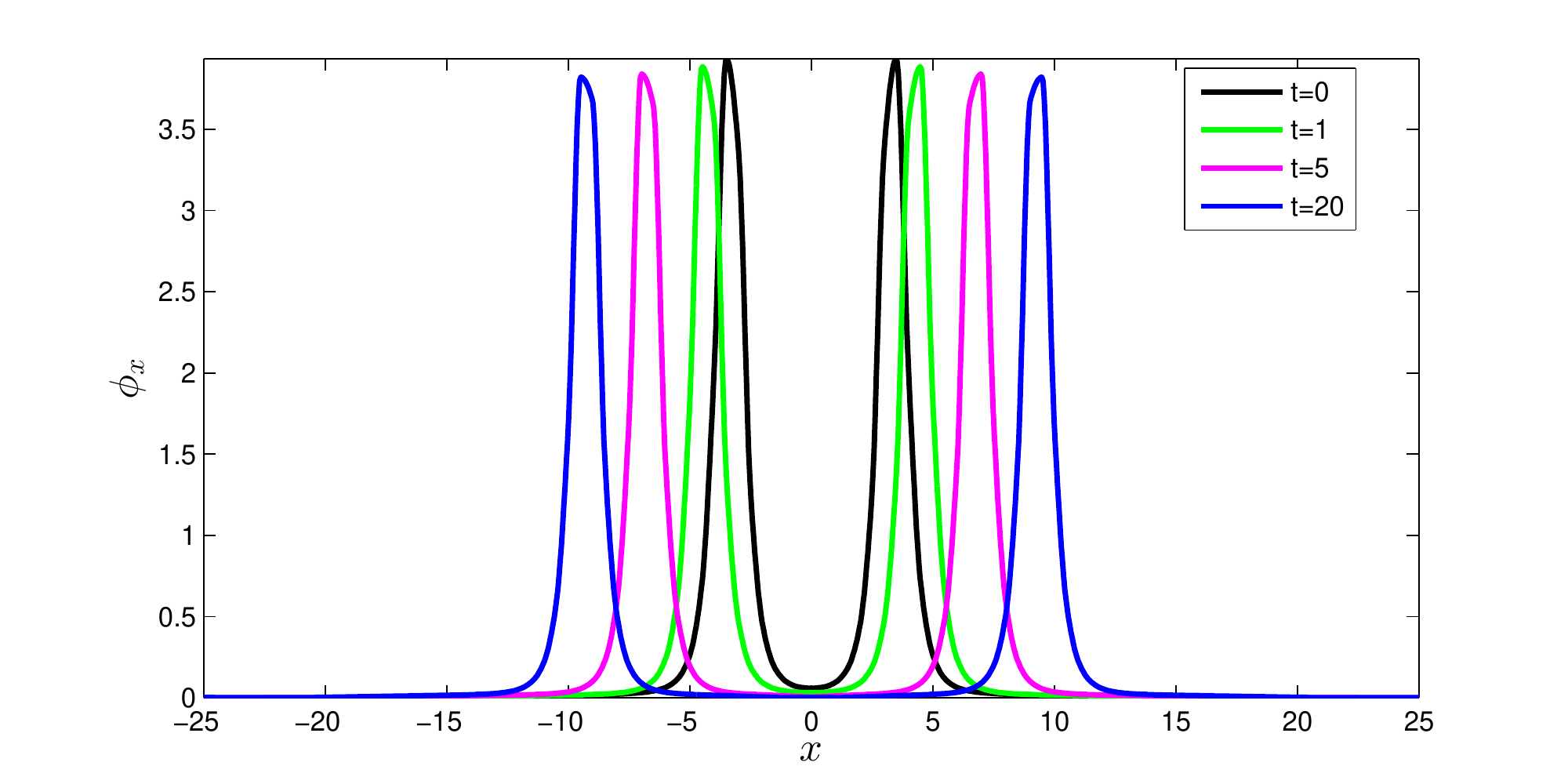}
\caption{$\phi_ x$ snapshots at different time steps. The disclination dipole moves apart.}
\label{fig:dynamic_rep_2}
\end {figure}

The difference between the disclination repulsion and annihilation is that now two disclinations with the same sign are used in the initial condition, as shown in Figure \ref{fig:dynamic_rep_1a}.  Figure \ref{fig:dynamic_rep_1b} shows the director field corresponding to the initial $\phi$ prescription. Figure \ref{fig:dynamic_rep_2} represents the motions of the disclination cores during the dynamic simulation. We observe that the two disclinations move apart due to the repulsive force of elastic origin between them. 

\begin{figure}[H]
\centering
\subfigure[Director snapshot at t=0.]{
\includegraphics[width=0.45\textwidth]{figure/repel/1.pdf}
}
\subfigure[Energy density plot at t=0.]{
\includegraphics[width=0.45\textwidth]{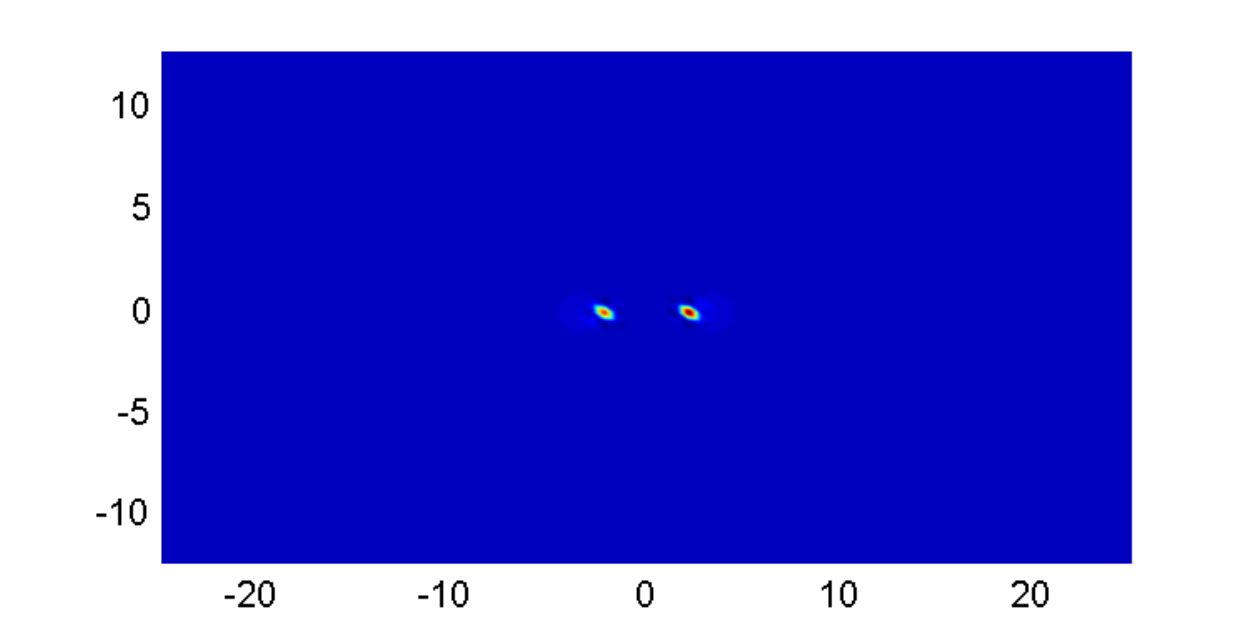}
}
\subfigure[Director snapshot at t=1.]{
\includegraphics[width=0.45\textwidth]{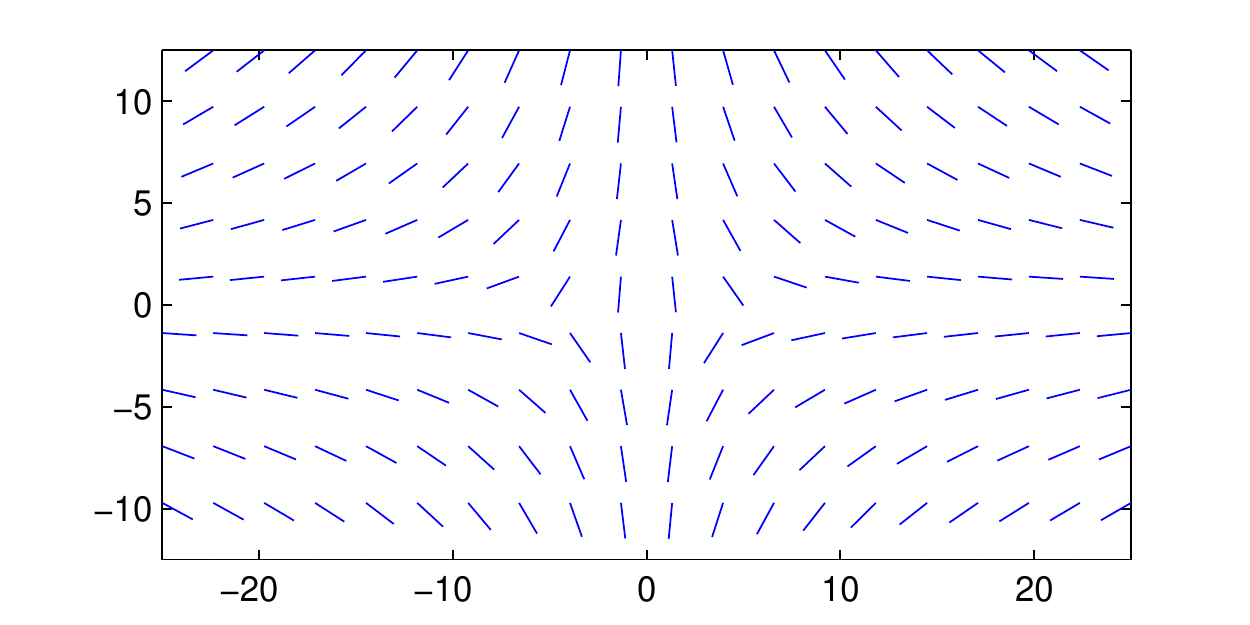}
}
\subfigure[Energy density plot at t=1.]{
\includegraphics[width=0.45\textwidth]{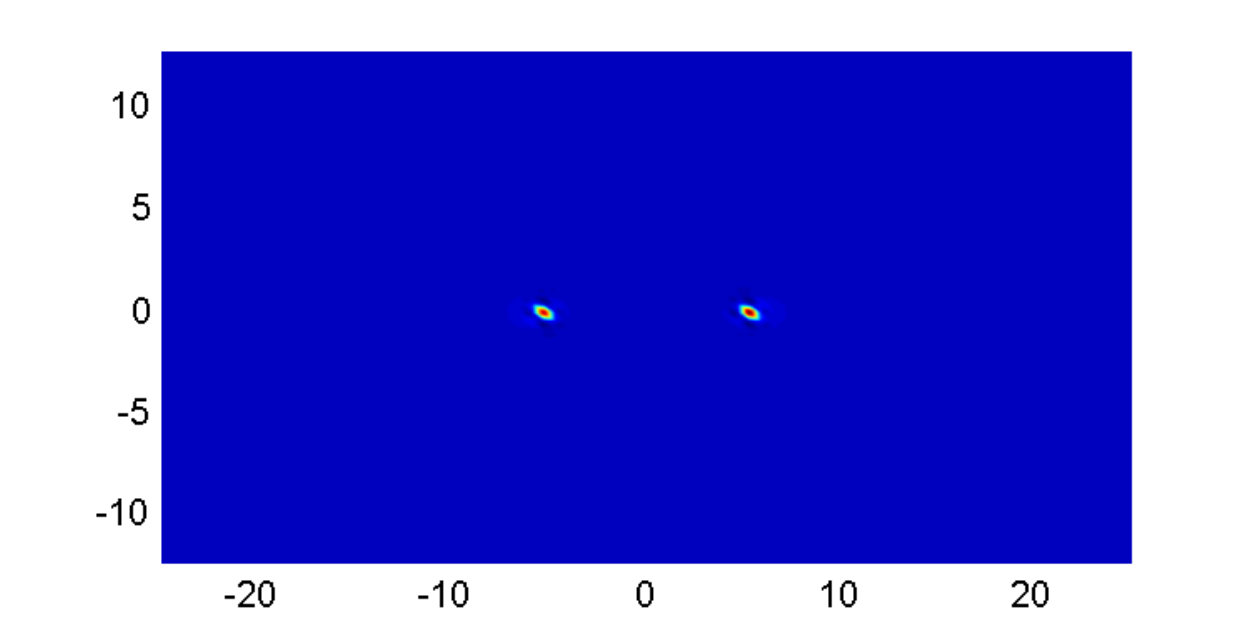}
}
\subfigure[Director snapshot at t=5.]{
\includegraphics[width=0.45\textwidth]{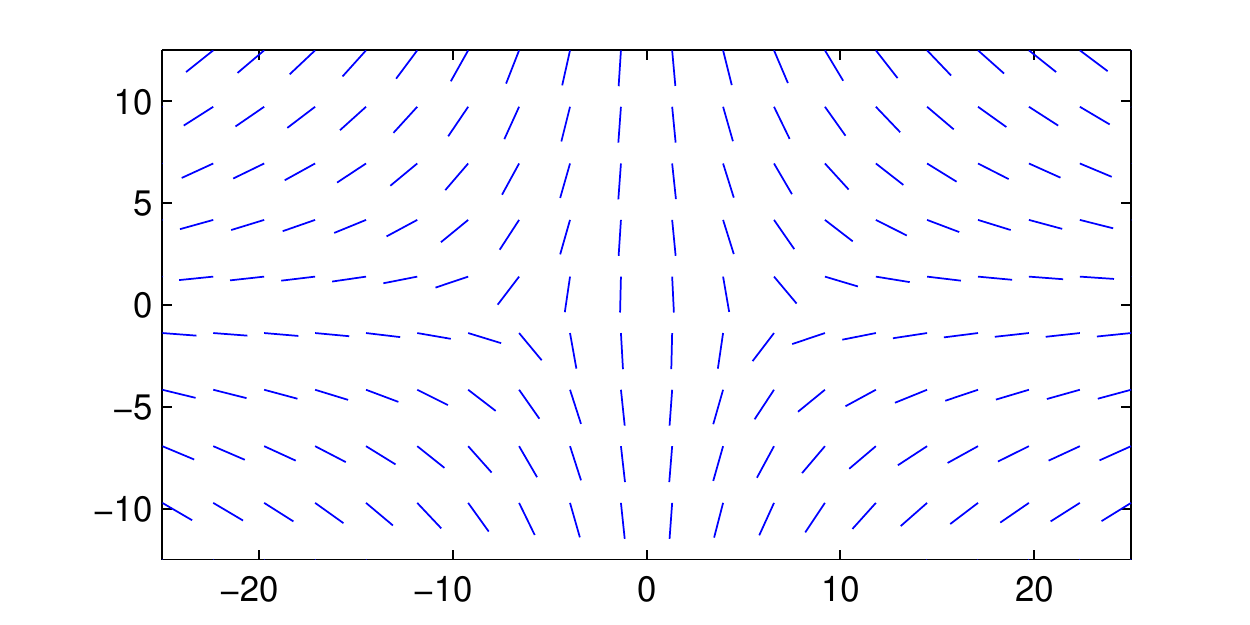}
}
\subfigure[Energy density plot at t=5.]{
\includegraphics[width=0.45\textwidth]{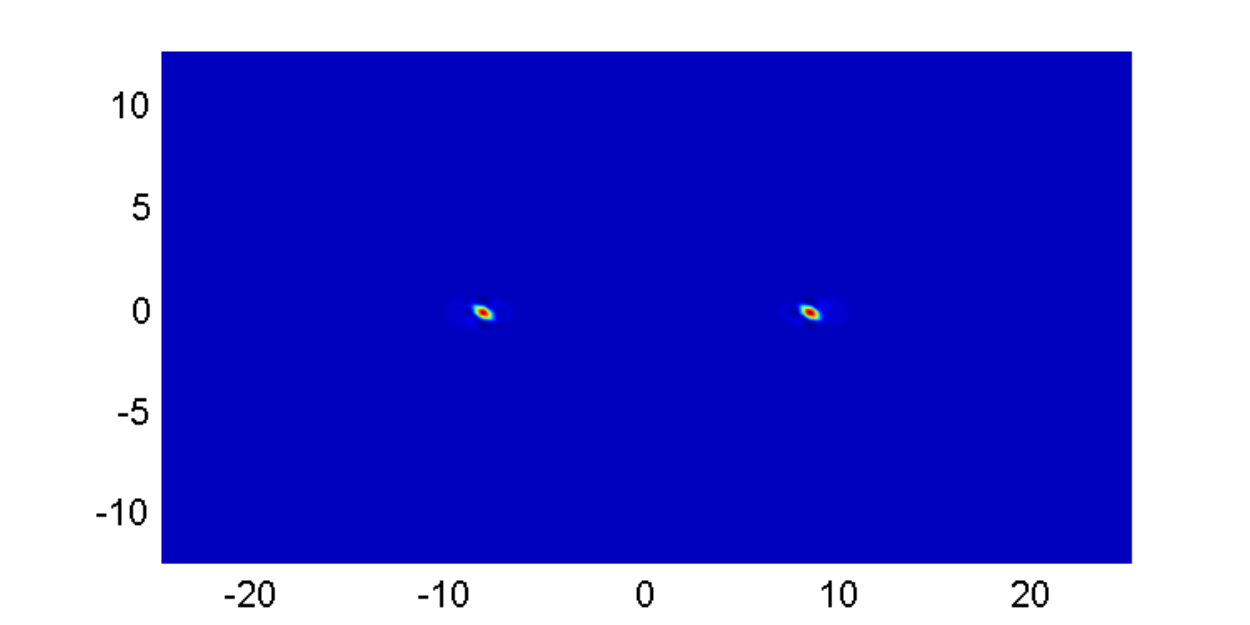}
}
\subfigure[Director snapshot at t=20.]{
\includegraphics[width=0.45\textwidth]{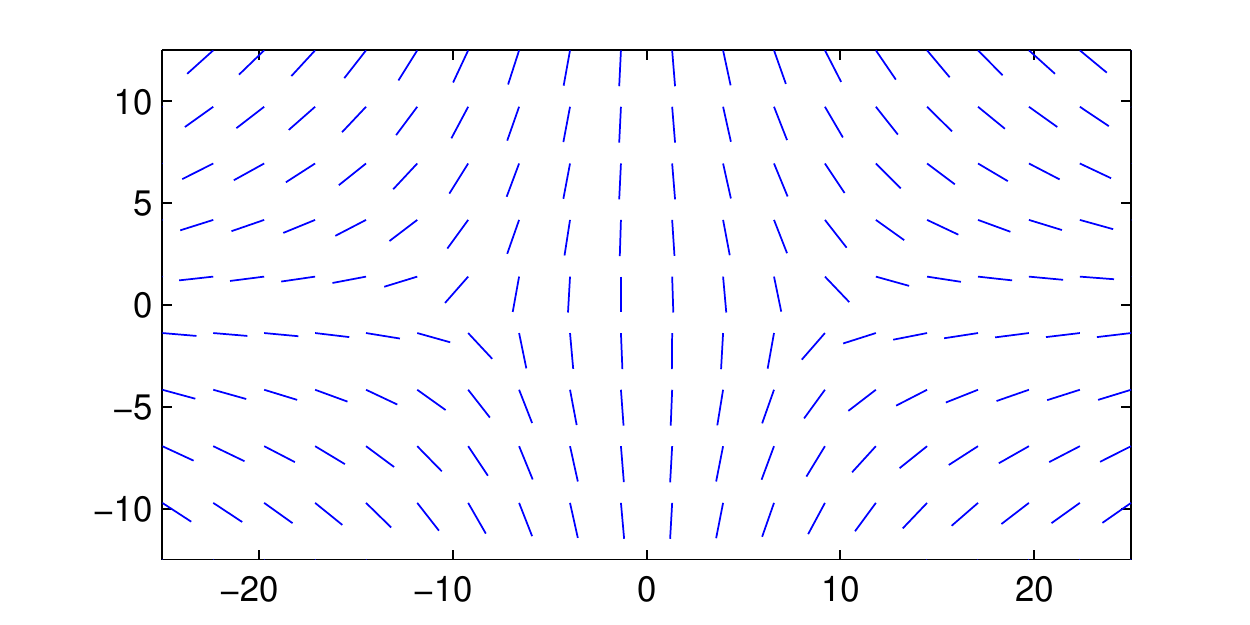}
}
\subfigure[Energy density plot at t=20.]{
\includegraphics[width=0.45\textwidth]{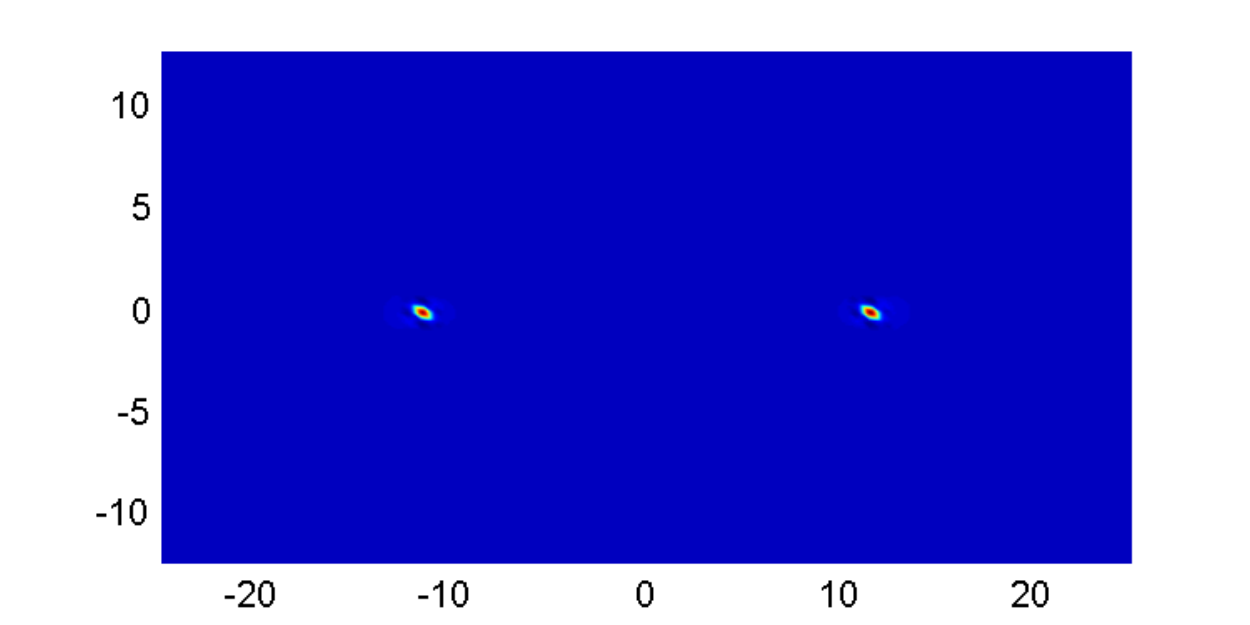}
}
\caption{Snapshots for the director field and energy density at different time steps. The two disclinations move apart and repel each other.}\label{fig:dynamic_rep_3}
\end {figure}

Figure \ref{fig:dynamic_rep_3} shows the director snapshots and energy density plots for disclination repulsion at different time steps. 

\subsection{Velocity profiles with separation distance in different $m$ cases}

\begin{figure}[H]
\centering
\subfigure[Velocity variation of one disclination in a dipole as a function of separation distance. $m=1$.]{
\includegraphics[width=0.6\linewidth]{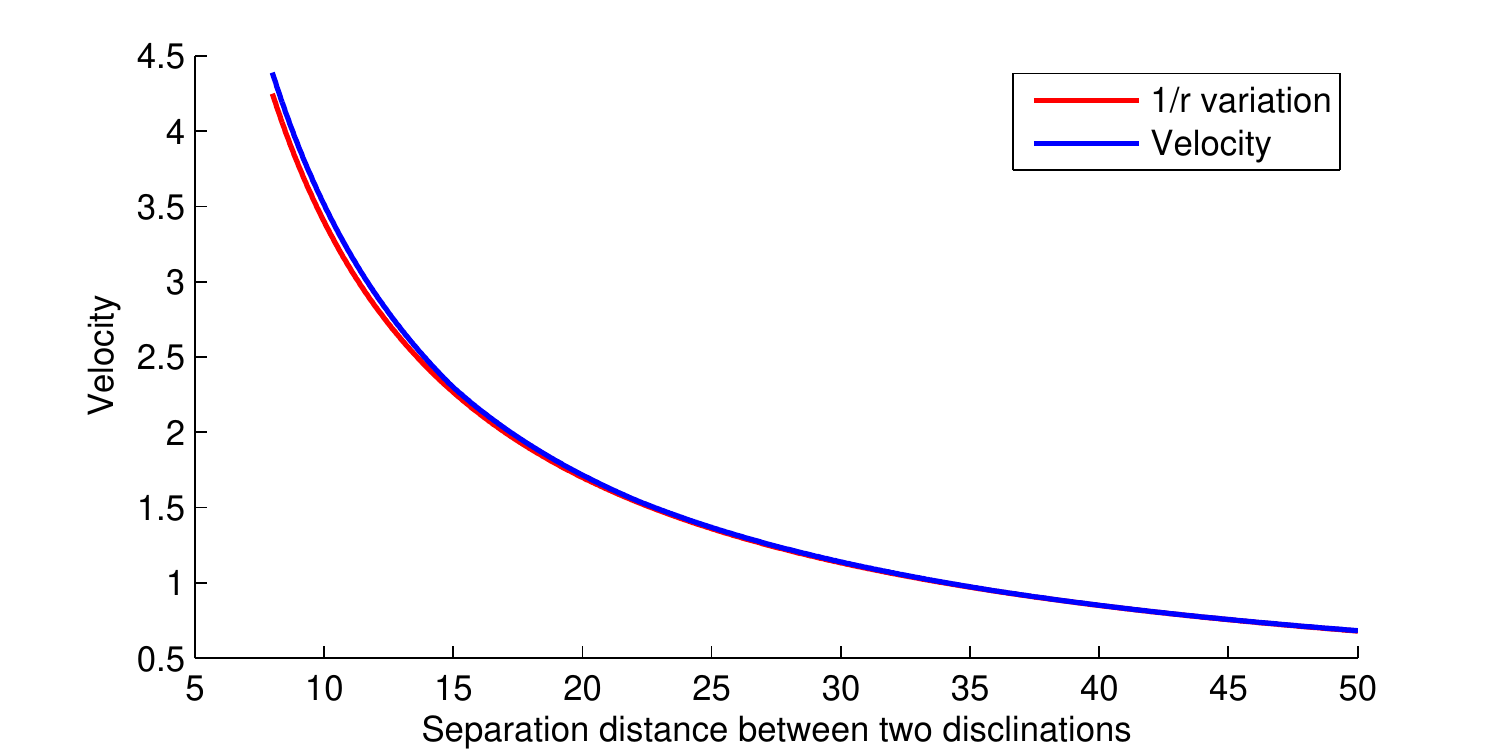}
\label{fig:dynamic_rep_velocity_m_1}}
\subfigure[Velocity variation of one disclination in a dipole as a function of separation distance. $m=2$.]{
\includegraphics[width=0.6\linewidth]{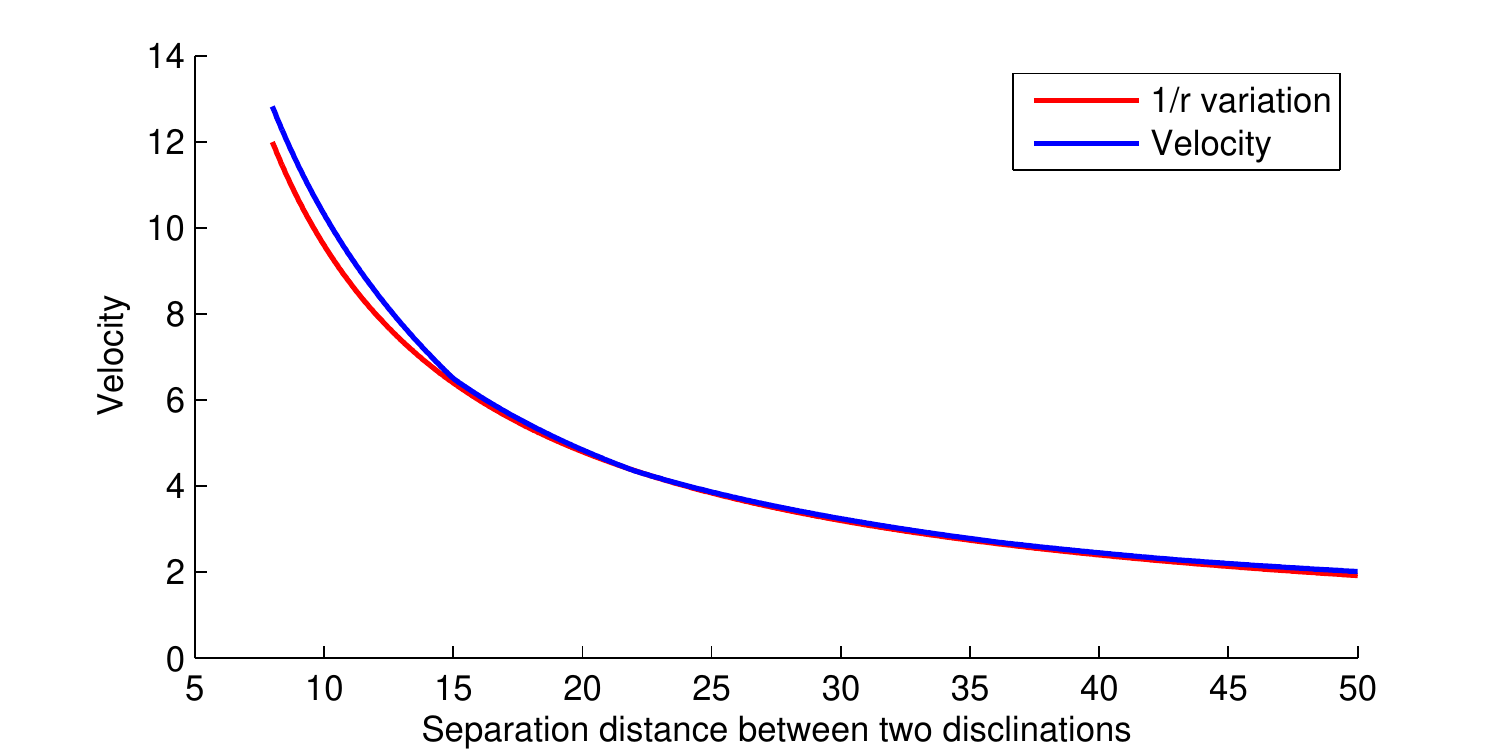}
\label{fig:dynamic_rep_velocity_m_2}}
\caption{Relation between velocity and the separation distance for $m=1$ and $m=2$.}
\label{fig:dynamic_rep_velocity} 
\end {figure}

\begin{figure}[H]
\centering
\includegraphics[width=0.6\linewidth]{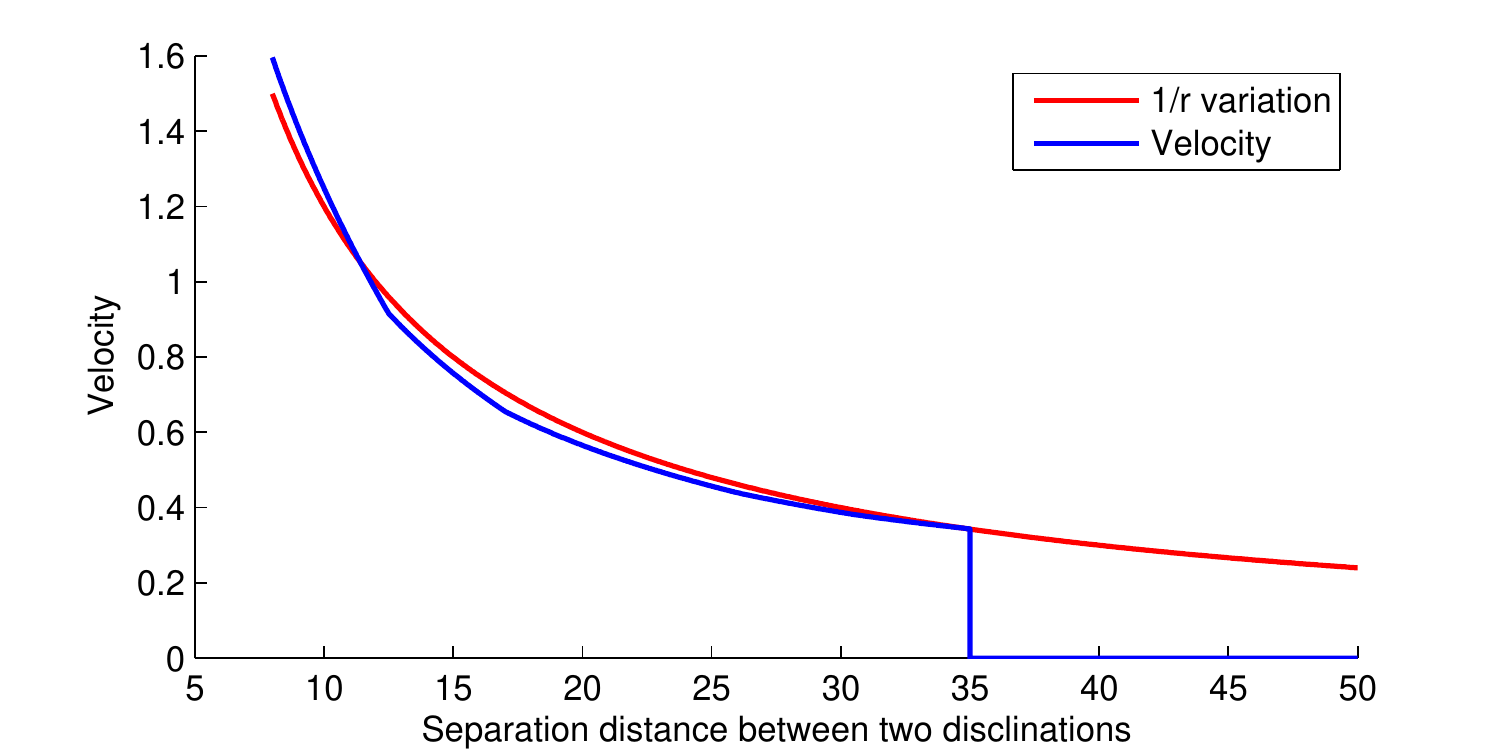}
\caption{Velocity variation of one disclination in a dipole as a function of separation distance. $m=0$.}
\label{fig:dynamic_rep_velocity_m0} 
\end {figure}

Figure \ref{fig:dynamic_rep_velocity} compares the the velocity versus dipole separation relationship of a single disclination in a dipole field, with the expected result from the linear theory of defects \cite{eshelby1980force}. In each case, the two disclinations are initialized with a separation distance of 50 in a body of $100 \times 100$. $a$ is set to be 0.5 in these cases. The core locations are marked at every 200 time steps and the physical discrete time at these instants is recorded. This allows the determination of the (absolute) velocity of (any) one disclination in the dipole pair as a function of the separation distance, as shown in Figure \ref{fig:dynamic_rep_velocity}. In (\ref{eq:sum_gov_eq}), $\tau$ serves as the driving force for the disclination motion. The driving force on one disclination of the dipole core is generated from the elastic interaction with the other disclination, which scales like the reciprocal of the separation distance according to the linear theory of defects. Hence, the motion of the disclinations slows down as the separation distances increases. In Figure \ref{fig:dynamic_rep_velocity}, the red line presents a trend of $1/r$ while the blue line represents the velocity of one disclination. Figure \ref{fig:dynamic_rep_velocity_m_1} shows the relationship between velocity and separation distance in the $m=1$ case. Thus, the velocity matches with $1/r$ trend very well in this case. Figure \ref{fig:dynamic_rep_velocity_m_2} shows the relationship  between velocity and separation distance in $m=2$ case. In this case, the velocity is the largest of all the three cases and matches $1/r$ trend in the large separation distance range. Within the separation distance from 5 to 15, the disclinations begin to annihilate.  For $m=0$ case, the disclinations are found not to move until the separation distance is less than $35$. Figure \ref{fig:dynamic_rep_velocity_m0} shows the velocity profile of the $m=0$ case. It shows that the velocity does not match the $1/r$ very well when the separation distance is small but has a better agreement with $1/r$ trend as far away separation distance.

We note here that there is no reason, \emph{a-priori}, for the velocity in our nonlinear, \emph{dynamic} model to match the expected result from the notion of `non-Newtonian forces' of static defect theory \cite{eshelby1980force, ericksen1995remarks} but our results demonstrate that to a large extent there is consonance between our results and that of traditional defect theory. However, the differences are noteworthy as well - in particular the emergence of apparent `intrinsic pinning' in a translationally-invariant pde model (cf.\cite{zhang2015single} where the details of this phenomena are investigated in greater detail) for the case $m = 0$, the most natural kinetic model in our setting.

% Figure \ref{fig:dynamic_rep_velocity_small} shows the relationship between velocity and separation distance within a small finite body. The velocity is higher as the disclinations get closer to the boundary as the boundary with $0$ moment boundary condition `attracts' the disclination due to an `image force' effect. Figure \ref{fig:dynamic_rep_velocity_large} shows the relationship within a large body, which shows a good agreement with a $1/r$ trend.

\subsection{Disclination dissociation}\label{sec:one_split}

We model the process of a strength-one disclination dissociating into two strength-half disclinations. Dissociations of a positive and a negative strength-one disclination are simulated.

\begin{figure}[H]
\centering
\subfigure[Initialization of $\phi$ for strength $+1$ disclination splitting. A $\phi$ field corresponding to strength $+1$ disclination is prescribed.]{
\includegraphics[width=0.45\textwidth]{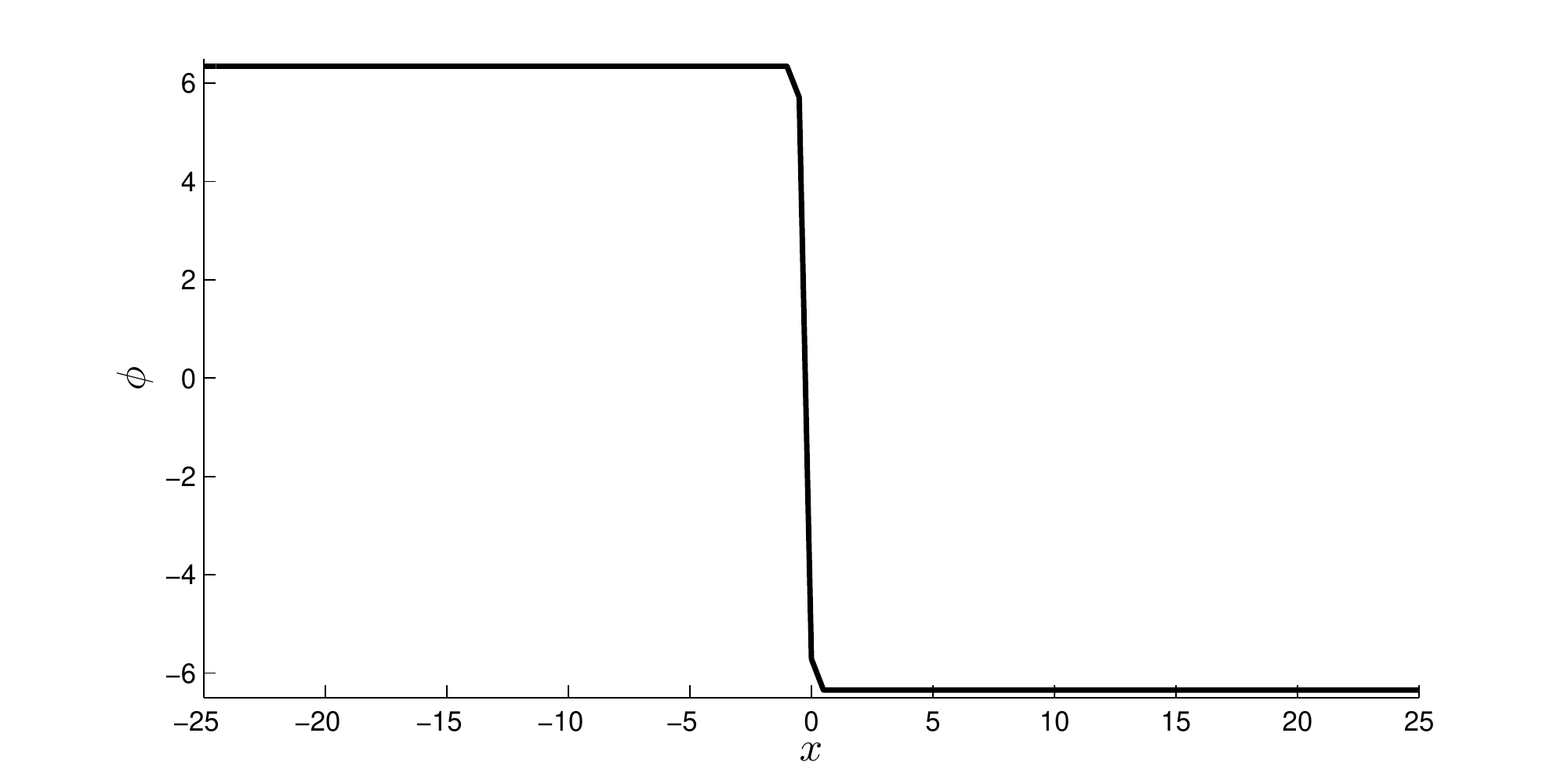}
\label{fig:split_positive_1a}}\qquad
\subfigure[Director field corresponding to initialized $\phi$. ]{
\includegraphics[width=0.45\textwidth]{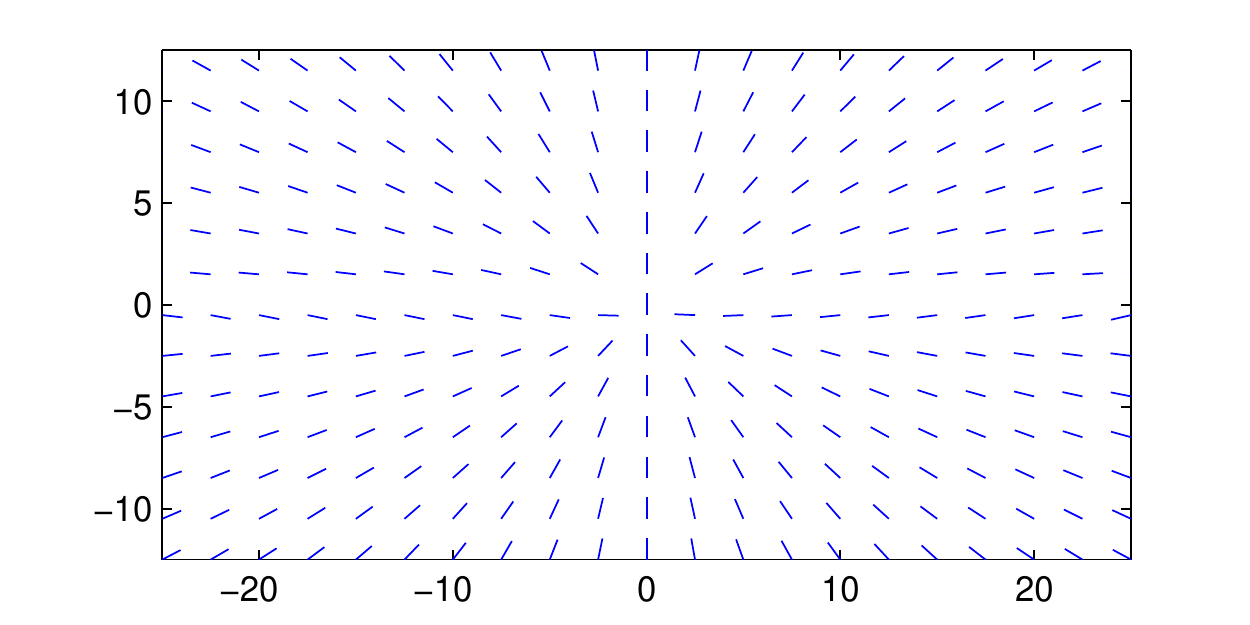}
\label{fig:split_positive_1b}}
\caption{Initialization for strength $+1$ disclination dissociation.}
\end {figure}

\begin{figure}[H]
\centering
\includegraphics[width=0.8\linewidth]{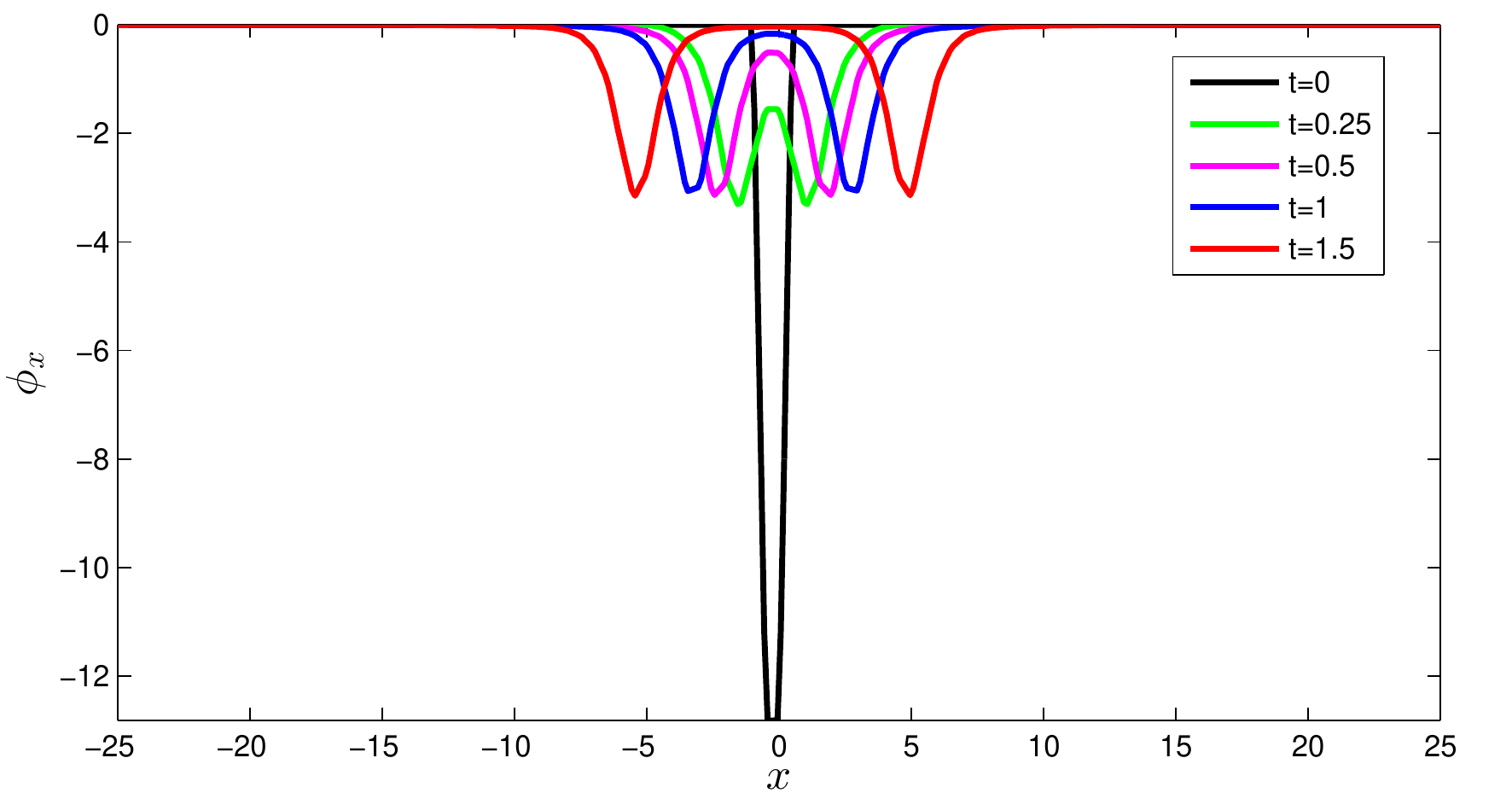}
\caption{$\phi_x$ snapshots at different time steps. It shows one $+1$ disclination splits into two $+1/2$ disclinations and these two disclinations repel each other.}
\label{fig:split_positive_2}
\end {figure}

We prescribe a strength $+1$ disclination at the center of the body as shown in Figure \ref{fig:split_positive_1a}. Figure \ref{fig:split_positive_1b} shows the director field corresponding to the initial $\bflambda$. The initial condition on the $\theta$ field is generated by solving for moment equilibrium using the Neumann boundary condition on the director field corresponding to the moment distribution on the boundary generated from the exact solution for a strength $+1$ disclination in an infinite medium. During evolution, a 0-moment Neumann boundary condition is imposed. Figure \ref{fig:split_positive_2} shows how the strength $+1$ disclination splits into two $+1/2$ disclinations. We observe that the strength $+1$ disclination first splits into two strength $+1/2$ disclinations and then these two strength $+1/2$ disclinations move apart and repel each other. 

\begin{figure}[H]
\centering
\subfigure[Director snapshot at t=0.]{
\includegraphics[width=0.45\textwidth]{figure/splitting_positive/1.pdf}
}
\subfigure[Energy density plot at t=0.]{
\includegraphics[width=0.45\textwidth]{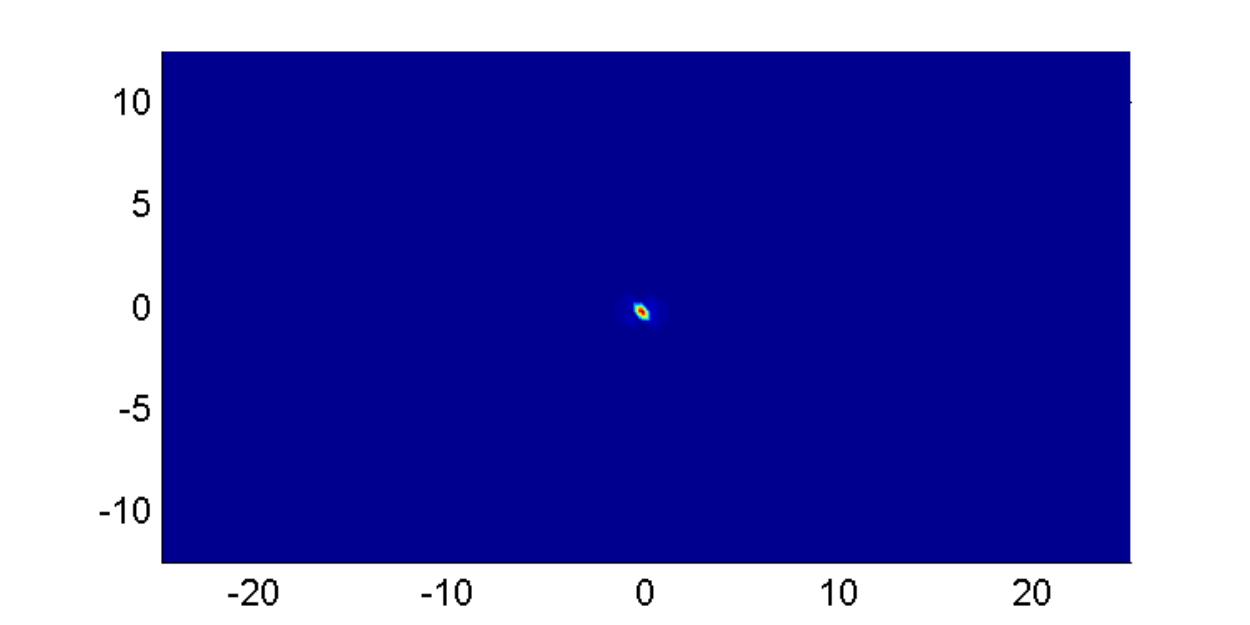}
}
\subfigure[Director snapshot at t=0.25.]{
\includegraphics[width=0.45\textwidth]{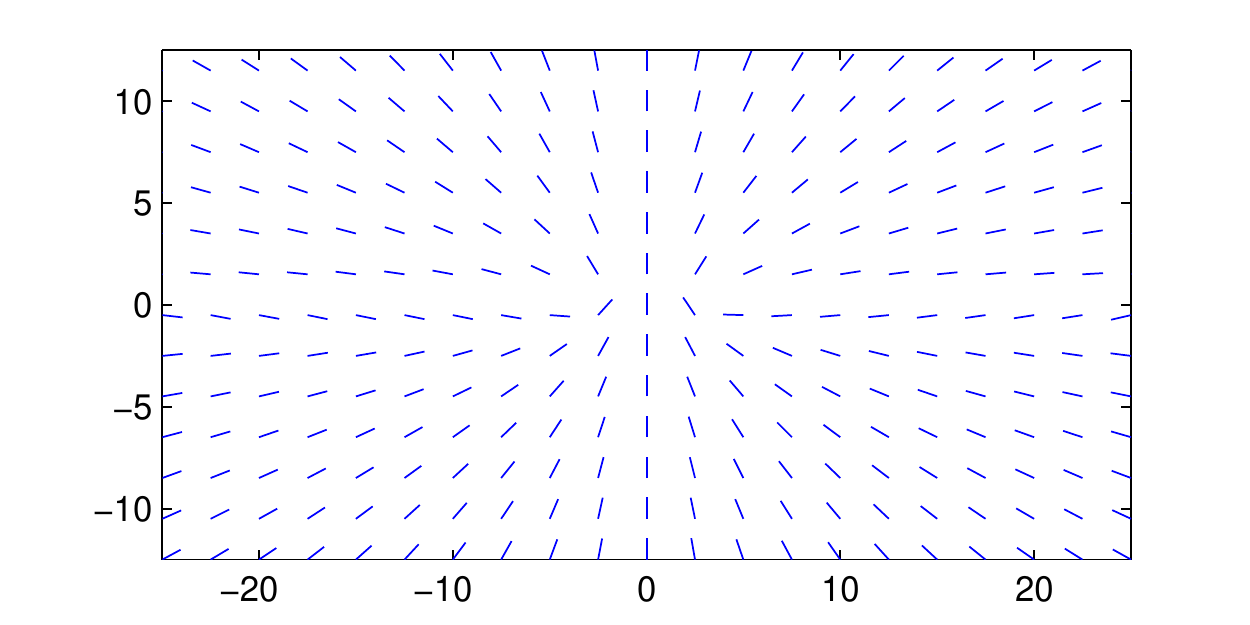}
}
\subfigure[Energy density plot at t=0.25.]{
\includegraphics[width=0.45\textwidth]{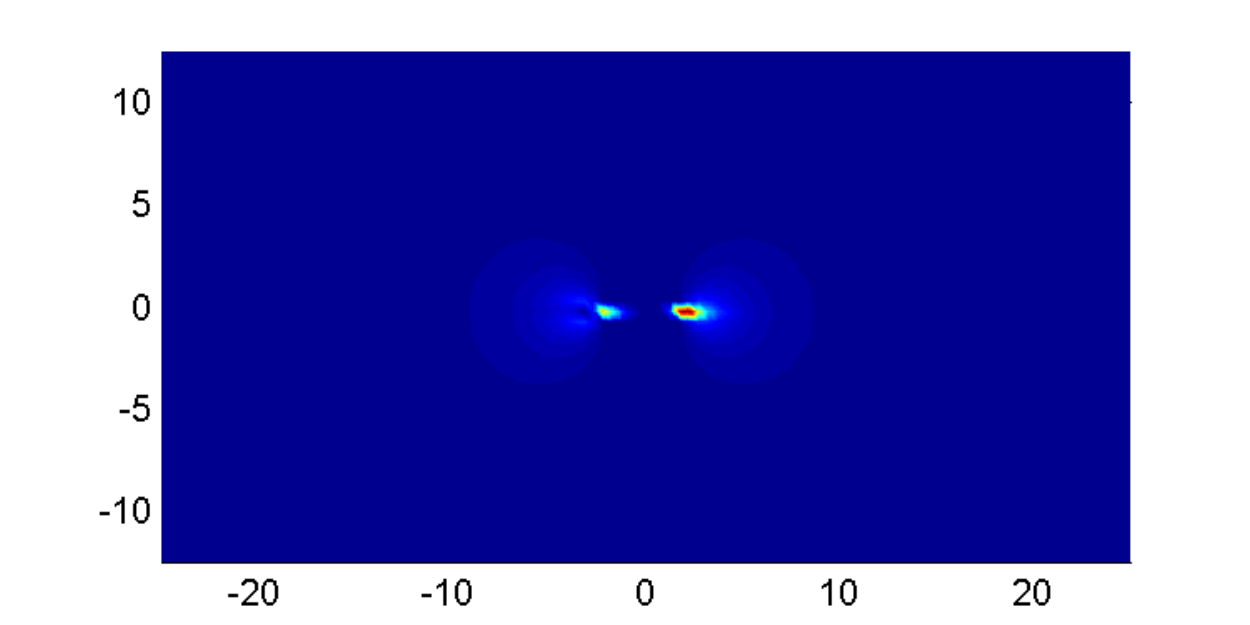}
}
\subfigure[Director snapshot at t=1.]{
\includegraphics[width=0.45\textwidth]{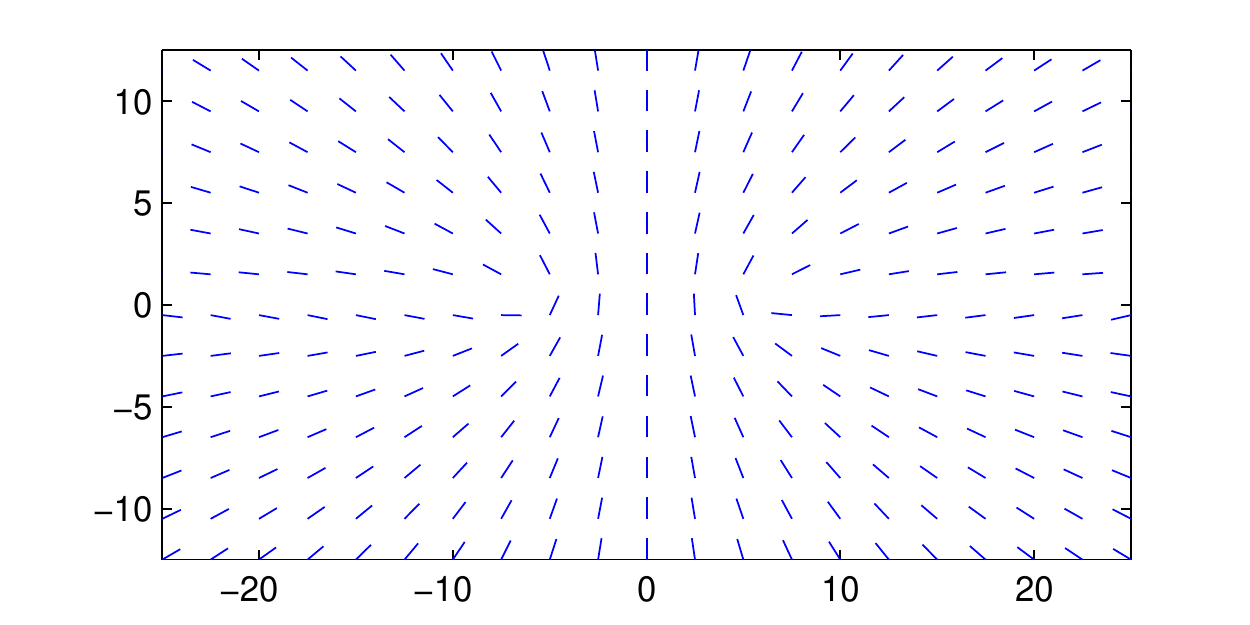}
}
\subfigure[Energy density plot at t=1.]{
\includegraphics[width=0.45\textwidth]{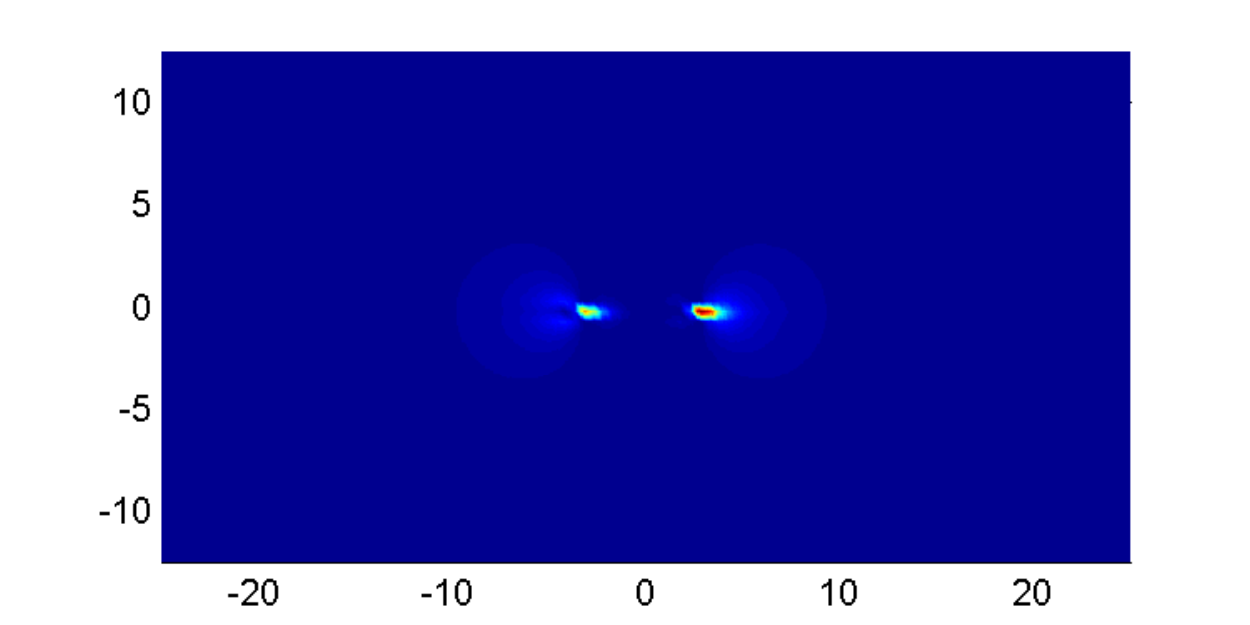}
}
\subfigure[Director snapshot at t=1.5.]{
\includegraphics[width=0.45\textwidth]{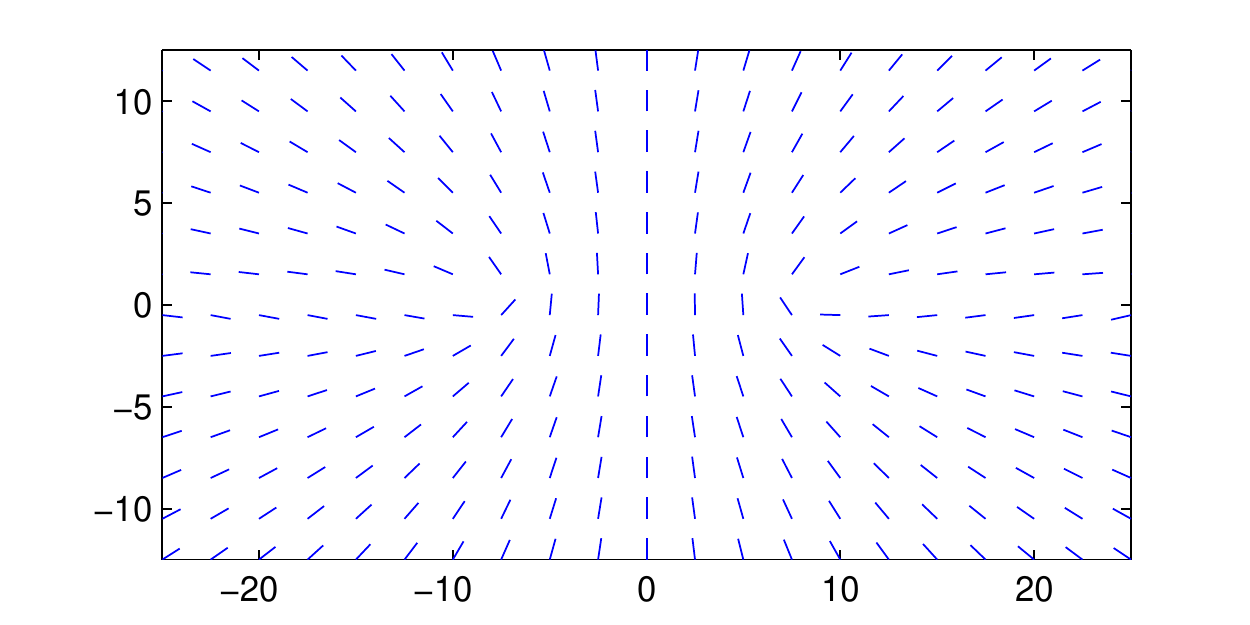}
}
\subfigure[Energy density plot at t=1.5.]{
\includegraphics[width=0.45\textwidth]{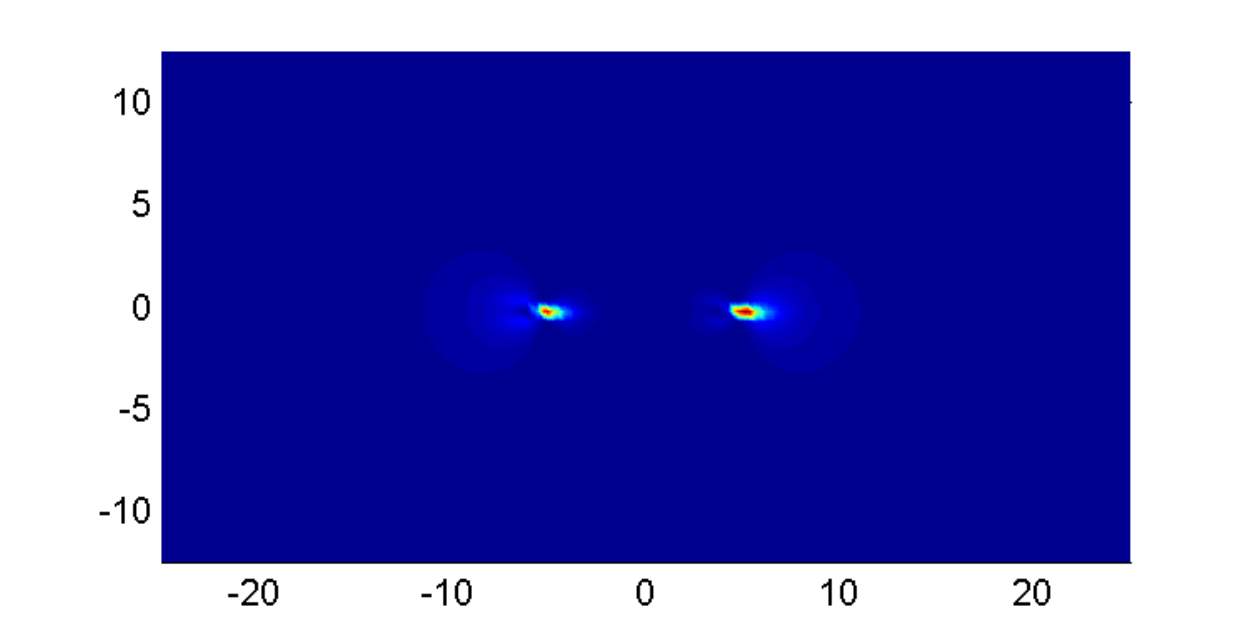}
}
\caption{Snapshots for director field and energy density at different time steps. One disclination splits into two half disclinations and these two disclinations repel each other.}\label{fig:split_positive_3}
\end {figure}

Similarly, the director field behaviors are shown in Figure \ref{fig:split_positive_3}. Initially, the director field represents a strength $+1$ disclination. And then it splits into two strength $+1/2$ disclinations from the core and these two disclinations are both subject to repulsion. From the energy density plots, we can also see that the energy core splits into two cores and these two energy cores repel each other.  

\begin{figure}[H]
\centering
\subfigure[Initialization of $\phi$ for $-1$ disclination dissociation. $\phi$ fields corresponding to strength $-1$ disclination is prescribed.]{
\includegraphics[width=0.45\textwidth]{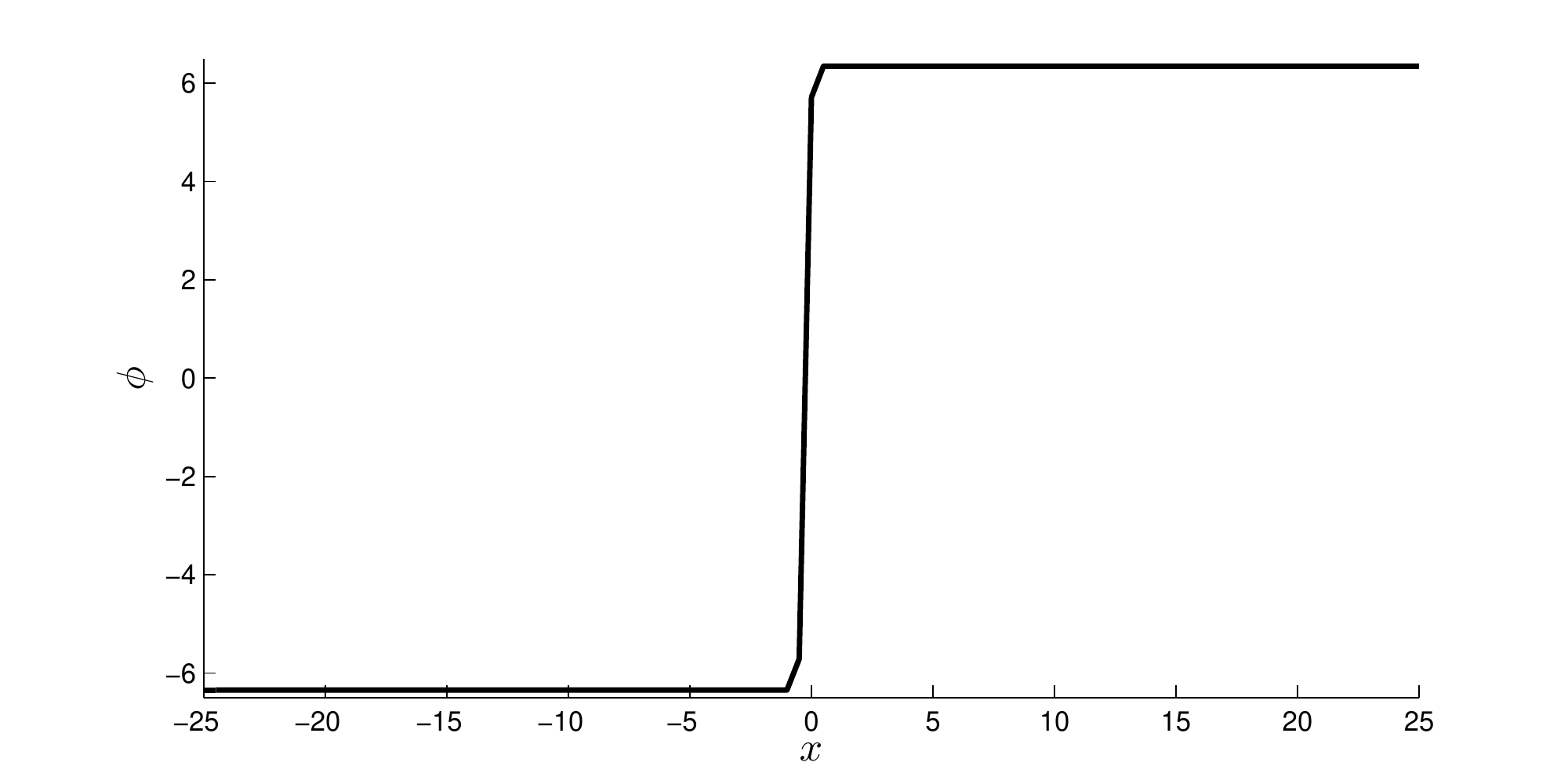}
\label{fig:split_negative_1a}} \qquad
\subfigure[Director field corresponding to initialized $\phi$.]{
\includegraphics[width=0.45\textwidth]{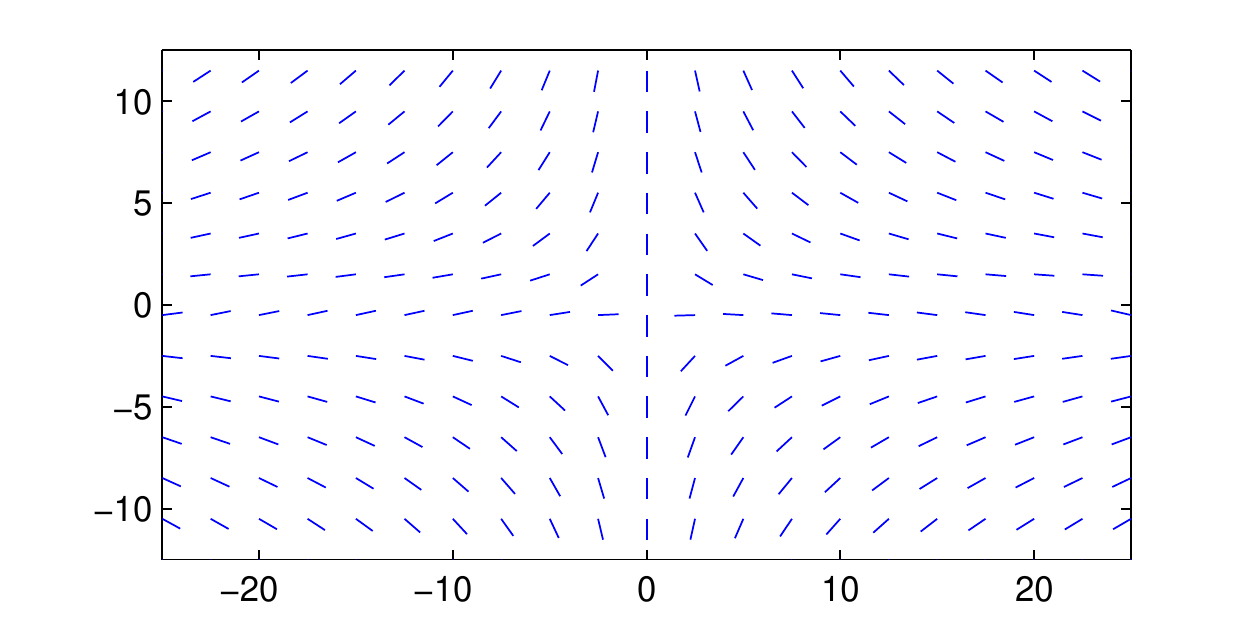}
\label{fig:split_negative_1b}}
\caption{Initialization for strength $-1$ disclination dissociation.}
\end {figure}

\begin{figure}[H]
\centering
\includegraphics[width=0.7\linewidth]{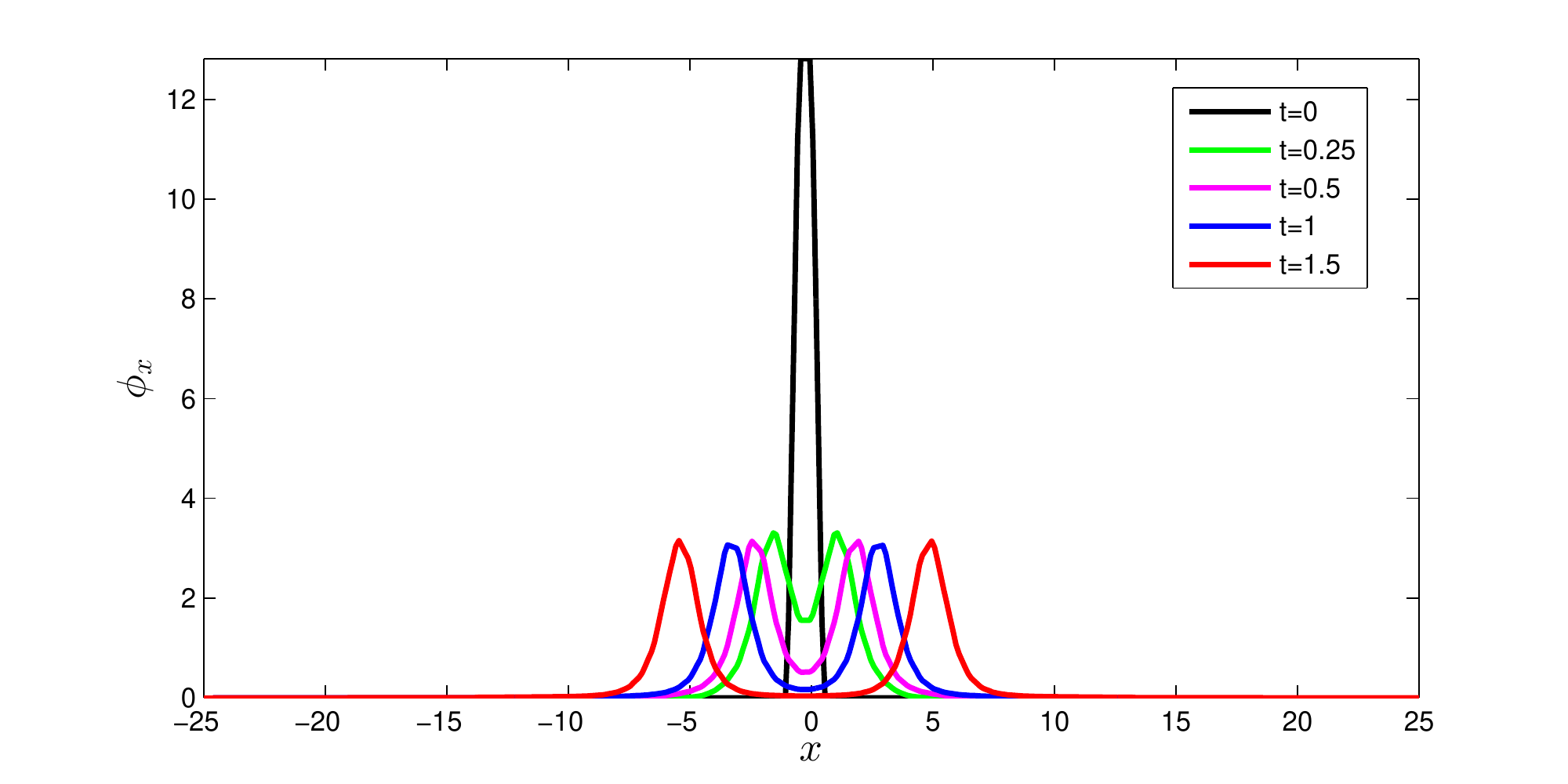}
\caption{$\phi_x$ snapshots at different time steps. It shows $-1$ disclination splits into two $-1/2$ disclinations and these two disclinations repel each other.}
\label{fig:split_negative_2}
\end {figure}

The splitting of a strength $-1$ disclination is similar. We prescribed a strength $-1$ disclination at the center of the body as shown in Figure \ref{fig:split_negative_1a} and the boundary conditions were set up following exactly the procedure for the previous case, accounting for the change in strength of the disclination. Figure \ref{fig:split_negative_2} shows the process of the strength $-1$ disclination splitting into two strength $-1/2$ ones. 

\begin{figure}[H]
\centering
\subfigure[Director snapshot at t=0.]{
\includegraphics[width=0.45\textwidth]{figure/splitting_negative/1.pdf}
}
\subfigure[Energy density plot at t=0.]{
\includegraphics[width=0.45\textwidth]{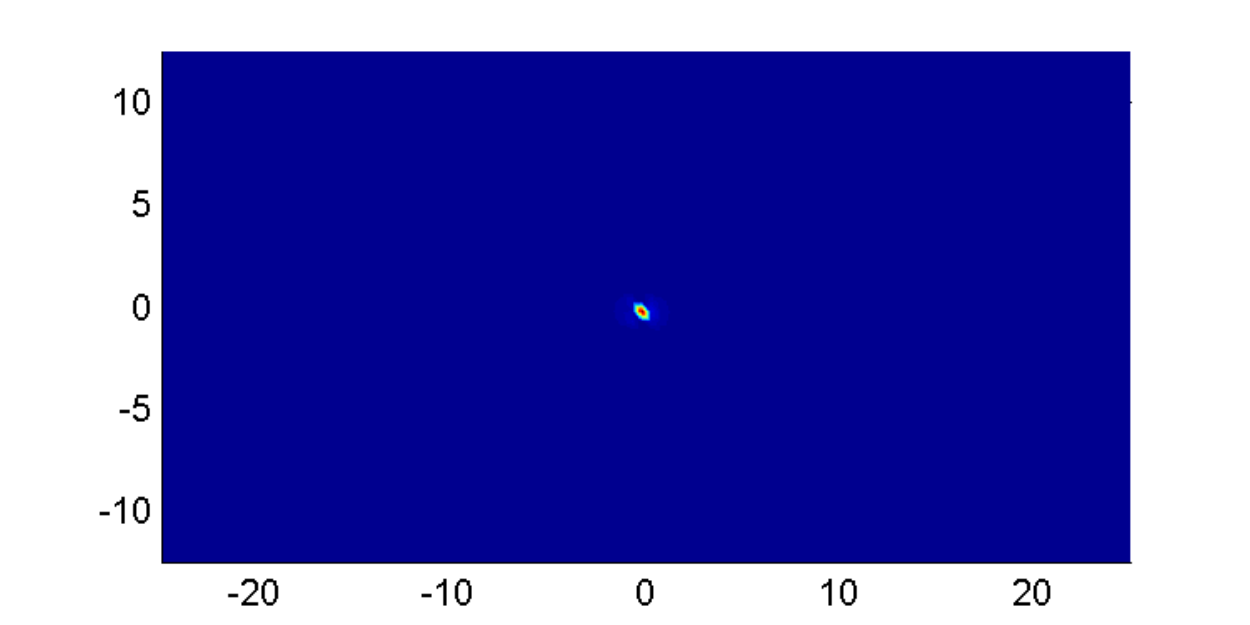}
}
\subfigure[Director snapshot at t=0.25.]{
\includegraphics[width=0.45\textwidth]{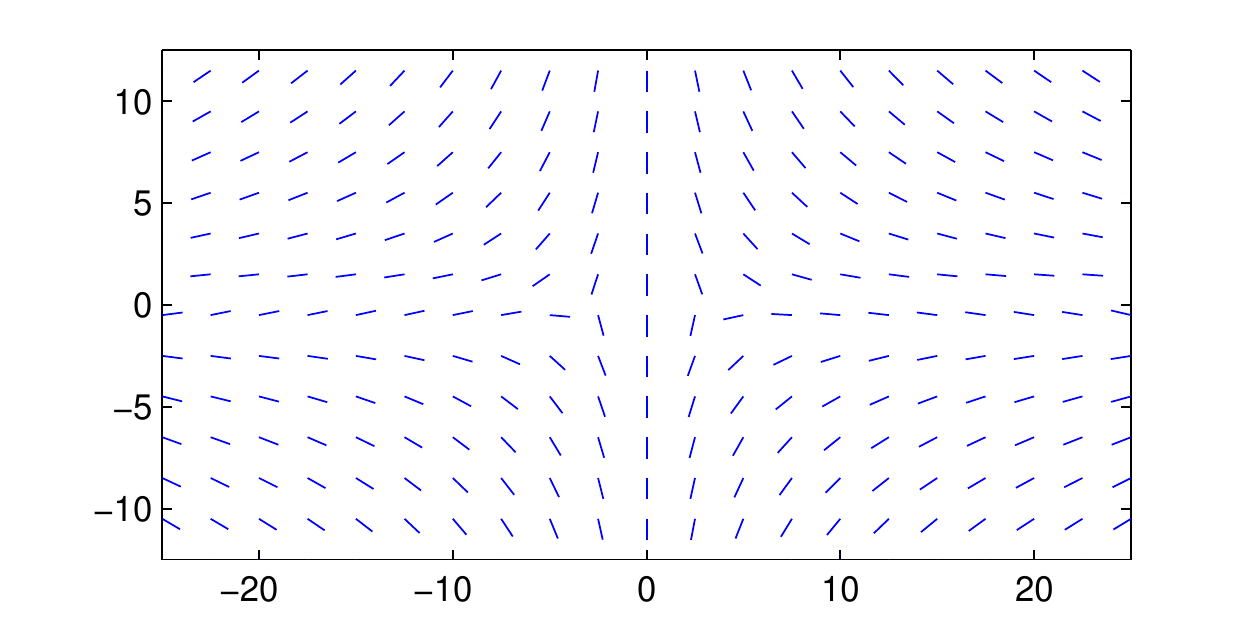}
}
\subfigure[Energy density plot at t=0.25.]{
\includegraphics[width=0.45\textwidth]{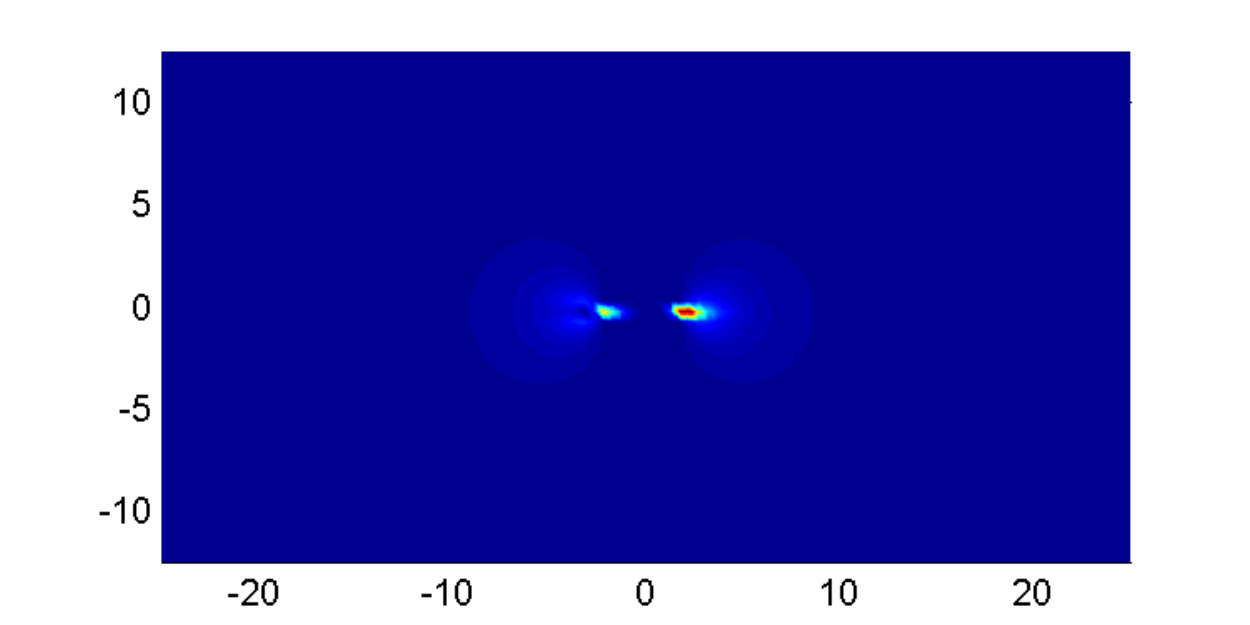}
}
\subfigure[Director snapshot at t=1.]{
\includegraphics[width=0.45\textwidth]{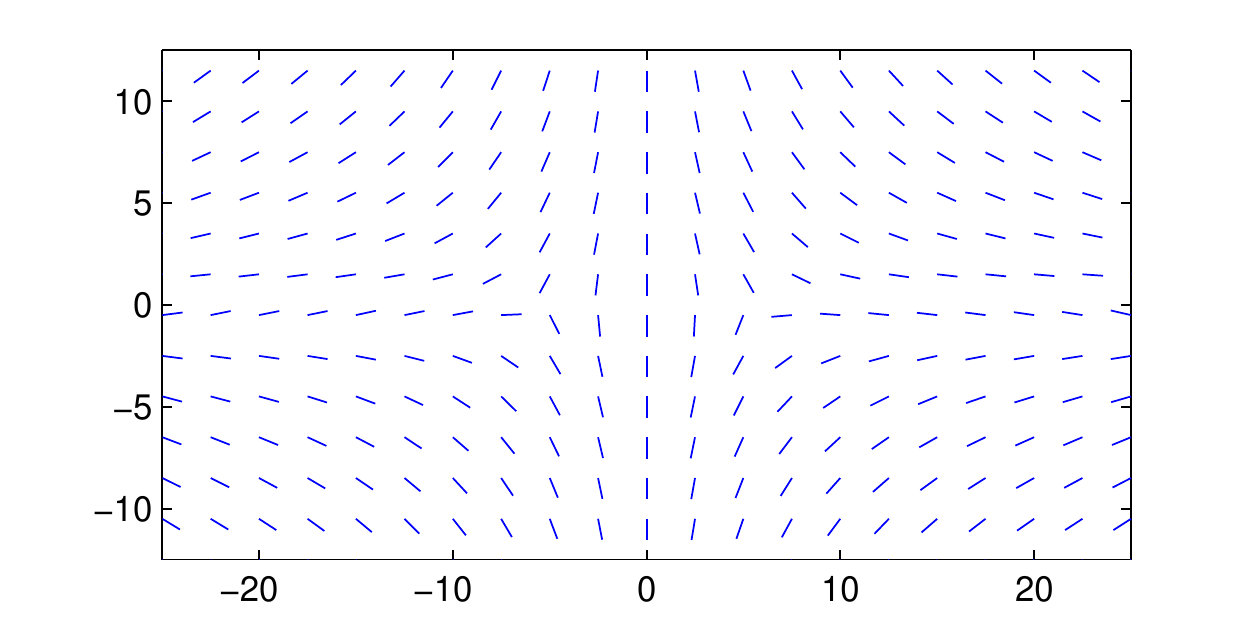}
}
\subfigure[Energy density plot at t=1.]{
\includegraphics[width=0.45\textwidth]{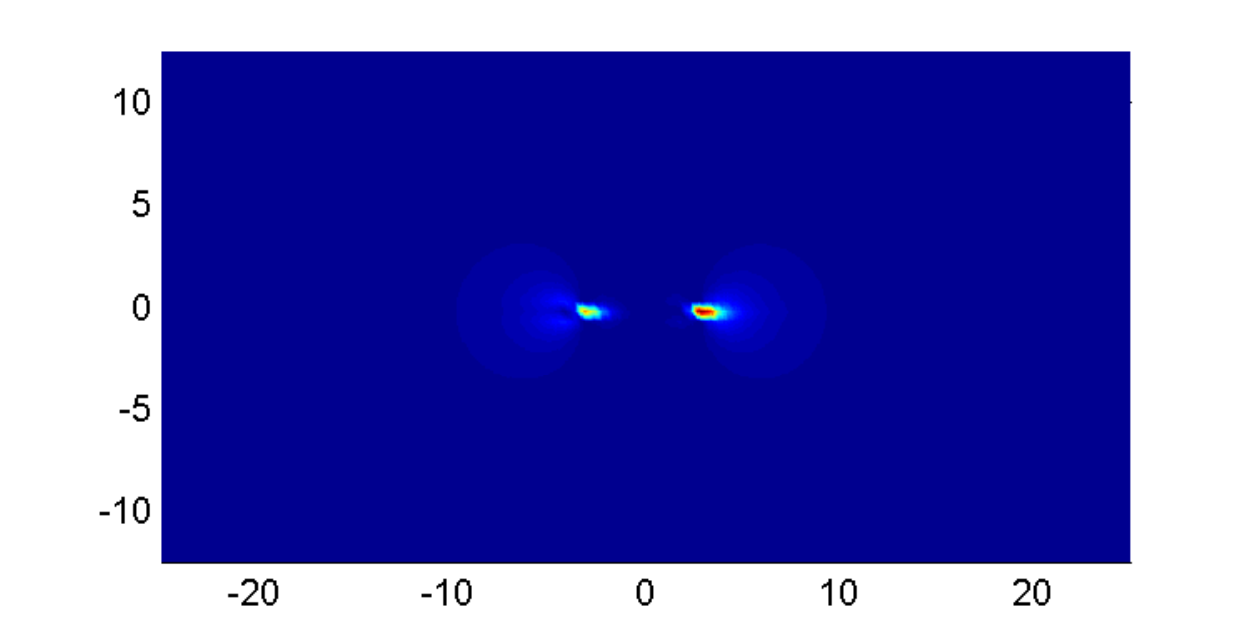}
}
\subfigure[Director snapshot at t=1.5.]{
\includegraphics[width=0.45\textwidth]{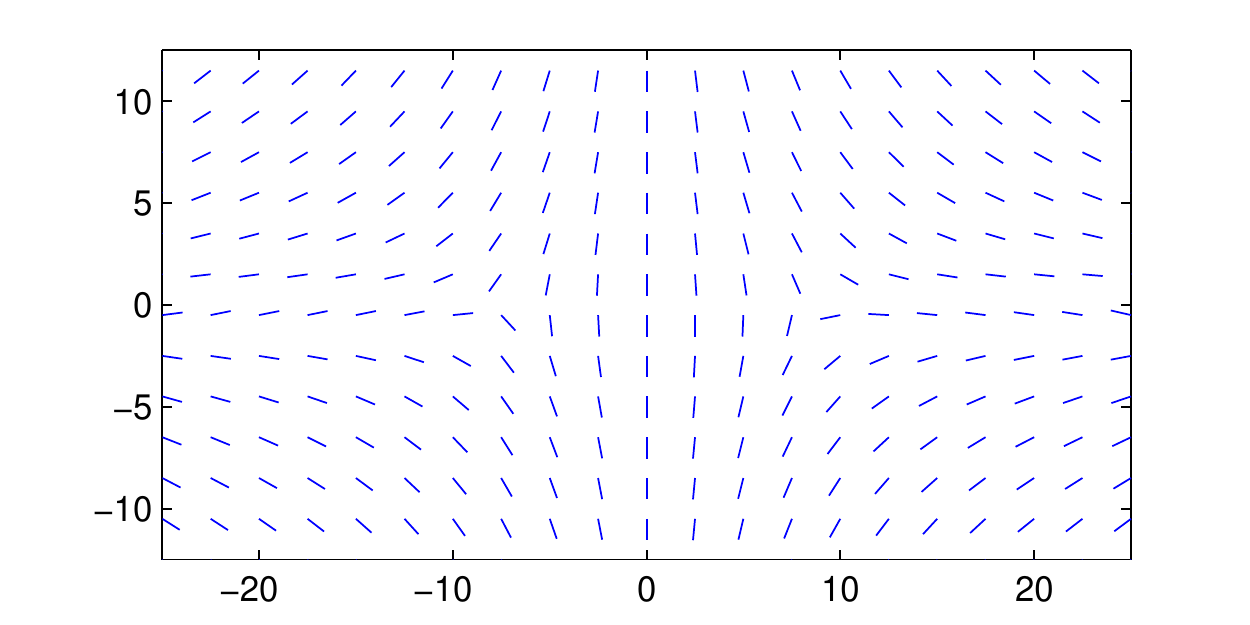}
}
\subfigure[Energy density plot at t=1.5.]{
\includegraphics[width=0.45\textwidth]{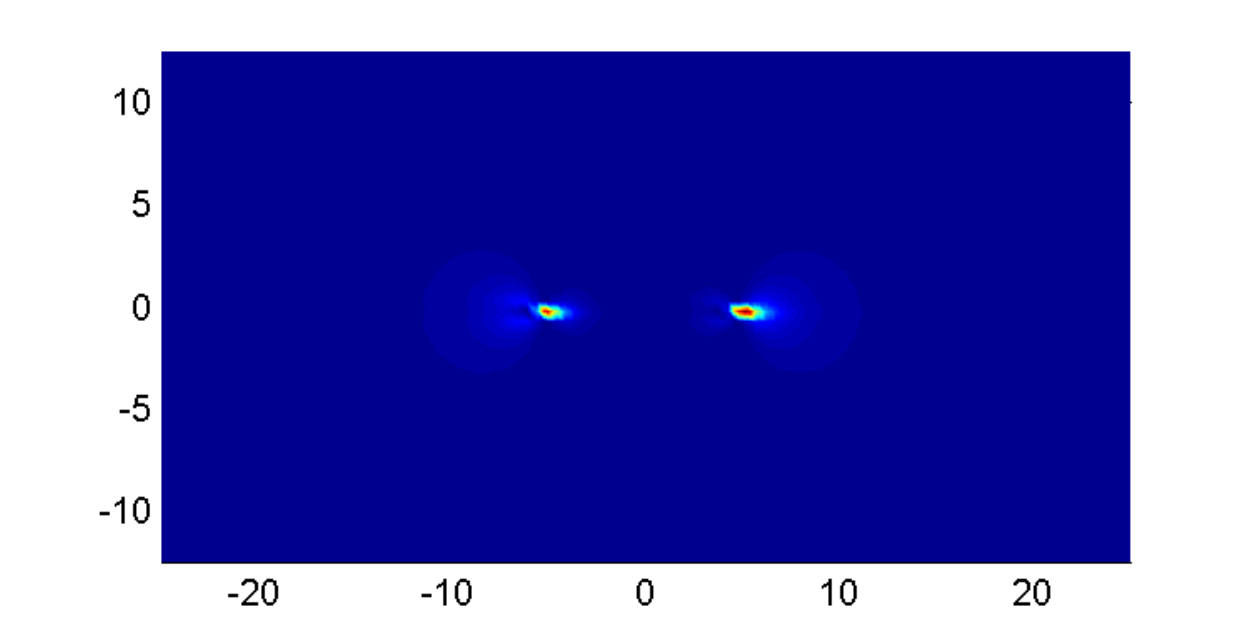}
}
\caption{Snapshots for director field and energy plot at different times. One disclination splits into two half disclinations and these two disclinations repel each other.}\label{fig:split_negative_3}
\end {figure}      
       
 Figure \ref{fig:split_negative_3} shows the process of the strength $-1$ disclination dissociating into two strength $-1/2$ disclinations in terms of the director field. From both the director field snapshots and energy density plots, we can see that the dissociation process is qualitatively similar to the strength $+1$ disclination dissociation. 
 
\subsection{Invariance of disclination dissociation with different $\bflambda$ initializations}\label{inv_lambda}

In Section \ref{sec:one_split}, the initialization of $\bflambda$ for the $k$-strength disclination dissociation is given as 
 \begin{equation*}
\bflambda = \begin{cases}
\frac{-k\pi}{a}\bfe_2, & \text{ if $|y|<{\frac{a}{2}}$  and $x \ge 0$} \\
\frac{k\pi}{a}\bfe_2, & \text{ if $|y|<{\frac{a}{2}}$  and $x < 0$} \\
0, & \text{otherwise}.
\end{cases}
\end{equation*}
 
In this Section, we will consider a different $\bflambda$ initialization as follows
  \begin{equation*}
\bflambda = \begin{cases}
\frac{-2k\pi}{a}\bfe_2, & \text{ if $|y|<{\frac{a}{2}}$  and $x \ge 0$} \\
0, & \text{otherwise},
\end{cases}
\end{equation*}
in order to probe the extent of the dependence of the dissociation phenomena, i.e. defect core dynamics affected by the evolution of the couple stress field in the body, on the fine details of the layer field ($\bflambda$) evolution.

\begin{figure}[H]
\centering
\subfigure[Initialization of $\phi$ for $+1$ disclination dissociation. Difference with the initialization shown in Figure \ref{fig:split_positive_1a} is to be noted. ]{
\includegraphics[width=0.45\textwidth]{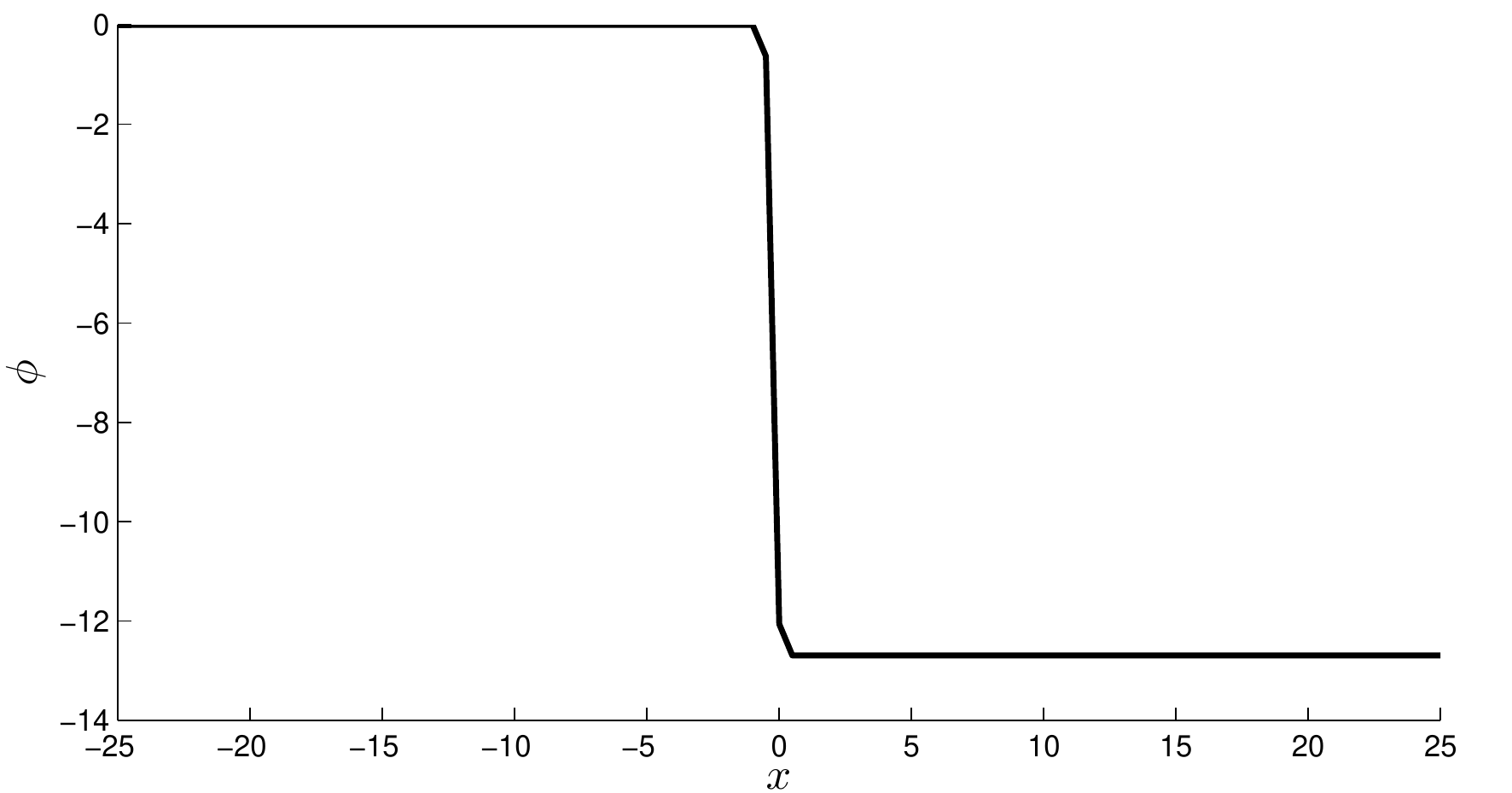}
\label{fig:split_positive_new_1}
}\qquad
\subfigure[Director field corresponding to the initialized $\phi$. The result is identical to that shown in Figure \ref{fig:split_positive_1b}.]{
\includegraphics[width=0.45\textwidth]{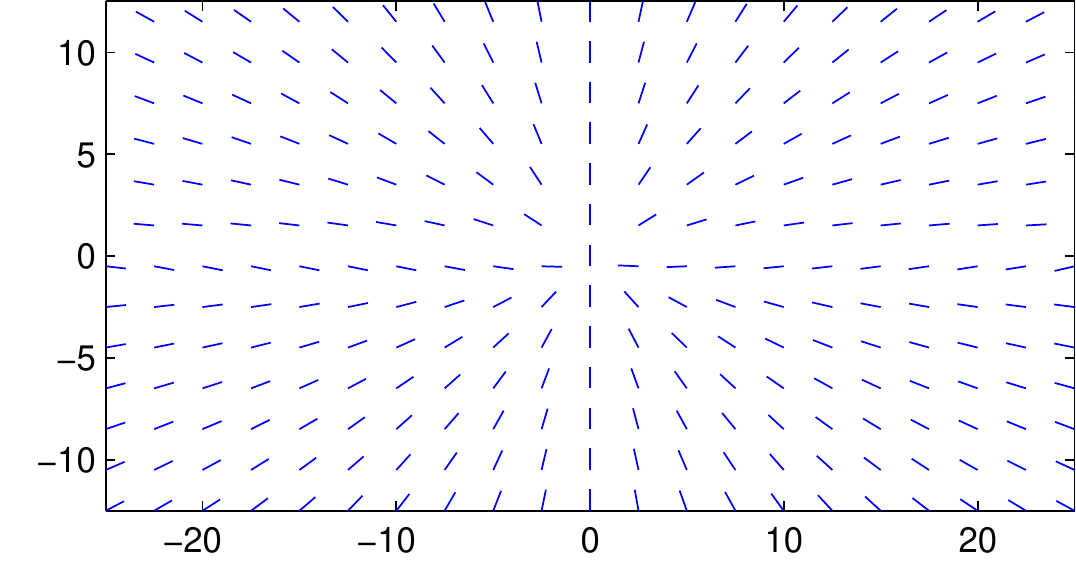}
\label{fig:split_positive_new_2}
}
\subfigure[$\phi_x$ snapshots at different time steps, showing the splitting of the $+1$ disclination.]{
\includegraphics[width=0.45\textwidth]{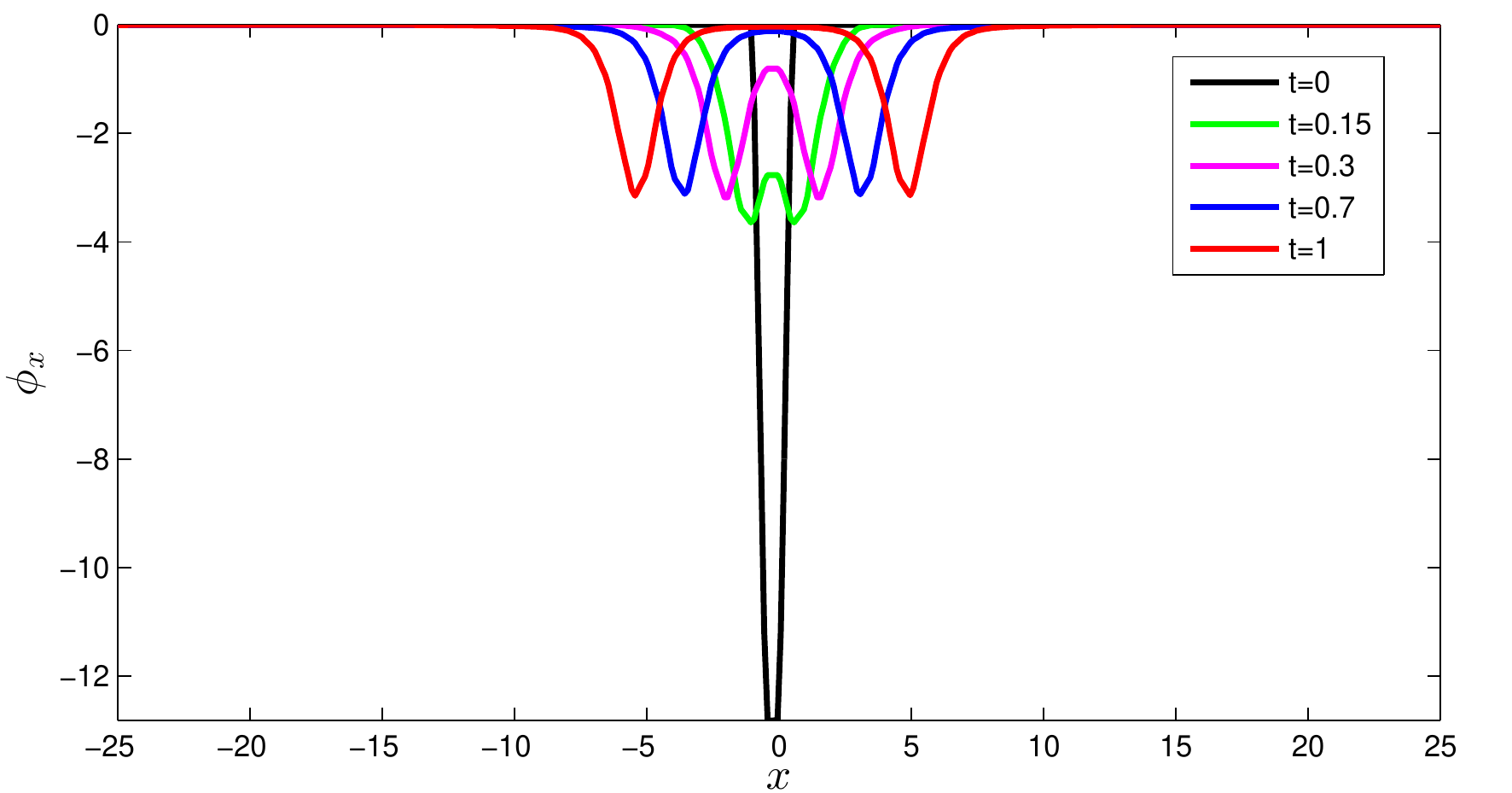}
\label{fig:split_positive_new_3}
}\qquad
\subfigure[A director snapshot after a $+1$ disclination has dissociated into two $+1/2$ disclinations.]{
\includegraphics[width=0.45\textwidth]{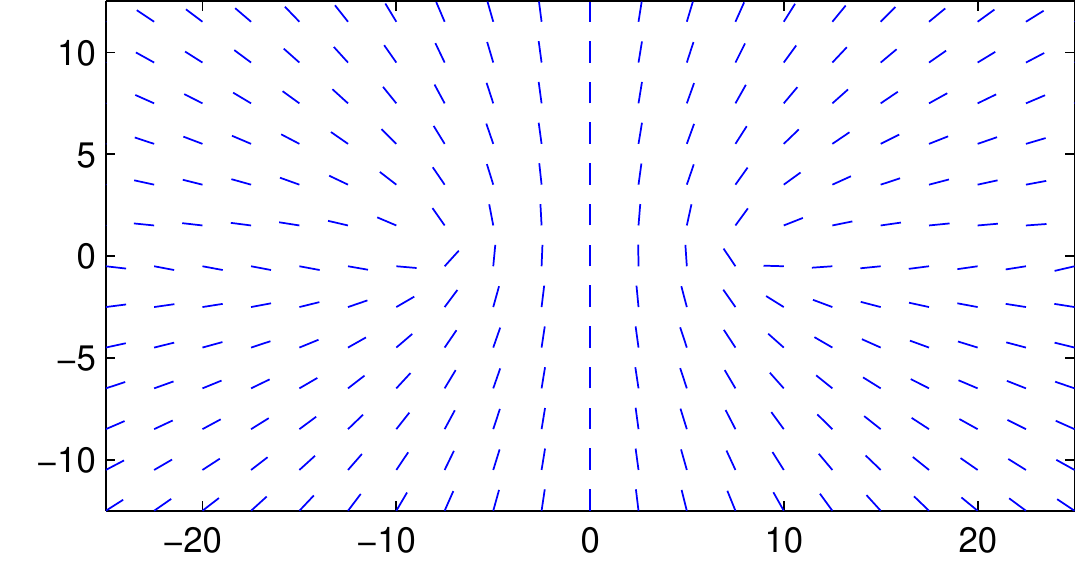}
\label{fig:split_positive_new_4}
}
\caption{Initialization and results for $+1$ disclination dissociation with the new $\bflambda$ initialization. The dissociation process is the same as the one in Section \ref{sec:one_split}.}\label{fig:split_positive_new}
\end {figure}

\begin{figure}[H]
\centering
\subfigure[$|\bflambda|$ evolution during disclination dissociation with $\phi$ initialization defined in Section \ref{sec:one_split}.]{
\includegraphics[width=0.45\textwidth]{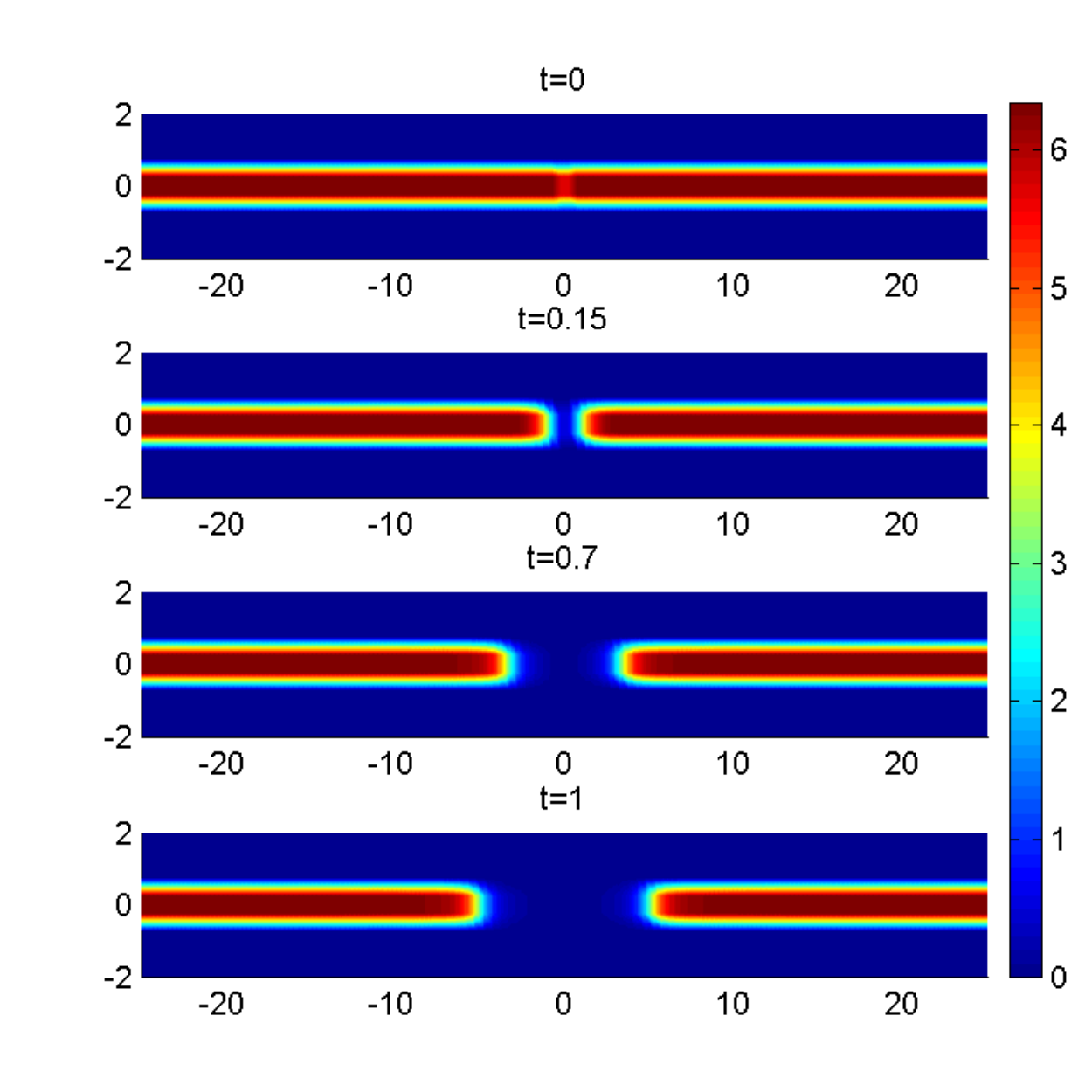}
\label{fig:split_positive_comp_1}
}\qquad
\subfigure[$|\bflambda|$ evolution during dissociation of a $+1$ disclination  from the new $\phi$ initialization. ]{
\includegraphics[width=0.45\textwidth]{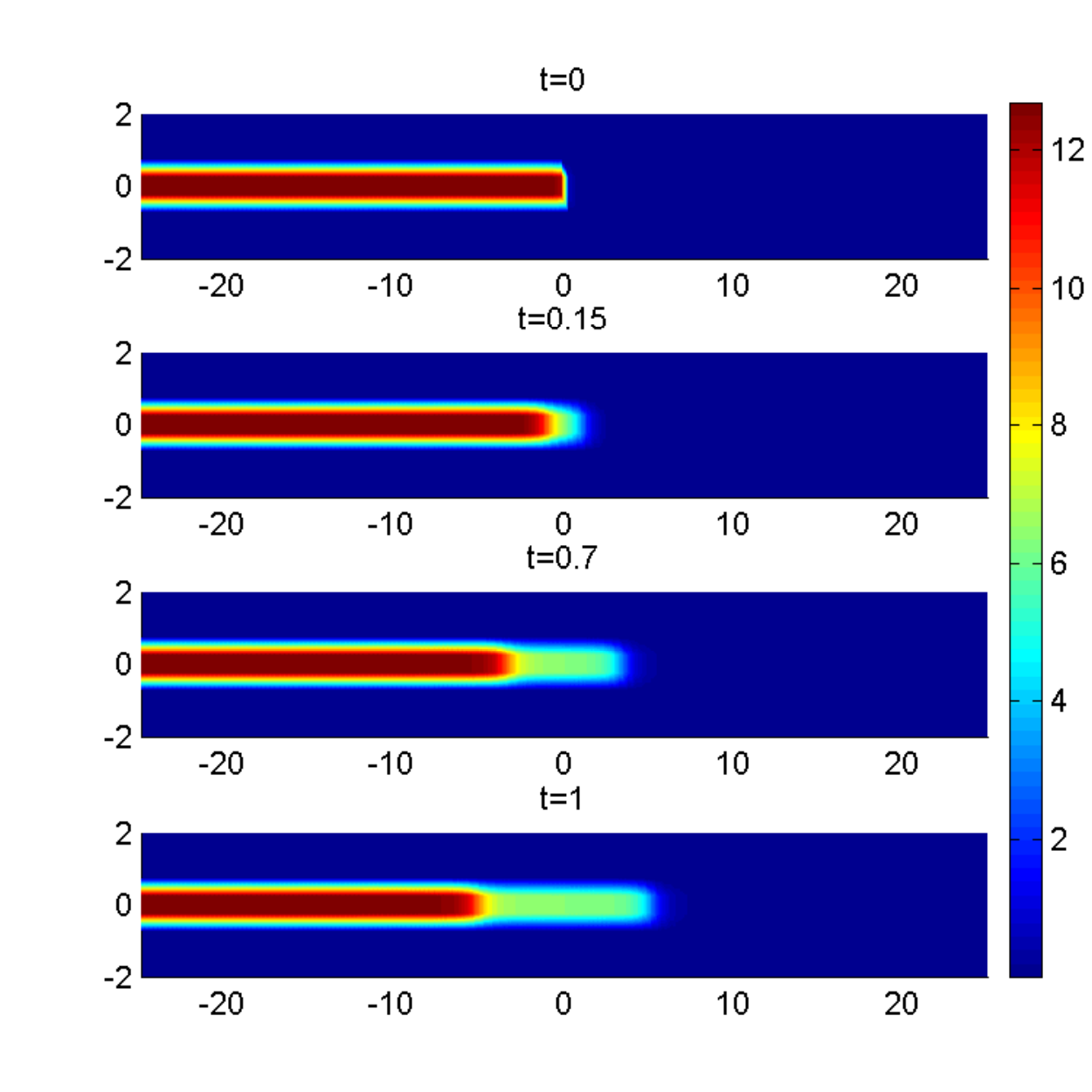}
\label{fig:split_positive_comp_2}
}
\subfigure[$\phi$ comparison for $+1$ disclination dissociation.]{
\includegraphics[width=0.45\textwidth]{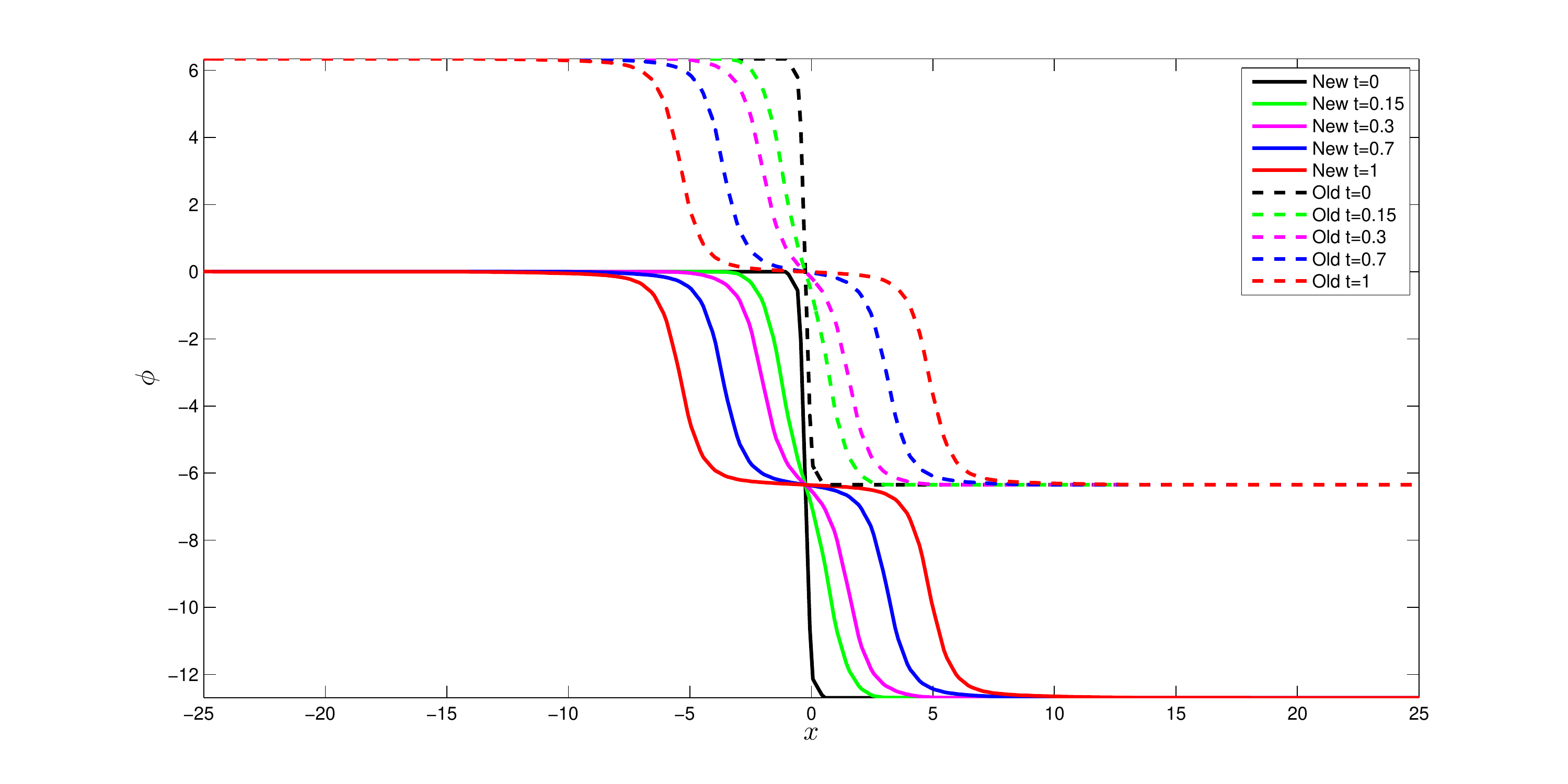}
\label{fig:split_positive_comp_3}
}\qquad
\subfigure[$\phi_x$ comparison for $+1$ disclination dissociation.]{
\includegraphics[width=0.45\textwidth]{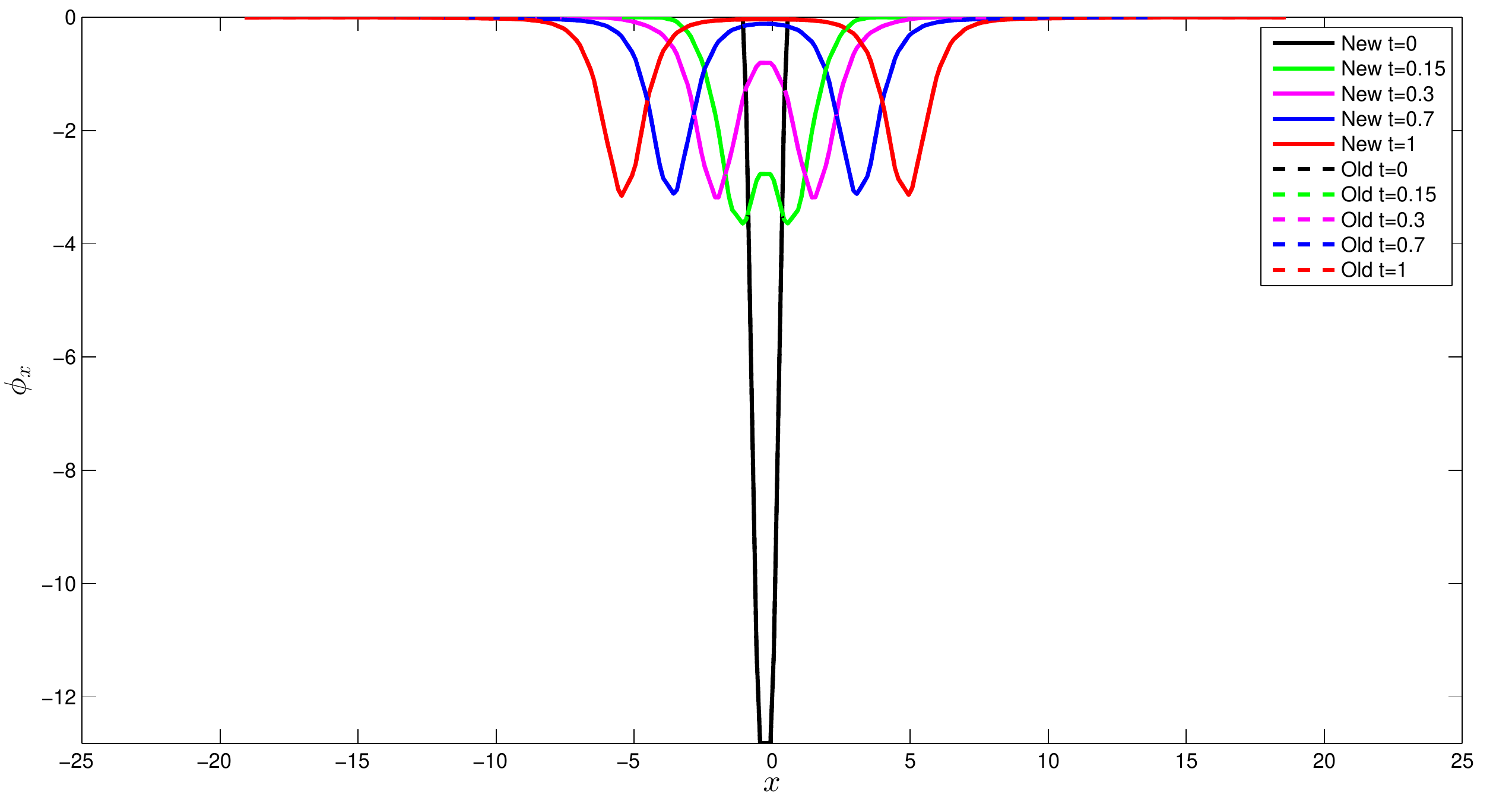}
\label{fig:split_positive_comp_4}
}
\caption{The comparisons of $|\bflambda|$, $\phi$ and $\phi_x$ for a strength $+1$ disclination dissociation. The dashed lines are the results from the ``old'' initialization defined in Section \ref{sec:one_split} and Figures \ref{fig:split_positive_1a},\ref{fig:split_positive_1b} and \ref{fig:split_positive_2}. Although the $|\bflambda|$ evolutions and $\phi$ are different, the $\phi_x$ prfiles are identical during the whole process. }\label{fig:split_positive_comp}
\end {figure}

This new initialization can be achieved by applying a $\frac{k\pi}{a}$ shift on the original $\bflambda$ field within the layer, while keeping the `jump' of $\bflambda$ within the layer same equal to $\frac{2k\pi}{a}$. Figure \ref{fig:split_positive_new_1} shows the new $\phi$ initialization for $+1$ disclination dissociation, and Figure \ref{fig:split_positive_new_2} is the director field corresponding to the $\phi$ initialization. Figure \ref{fig:split_positive_new_3} and Figure \ref{fig:split_positive_new_4} show the results of dissociations of strength $+1$ disclination. With the new initialization, the single strength $+1$ disclination still dissociates into two $+\frac{1}{2}$ disclinations. Figure \ref{fig:split_positive_comp} shows the $|\bflambda|$ evolutions with two different $\phi$ initializations, and the comparison of $\phi$ and $\phi_x$ at different time steps during the $+1$ disclination dissociation. In Figure \ref{fig:split_positive_comp}, the solid lines represent the results from the ``old'' initialization applied in Section \ref{sec:one_split}, while the broken lines represent the results from the new initialization. Although $|\bflambda|$ and $\phi$ are different at every time step, $\phi_x$ maintains the same profile during the whole dissociation process, which shows that the dissociations are the same with these two different $\bflambda$ initializations. The $-1$ disclination dissociation shows the same results. The $+1$ disclination splits into two $+1/2$ disclinations and the $-1$ disclination splits into two $-1/2$ disclinations. Thus, although the initializations are different, the dissociation processes of $\pm 1$ disclinations are same as before.

This example shows that to the extent that two $\bflambda$ evolutions maintain identical disclination fields, the dynamics and energetics of the defect field, at least at an overall `macroscopic' observational level, appears to be unaltered. This fact has important modeling implications, as will be discussed in the last Section \ref{conclude}. 

\section{Modification of the gradient flow dynamics to deal with disclination motion}
 
In Section \ref{sec:shortcoming}, we have shown that the gradient flow dynamics cannot deal with disclination motion. In this section, motivated by the insights gained from the disclination dynamics model in Sections \ref{sec:theory_layer} and \ref{sec:app_layer}, we suggest a modification to the gradient flow dynamics to enable it to solve physically realistic disclination dynamics problems. Recall the evolution equation (\ref{eqn:general_eqns_lc}) in the general disclination dynamic theory:
\begin{gather*}
\frac{\partial \bflambda}{\partial t} =\frac{1}{B_m |\curl \bflambda|^m} \curl\bflambda \times \left[ \left(\grad \theta - \bflambda + \curl\left(\frac{\partial \psi}{\partial \curl \bflambda}\right)- \gamma\frac{\partial f}{\partial \bflambda}\right) \times \curl \bflambda \right] \\
=\frac{|\curl \bflambda|^{2-m}}{B_m} \frac{\curl\bflambda}{|\curl \bflambda|} \times \left[ \left(\grad \theta - \bflambda + \curl\left(\frac{\partial \psi}{\partial \curl \bflambda}\right)- \gamma \frac{\partial f}{\partial \bflambda}\right) \times \frac{\curl \bflambda}{|\curl \bflambda|} \right].
\end{gather*}
We notice that the term $\grad \theta - \bflambda + \curl\left(\frac{\partial \psi}{\partial \curl \bflambda}\right)- \gamma \frac{\partial f}{\partial \bflambda}$ in this evolution equation is the same as the right-hand-side of the gradient flow dynamics (\ref{eqn:grad_dimen}). As mentioned earlier, a salient feature of the $\curl \bflambda$ multiplier allows evolution only at points where $\curl \bflambda$ is non-zero, i.e. in the core and immediate vicinity of the core. Thus, instead of using the regular gradient flow evolution of Section \ref{sec:shortcoming}, we modify the $\bflambda$ evolution equation as follow:
\begin{gather*}
\frac{\partial \bflambda}{\partial s} = H(|\curl \bflambda|-T) \left[\grad \theta - \bflambda + \curl\left(\frac{\partial \psi}{\partial \curl \bflambda}\right)- \gamma\frac{\partial f}{\partial \bflambda}\right].
\end{gather*}
where $T$ is a prescribed threshold and the Heaviside step function is set to be
\begin{gather*}
H(x) = 
\begin{cases} 
      0 & x < 0 \\
      1 & x \ge 0.
   \end{cases}
\end{gather*}
In other words, the layer field is evolved according to
\begin{equation*}
\frac{d \lambda_k}{ds} = \begin{cases}
- \frac{\delta W}{\delta \lambda_k} =- \gamma\frac{\partial f}{\partial |\lambda|}\frac{1}{|\lambda|}\lambda_k + \theta_{,k} - \lambda_k + \epsilon e_{ijk}e_{irs}\lambda_{s,rj}
& \text{if $|\curl \bflambda| \ge T$} \\
0 & \text{otherwise}.
\end{cases}
\end{equation*}

\begin{figure}[H]
\centering
\subfigure[Energy density plot from the regular gradient flow method.]{
\includegraphics[width=0.45\textwidth]{figure/gradient/annihilation_free_e}
}\qquad
\subfigure[Director field result from regular gradient flow method.]{
\includegraphics[width=0.45\textwidth]{figure/gradient/annihilation_free_theta}
}
\subfigure[Energy density plot from the modified gradient flow method.]{
\includegraphics[width=0.45\textwidth]{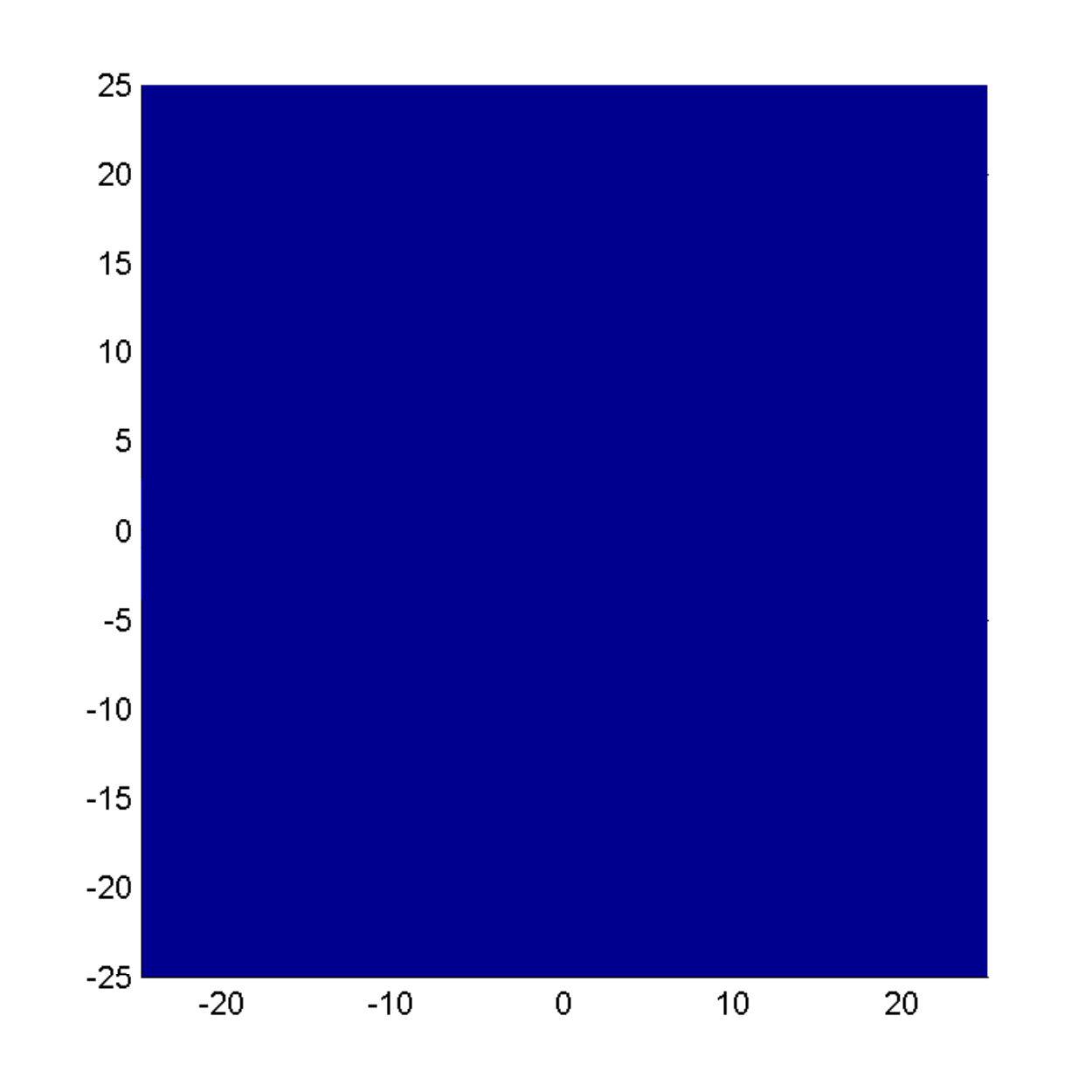}
}\qquad
\subfigure[Director field result from the modified gradient flow method.]{
\includegraphics[width=0.45\textwidth]{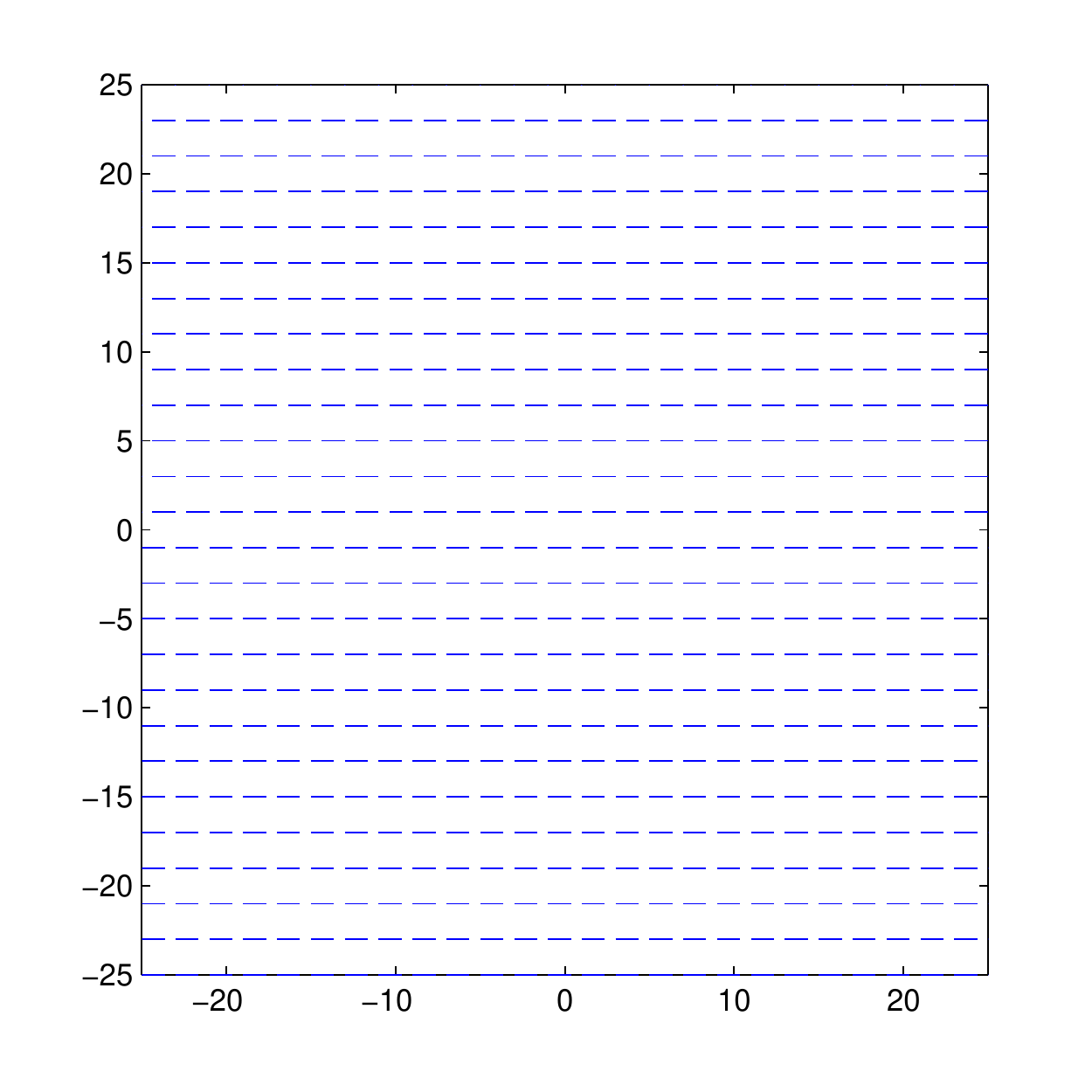}
}
\caption{Top: the energy density and director field results from regular gradient flow method. Bottom: the energy density and director field results from the modified gradient flow method, which match physical expectation.}\label{fig:gradient_mod}
\end {figure}

Based on the above modified evolution equation, we recalculate the disclination annihilation case. The results are shown in Figure \ref{fig:gradient_mod}(c) and (d). Compared to the energy density and director results from Section \ref{sec:shortcoming}, the energy density as well as the director results obtained from the modified evolution equations, shown in Figure \ref{fig:gradient_mod}(c) and (d), are much more reasonable, matching physical expectation.

\section{Some observations}\label{conclude}

We conclude with three further observations.

\begin{figure}[H]
\centering
\subfigure[The norm of a distinct $\bflambda$ field from that in Fig. \ref{figure_3}, but with identical $\curl \bflambda$ field, for a $+1/2$ disclination.]{
\includegraphics[width=0.45\textwidth]{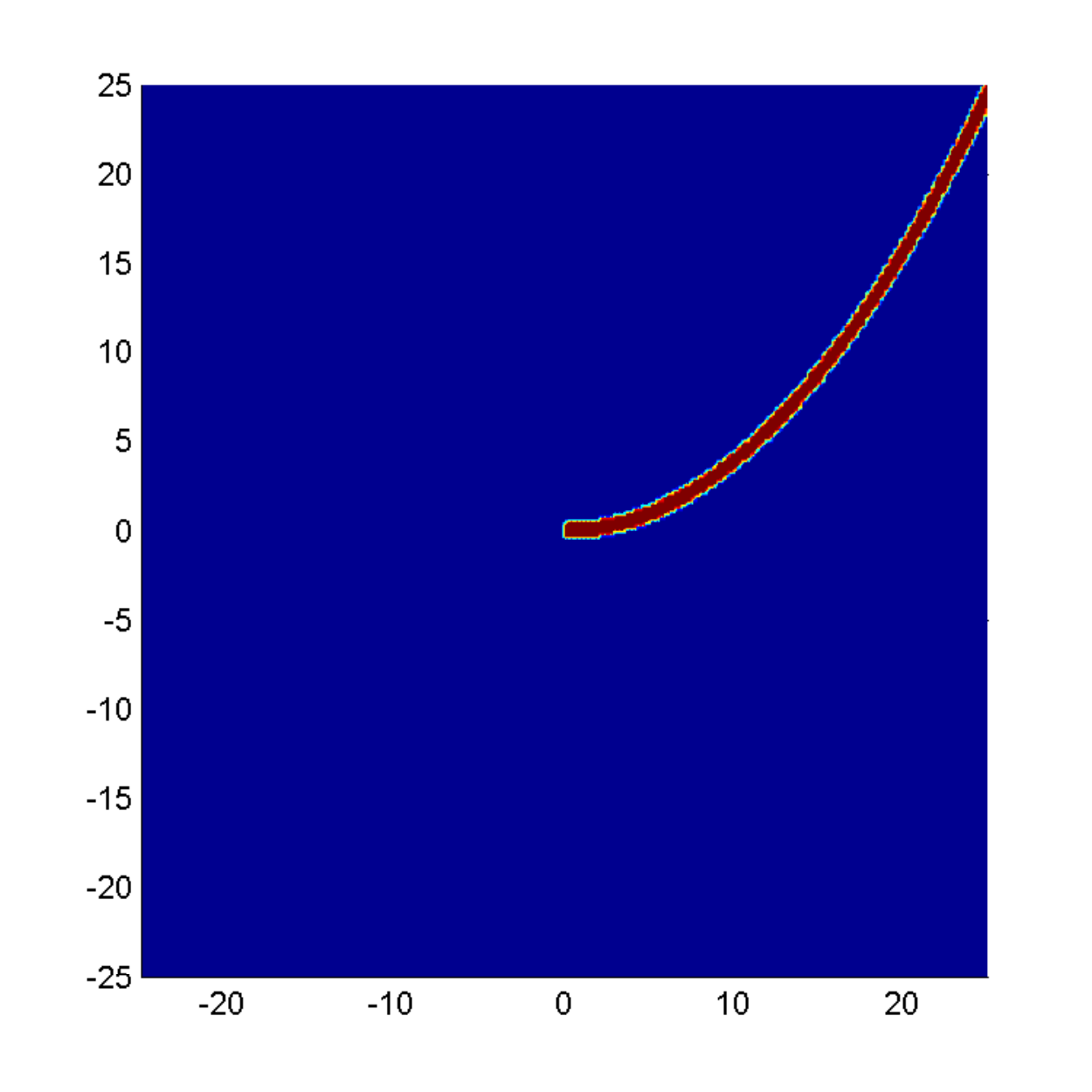}
\label{fig:lambda_distinct_norm}}
\subfigure[The norm of the $\bflambda$ field given in Figure \ref{figure_3}.]{
\includegraphics[width=0.45\textwidth]{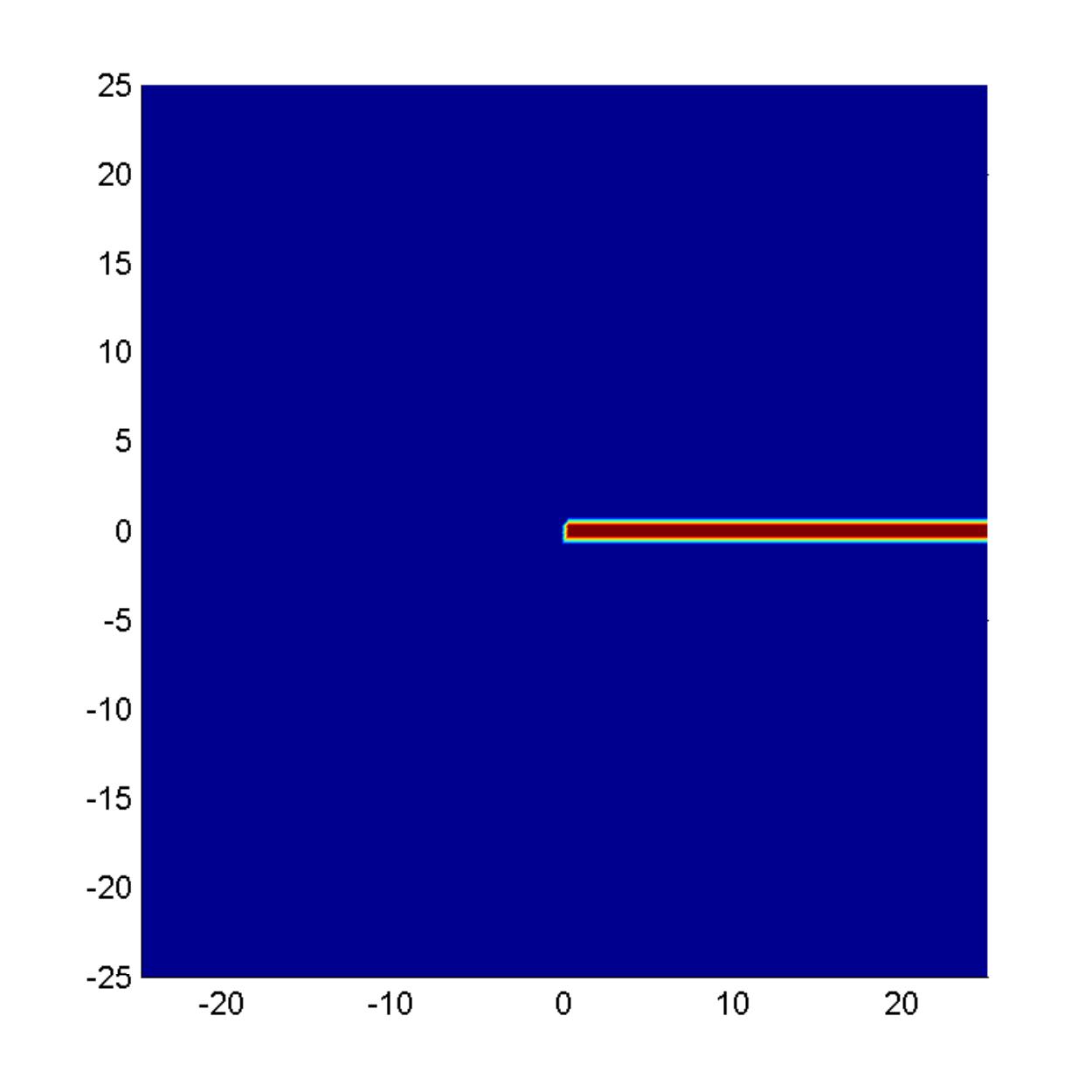}
\label{fig:lambda_norm}}
\caption{The comparison of two $|\bflambda|$ fields with identical $\curl \bflambda$ fields for a $+1/2$ disclination. The direction of $\bflambda$ is perpendicular to the layer at each point along the layer.}\label{fig:lambda_distinct}
\end{figure}

\begin{figure}[H]
\centering
\subfigure[Director field at $l/L=0.005$ with the $\bflambda$ initialization given in Figure \ref{figure_3}. ]{
\includegraphics[width=0.38\textwidth]{figure/single/positive_half_smallest}
\label{fig:lambda_old_director}}\qquad
\subfigure[The energy density plot for the $+1/2$ disclination with the $\bflambda$ initialization given in Figure \ref{figure_3}.]{
\includegraphics[width=0.4\linewidth]{figure/single/positive_half_e}
\label{fig:lambda_old_energy}}
\subfigure[Director field at $l/L=0.005$ with the $\bflambda$ initialization given in Figure \ref{fig:lambda_distinct_norm}. ]{
\includegraphics[width=0.38\textwidth]{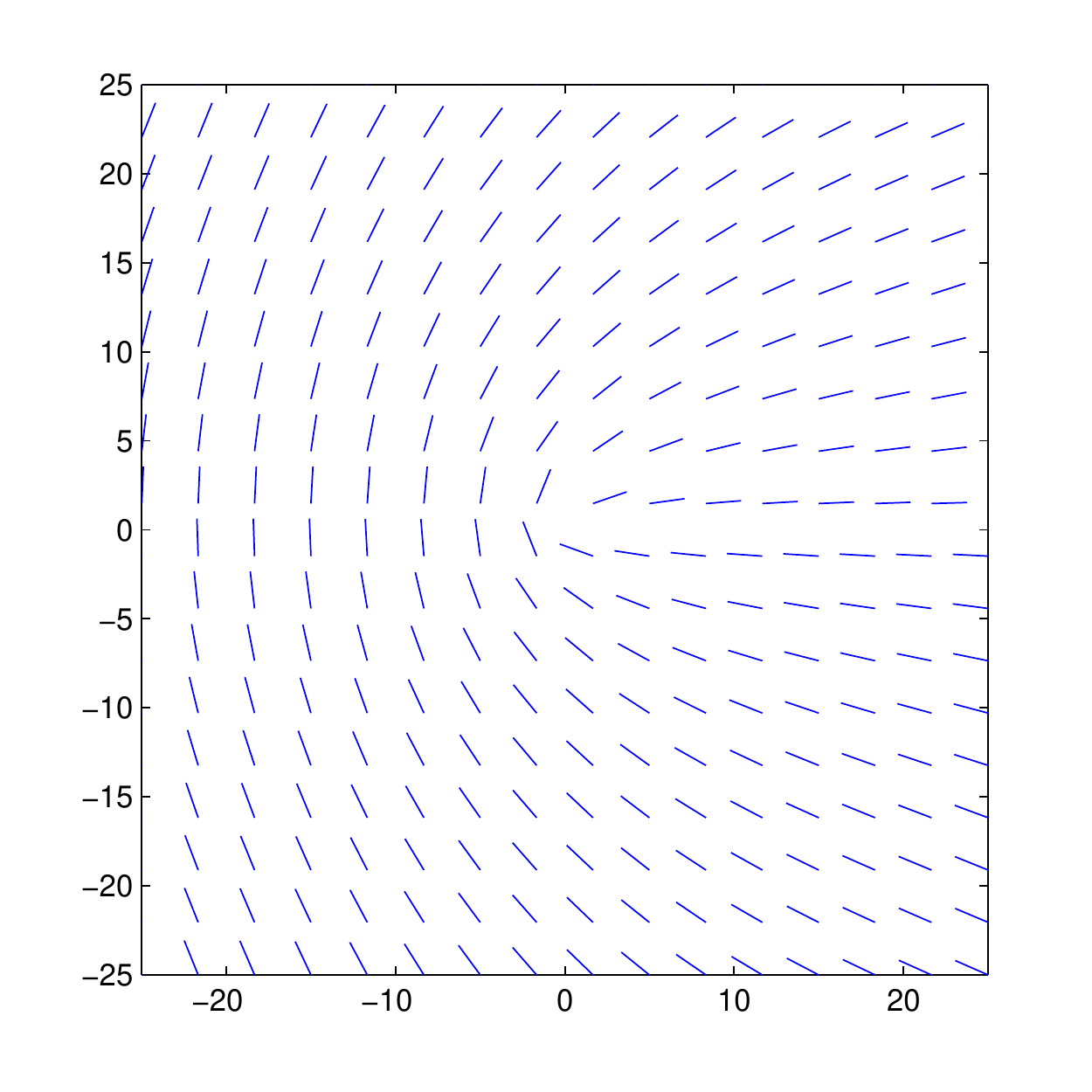}
\label{fig:lambda_distinct_director}}\qquad
\subfigure[The energy density plot for the $+1/2$ disclination with the $\bflambda$ initialization given in Figure \ref{fig:lambda_distinct_norm}.]{
\includegraphics[width=0.4\linewidth]{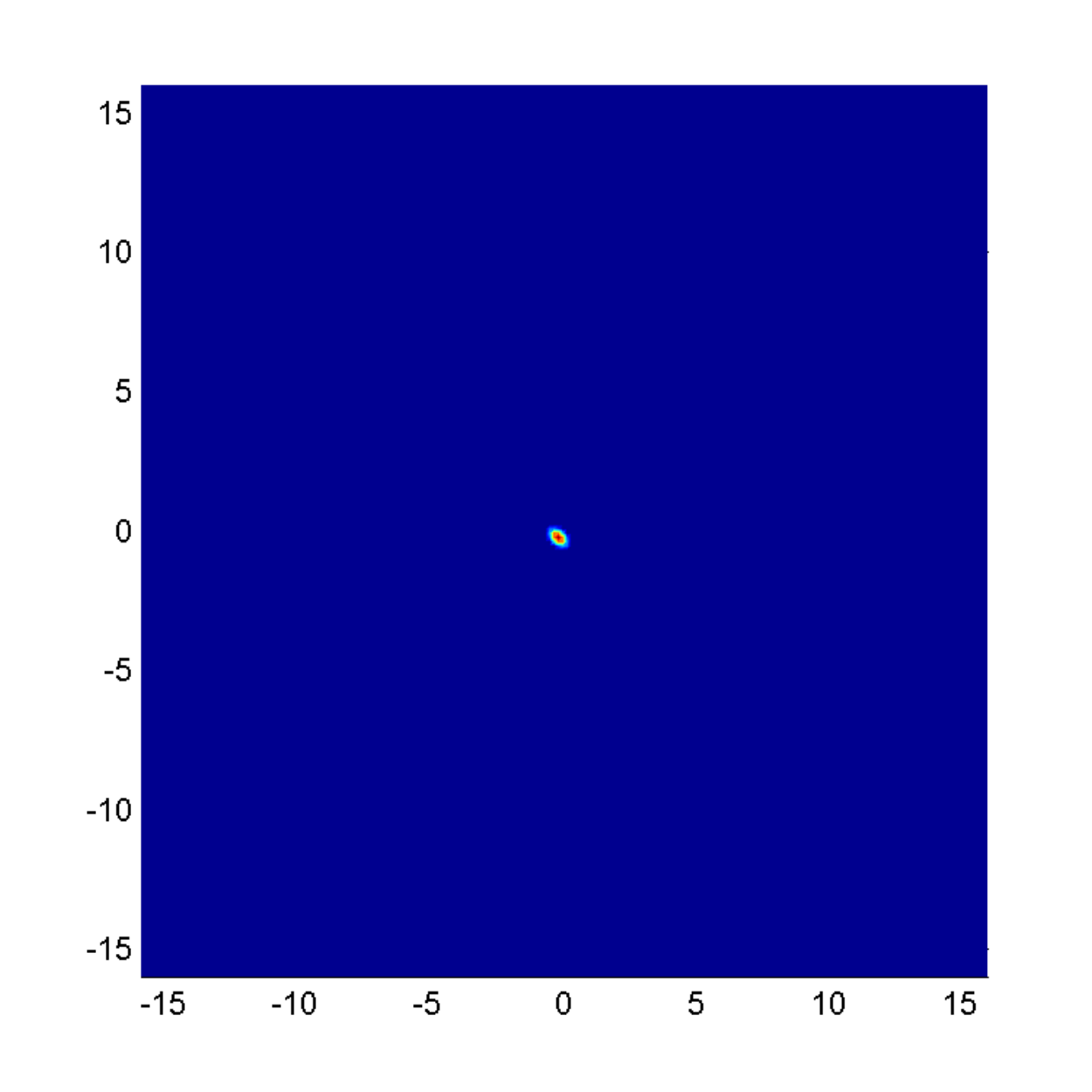}
\label{fig:lambda_distinct_energy}}
\caption{Director and energy density for the $+\frac{1}{2}$ disclination with two different $\bflambda$ initializations. The results show that two distinct $\bflambda$ fields with identical $\curl \bflambda$ fields give identical director and energy distributions.}\label{fig:lambda_distinct_compare}
\end {figure}

Figure \ref{fig:lambda_distinct} shows a comparison of the norm of two different $\bflambda$ fields for the $+1/2$ disclination. Figure \ref{fig:lambda_distinct_norm} is the norm field of the new $\bflambda$ prescription and Figure \ref{fig:lambda_norm} shows the norm of the $\bflambda$ field given in Figure \ref{figure_3}. The $\curl \bflambda$ fields are identical in these two $\bflambda$ settings. Figure \ref{fig:lambda_distinct_compare} shows the director pattern and energy density results based on the two $\bflambda$ prescriptions given in Figure \ref{fig:lambda_distinct}. As theoretically explained in \cite{pourmatin2012fundamental}, the comparison shows that as far as static director and energy distributions are concerned,  an identical `one-point' specification of the director on the domain renders two distinct $\bflambda$ fields with identical $curl \bflambda$ fields, indistinguishable. Furthermore, the results of Section \ref{inv_lambda} show that this invariance is carried over to the disclination dynamics as well. Of course, this invariance is not to be mistaken with `gauge-invariance' in the sense that, for a fixed $\bfb$ field, the theory requires the use of at least one, non-divergence-free $\bflambda$ field of the `layer-type' consistent with $\bfb = - \curl \bflambda$ for the correct prediction of the director distribution, i.e. not employing a $\bflambda$ field and insisting on just the use of the $\bfb$ field is not feasible (even though, in some instances, not introducing a layer-type $\bflambda$ field can suffice for the correct prediction of \emph{only} the energy density field). This partial invariance of the results of our model with respect to the precise details of the $\bflambda$ field suggests a useful freedom in numerical simulations. Essentially, in principle, the $\bflambda$ field can be reinitialized at every instant of time, consistent with the evolving $\bfb$ field. Thus, a strategy may be to evolve the $\bfb$ field instead of $\bflambda$ and use the $\bflambda$ construct to simply facilitate the calculation of the energetics and the director distribution at each instant of time. Moreover, our demonstration that a `layer' of finite thickness is merely a geometric approximation, without energetic consequences, of a surface of director-vector discontinuity suggests natural ways of associating a `non-turning' director distribution within the layer for calculations of Leslie-Ericksen viscous stresses. We shall demonstrate such features in future work.

Second, Appendix \ref{App:2} shows the energetic, dynamic, and topological interoperability between our model of disclination dynamics in nematic liquid crystals with orientational order and that of screw dislocation dynamics in elastic solids with positional order. We consider this as a positive development that can only benefit the understanding of defect dynamics in liquid crystals and elastic solids, leading to their plasticity.

Third, there are very interesting similarities and contrasts between the models and results of energy driven pattern formation discussed in the papers \cite{kohn2006energy, jerrard1998dynamics, alicandro2014metastability, ercolani2009variational, newell2012pattern} and our model. A comparative study is an undertaking in its own right that will form the subject of future study.

\section*{Acknowledgments}
Support from the NSF DMREF program through grant DMS1434734 is gratefully acknowledged.
\appendix
\section{Numerical Schemes for $\phi$ evolution equations in 2D layer model}\label{App:1}

We adapt the computational scheme developed in \cite{zhang2015single} for an exactly similar problem in the context of dislocation dynamics to solve our equations. 

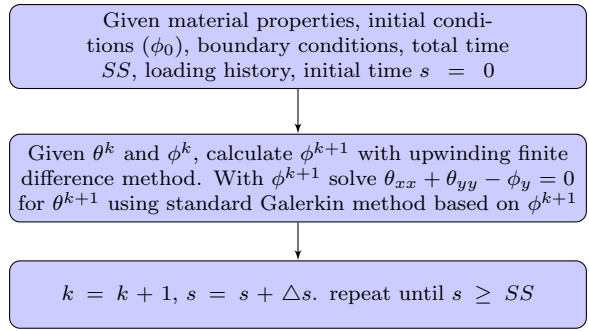
\begin{figure}[H]
\centering
\begin{tikzpicture}[node distance = 1.75cm, auto]
    % Place nodes
    \scriptsize
    \node [block] (init) {{Given material properties, initial conditions ($\phi_0$), boundary conditions, total time $SS$}, loading history, initial time $s=0$};  
    \node [block, below of=init, node distance=1.75cm] (divT) {Given $\theta^k$ and $\phi^k$, calculate $\phi^{k+1}$ with upwinding finite difference method. With $\phi^{k+1}$ solve $\theta_{xx}+\theta_{yy}-\phi_{y}=0$ for $\theta^{k+1}$ using standard Galerkin method based on $\phi^{k+1}$};    
    \node [block, below of=divT, node distance=1.55cm] (decide) {$k=k+1$, $s=s+\triangle s$. repeat until $s\geq SS$};
    % Draw edges
    \path [line] (init) -- (divT);
    \path [line] (divT) -- (decide);
    \end{tikzpicture}
%}
\caption{Flow charts for computing scheme: $\phi$ and $\theta$ are basic unkonwn fields. }
\label{fig:flowchart}
\end{figure}

For the sake of completeness, we reproduce the details of the scheme from that paper with appropriate adjustments for field variable names and parameters. The numerical scheme we adopt to solve the problem is identical to that in \cite{zhang2015single}, again with just a translation of the names of the field variables. We include the details here for completeness. In general, the Finite Element Method (FE) is used to solve the equation for balance of linear momentum in a staggered scheme that utilizes $\phi$ as a given quantity obtained by evolving it in the remaining part of the scheme. The general computing flow is shown in Figure \ref{fig:flowchart}. 

An FE mesh with an embedded 1-d finite difference grid is used. We use linear quadrilateral elements, with  $5\times 5$ Gauss quadrature points. All elements are of uniform size over the whole domain.

The 1-d, finite difference grid is embedded in the layer, coincident with the line $y=0$. Suppose that the layer is meshed into $M$ rows and $N$ columns, where $N$ is the total number of 1-d grid points and $M$ is always an odd number so that the middle row of elements always have centers on $y=0$. Each column of FE elements in the layer correspond to exactly one grid point. Let $x_k$ be the $x$ coordinate of the $k^{th}$ 1-d grid point, which is at the center of the $k_{th}$ element in the $(M+1)/2$ row of layer elements.  Recall that $\tau(x_k)$ is defined as $\frac{1}{a}\int^{a/2}_{-a/2}\left(\theta_{y}-\phi\right)dy$. Let  $(\theta_{y}-\phi )(I,k)$ denotes the integrand evaluated at the $I^{th}$ Gauss point whose $x$ coordinate is $x_k$, and let $N_{k}$ be the total number of such Gauss points. Then $\tau(x_k)$ is calculated as
\begin{equation*}
\tau(x_k) = \frac{1}{N_{k}}\left(\sum^{N_{k}}_{I=1} (\theta_{y}-\phi)(I,k)\right).
\end{equation*}

The numerical scheme developed in \cite{das2013can} is adopted and improved to solve (\ref{eq:sum_gov_eq})$_2$, the $\phi$ evolution. The scheme is called upwinding as the basic idea is to infer the direction of wave propagation from the linearization of (\ref{eq:sum_gov_eq})$_2$ and use this direction in the actual nonlinear equation. Let us denote time step with $\triangle t$ and spatial grid size of the finite difference grid with $\triangle h$. Due to the necessity of very small element sizes to demonstrate convergence, an explicit treatment of the diffusion term in (\ref{eq:sum_gov_eq})$_2$ becomes prohibitive because of a $\triangle t = \mathcal{O} (\triangle h^2 )$ scaling. This is circumvented by treating the $\phi_{xx}$ term implicitly, resulting in a linearly implicit scheme as follows. We first linearize (\ref{eq:sum_gov_eq})$_2$ and discretized:
\begin{equation}
\label{eqn:first_variation}
\begin{aligned}
\delta\phi^k _t (x_h ) &= -(2-m)\left(-sgn\left(\phi^k_x \left(x_h\right)\right)\right) \left|\phi_x^k(x_h)\right|^{1-m} \left[\tau^k\left(x_h\right) + a\phi^{k+1}_{xx}\left(x_h\right) - \left(\tau^{b}\left(x_h\right)\right)^k\right]\delta\phi^k_x\left(x_h\right)\\
&+\left|\phi^k_x\left(x_h\right)\right|^{2-m}\left[\epsilon\delta\phi_{xx}^k\left(x_h\right)\right] \\
&+\left|\phi^k_x\left(x_h\right)\right|^{2-m}\left[ \tau^{b'}\left(x_h \right)\delta\phi^k (x_h)  \right],
\end{aligned}
\end{equation}
where a quantity such as $\phi^k_x(x_h)$ implies the value of $\phi_x(x)$ evaluated at $h^{th}$ grid point at $k^{th}$ time step. The first term in (\ref{eqn:first_variation}) provides an advection equation with wave speed
\begin{equation}\nonumber
c^k (x_h )= (2-m)\left(-sgn\left(\phi^k_x \left(x_h\right)\right)\right) \left|\phi_x^k(x_h)\right|^{1-m} \left[\tau^k\left(x_h\right) + a\phi^{k+1}_{xx}\left(x_h\right) - \left(\tau^{b}\left(x_h\right)\right)^k\right].
\end{equation}
$\phi^k_x(x_h)$ and $\phi^k_{xx} (x_h)$ are obtained from central finite differences:
\begin{equation}\label{eq:central_diff}
\begin{aligned}
&\phi^k_x(x_h) = \frac{\phi^k(x_{h+1})-\phi^k(x_{h-1})}{2\triangle h}\\
&\phi^k_{xx}(x_h) = \frac{\phi^k(x_{h+1})-2\phi^k(x_{h})+\phi^k(x_{h-1})}{\triangle h^2}.
\end{aligned}
\end{equation}
Based on the sign of $c_k$, $\phi_x ^k$ is then computed by the following upwinding scheme:
\begin{equation}
\phi_x ^k = \begin{cases}
       \frac{\phi^k (x_{h+1}) - \phi^k (x_h )}{\triangle h} & \text{ if } c^k (x_h ) < 0\\
       \frac{\phi^k (x_{h}) - \phi^k (x_{h-1} )}{\triangle h} & \text{ if } c^k (x_h ) > 0\\
	   \frac{\phi^k (x_{h+1} ) - \phi^k (x_{h-1} )}{2\triangle h} & \text{ if } c^k (x_h ) = 0.
        \end{cases}
\end{equation}
The time step is governed by a combination of a CFL condition and a criterion for stability for an explicit scheme for a linear ordinary differential equation:
\begin{equation}
\label{eq:dt}
\triangle t^k = min \left( \frac{\triangle h}{c^k (x_h)}, \frac{1}{|\phi_x ^k (x_h)|^{2-m} (-(\tau^{b'}(x_h) )^{k}}\right).
\end{equation}
Note that if $\phi_{xx}$ was evaluated at $k$, then the step size would also be bounded by $\frac{\triangle h^2}{a\left|\phi_x^k(x_h)\right|}$, leading to a quadratic decrease in $\triangle t^k$ with element size. Treating $\phi_{xx}$ implicitly eliminates this constraint resulting in significant savings in computation time. $\phi^{k+1}_h$ is updated according to
\begin{equation}
\begin{aligned}
&\frac{\phi^{k+1} (x_h )-\phi^k (x_h )}{\triangle t^k} = |\phi_x ^k (x_h )|^{2-m} \left[\tau^k + a\phi_{xx}^{k+1} - (\tau^b (x_h) )^k \right]\\
\Rightarrow&\phi^{k+1}(x_h) - a\triangle t^k |\phi_x ^k (x_h )|^{2-m}\phi^{k+1} _{xx}(x_h ) = \phi^k(x_h) + \triangle t^k |\phi_x ^k (x_h )|^{2-m} \left[\tau^k -(\tau^b (x_h) )^k \right].
\end{aligned}
\end{equation}
The right hand side of the equation is known at current time $k$. But noting that $\phi^{k+1} _{xx}(x_h )$ is again computed from $\phi^{k+1}$ at $x_{h+1}$, $x_h$ and $x_{h-1}$, a system of linear equations of size $N$ has to be solved to get $\phi^{k+1}$. The computational expense of the linear solve is small compared to the savings obtained by relaxing $\triangle t^k$ corresponding to the explicit treatment of diffusion.

\section{Layer model for the screw dislocation case}\label{App:2}

Paralleling the development in Section \ref{sec:layer_model}, we define a layer model for straight screw dislocation dynamics in solids in this section. Consider the similar geometry as in Section \ref{sec:layer_model} as shown in Figure \ref{fig:dynamic_body}:
\begin{gather*}
\mathcal{V} = \{ (x,y):(x,y)\in [-W, +W] \times [-H, +H]\} \\
\mathcal{L} = \{ (x,y):(x,y)\in [-W, +W] \times [-l/2, +l/2]\} \\
0 \leq l < 2H, \quad W > 0.
\end{gather*}
The screw dislocation problem is one of anti-plane shear, i.e. there is only one non-vanishing displacement component of the solid, this being the out-of-plane one which is a function of the in-plane coordinates. In the present physical context, $\bfu$ is the displacement vector and $\bfU^P$ is the plastic distortion tensor, that plays an analogous role to $\bflambda$.
The  conservation law for the defect field for the screw dislocation \cite{zhang2015single} is given in the form 
\begin{gather*}
\dot{\bfalpha} = -\curl (\bfalpha \times \bfv).
\end{gather*}
The displacement is assumed to be of the form 
\[
\bfu = w(x, y) \bfe_3
\]
and $\bfU^P$ takes the form
\begin{gather*}
\bfU^P = \begin{cases}
\phi(x,t) \bfe_3 \otimes \bfe_2 & \text{in the layer} \\
0 & \text{otherwise}.
\end{cases}
\end{gather*}
We assume $\bfv$ to be of the form $\bfv = v(x,t)\bfe_1$.

Therefore, $\bfalpha$ is also non-zero only  in the layer, with component form
\begin{gather*}
\bfalpha = -\curl \bfU^P = - \phi_x \bfe_3 \otimes \bfe_3\\
\curl \bfalpha = \phi_{xx} \bfe_3 \otimes \bfe_2.
\end{gather*}
The stored energy function for the screw dislocation model is assumed as
\begin{gather*}
W = \int_\mathcal{V} \psi(\bfepsilon^e, \bfalpha)  + \gamma f(|\bfU^p|) dv  = \int_\mathcal{V} \left[ \frac{K}{2} |\grad \bfu -\bfU^P|^2 + \frac{\epsilon}{2}(\phi_x)^2 + \gamma f(|\phi|)\right]dv.
\end{gather*}
Here, $K$ is a shear modulus with dimension $Force \times Length^{-2}$; $\epsilon = KCa\xi^2$ is a parameter characterizing the energy density of the dislocation core;  $\xi$ is the Burgers vector magnitude of the dislocation, proportional to the lattice interatomic distance (with dimensions of $Length$), $C$ is a non-dimensional parameter to control the magnitude of the core energy, and the product $a\xi$ is the separation between two atomic layers with $a \geq 0$ a non-dimensional scaling factor. Unlike the nematic disclination case, the layer has a physical significance in the case of the crystal dislocation as does the $\bflambda$ field in predicting, often `stress-free,' permanent plastic deformation (with respect to a fixed reference) due to the motion of dislocations. The combination $\gamma f$ represents the `generalized stacking fault energy' reflecting lattice symmetries, and measurable from controlled computational atomistic experiments (\cite{vitek1968intrinsic}) and we assume the simple forms $\gamma = \frac{P K}{a}$ and 
\[
f =1-\cos\left( 2\pi |\phi|\left(\frac{\xi}{a \xi}\right)^{-1} \right)
 \]
where $P$ is a dimensionless parameter.

From $\dot{\bfalpha} = -\curl (\bfalpha \times \bfv)$, we have $\phi_t = - \phi_x v(x,t)$. Given the ansatz, only the layer is relevant for the dissipation and it can be written as 
\begin{gather*}
D = \int_\mathcal{L} v \{ [ T_{32} - \tau^b + \epsilon \phi_{xx} ] (-\phi_x)\} dv,
\end{gather*}
where 
\begin{gather*}
\tau^b := \gamma \frac{d f}{d \phi} = 2\pi P K \sin\left( 2\pi \phi\, a \right)
\end{gather*}
and 
\begin{gather*}
T_{32} = K(u_{3,2}-\phi) := K(\omega_y-\phi).
\end{gather*}

As in Section \ref{sec:layer_model} , we take the average of $(w_y-\phi)$ over the layer and requiring the dissipation $D\ge0$, the evolution equation for $\phi$ reads as
\begin{gather*}
\frac{\partial \phi}{\partial{t}} =\frac{|\phi_x|^{2-m}}{B_m}\left(\tau -\tau^b + \epsilon \phi_{xx}\right) \\
\text{where} \quad \tau = \frac{K}{a \xi} \int_{-\frac{a \xi}{2}}^{\frac{a \xi}{2}}(w_{y} - \phi) dy.
\end{gather*}
Again, $B_m$ is a non-negative coefficient characterizing energy dissipation with physical dimensions depending on $m$. The parameter $m$ can be chosen to probe different types of behavior. By introducing the following dimensionless variables,
\begin{gather*}
\tilde{x} = \frac{1}{\xi}x; \quad \tilde{y} = \frac{1}{\xi}y; \quad \tilde{\epsilon} = \frac{1}{K \xi^2}\epsilon=Ca; \quad
\tilde{\tau} = \frac{1}{K}\tau; \quad \tilde{\tau}^b = \frac{1 }{K}\tau^b; \\
\quad  \tilde{s}=\frac{K}{\xi^{2-m} B_m}t; \quad \tilde{w} = \frac{1}{\xi} w,
\end{gather*}
we obtain the dimensionless evolution equation for the layer model as described below:
\begin{equation*}
\frac{\partial {\phi}}{\partial{\tilde{s}}} ={|{\phi}_{\tilde{x}}|^{2-m}}\left(\tilde{\tau} - \tilde{\tau}^b + \tilde{\epsilon} {\phi}_{\tilde{x}\tilde{x}}\right).
\end{equation*}
\emph{After removing tildes for simplicity}, the dimensionless governing equations for the screw dislocation problem become

\boxalign{
\begin{eqnarray}
\label{eqn:dislocation_analog}
\left\{
\begin{aligned}
&w_{xx}+w_{yy}-\phi_{y} = 0 \quad \text{in } \mathcal{V} \\
&\frac{\partial \phi}{\partial{s}} ={|\phi_x|^{2-m}}\left(\tau - \tau^b + Ca \phi_{xx}\right) \quad \text{in } \mathcal{L}
\end{aligned}
\right.
\end{eqnarray}
where 
\begin{equation*}
\begin{aligned}
\tau = \frac{1}{a}\int^{a/2}_{-a/2} \left( w_{y} - \phi \right) dy, \quad
\tau^b = {2 \pi P} \sin\left(2\pi \phi \, a\right),
\end{aligned}
\end{equation*}}
and the first equation represents static balance of forces (balance of linear momentum), for the ansatz being considered here.

As can be seen from a comparison of (\ref{eqn:dislocation_analog}) and  (\ref{eq:sum_gov_eq}), the governing equations of the screw dislocation model are exactly analogous to the disclination model.

\newpage
\bibliography{proposal}

\begin{thebibliography}{10}

\bibitem{stewart2004static}
I.~W. Stewart, {\em The static and dynamic continuum theory of liquid crystals:
  a mathematical introduction}.
\newblock C{RC} Press, 2004.

\bibitem{acharya2013continuum}
A.~Acharya and K.~Dayal, ``Continuum mechanics of line defects in liquid
  crystals and liquid crystal elastomers,'' {\em Quarterly of Applied
  Mathematics}, vol.~72, no.~1, pp.~33--64, 2013.

\bibitem{pourmatin2012fundamental}
H.~Pourmatin, A.~Acharya, and K.~Dayal, ``A fundamental improvement to
  {E}ricksen-{L}eslie kinematics,'' {\em Quarterly of Applied Mathematics},
  vol.~LXXXIII, no.~3, pp.~435--466, 2015.

\bibitem{kleman1973defect}
M.~Kl{\'e}man, ``Defect densities in directional media, mainly liquid
  crystals,'' {\em Philosophical Magazine}, vol.~27, no.~5, pp.~1057--1072,
  1973.

\bibitem{de1995physics}
P.-G. de~Gennes and J.~Prost, ``The physics of liquid crystals (international
  series of monographs on physics),'' {\em Oxford University Press, USA},
  no.~0.10, pp.~0--20, 1995.

\bibitem{sonnet2012dissipative}
A.~M. Sonnet and E.~G. Virga, {\em Dissipative ordered fluids: theories for
  liquid crystals}.
\newblock Springer Science \& Business Media, 2012.

\bibitem{mottram2014introduction}
N.~J. Mottram and C.~J. Newton, ``Introduction to {Q}-tensor theory,'' {\em
  arXiv preprint arXiv:1409.3542}, 2014.

\bibitem{schopohl1987defect}
N.~Schopohl and T.~Sluckin, ``Defect core structure in nematic liquid
  crystals,'' {\em Physical Review Letters}, vol.~59, no.~22, p.~2582, 1987.

\bibitem{bauman2012analysis}
P.~Bauman, J.~Park, and D.~Phillips, ``Analysis of nematic liquid crystals with
  disclination lines,'' {\em Archive for Rational Mechanics and Analysis},
  vol.~205, no.~3, pp.~795--826, 2012.

\bibitem{ravnik2009landau}
M.~Ravnik and S.~{\v{Z}}umer, ``Landau--de {G}ennes modelling of nematic liquid
  crystal colloids,'' {\em Liquid Crystals}, vol.~36, no.~10-11,
  pp.~1201--1214, 2009.

\bibitem{di2014half}
G.~Di~Fratta, J.~Robbins, V.~Slastikov, and A.~Zarnescu, ``Half-integer point
  defects in the {Q}-{T}ensor theory of nematic liquid crystals,'' {\em Journal
  of Nonlinear Science}, pp.~1--20, 2014.

\bibitem{kralj1991nematic}
S.~Kralj, S.~{\v{Z}}umer, and D.~W. Allender, ``Nematic-isotropic phase
  transition in a liquid-crystal droplet,'' {\em Physical Review A}, vol.~43,
  no.~6, p.~2943, 1991.

\bibitem{macdonald2012robust}
C.~S. MacDonald, J.~A. Mackenzie, A.~Ramage, and C.~J. Newton, ``Robust
  adaptive computation of a one-dimensional-tensor model of nematic liquid
  crystals,'' {\em Computers \& Mathematics with Applications}, vol.~64,
  no.~11, pp.~3627--3640, 2012.

\bibitem{ignat2014uniqueness}
R.~Ignat, L.~Nguyen, V.~Slastikov, and A.~Zarnescu, ``{U}niqueness results for
  an {ODE} related to a generalized {G}inzburg--{L}andau model for liquid
  crystals,'' {\em SIAM Journal on Mathematical Analysis}, vol.~46, no.~5,
  pp.~3390--3425, 2014.

\bibitem{ignat2015stability}
R.~Ignat, L.~Nguyen, V.~Slastikov, and A.~Zarnescu, ``{S}tability of the
  melting hedgehog in the {L}andau--de {G}ennes theory of nematic liquid
  crystals,'' {\em Archive for Rational Mechanics and Analysis}, vol.~215,
  no.~2, pp.~633--673, 2015.

\bibitem{ignat2013stability}
R.~Ignat, L.~Nguyen, V.~Slastikov, and A.~Zarnescu, ``{S}tability of the vortex
  defect in the {L}andau--de {G}ennes theory for nematic liquid crystals,''
  {\em Comptes Rendus Math{\'e}matique}, vol.~351, no.~13, pp.~533--537, 2013.

\bibitem{nguyen2010refined}
L.~Nguyen and A.~Zarnescu, ``Refined approximation for a class of {L}andau-de
  {G}ennes energy minimizers,'' {\em arXiv preprint arXiv:1006.5689}, 2010.

\bibitem{cladis1972non}
P.~Cladis and M.~Kleman, ``Non-singular disclinations of strength s=+1 in
  nematics,'' {\em Journal de Physique}, vol.~33, no.~5-6, pp.~591--598, 1972.

\bibitem{bethuel1992bifurcation}
F.~Bethuel, H.~Brezis, B.~Coleman, and F.~H{\'e}lein, ``Bifurcation analysis of
  minimizing harmonic maps describing the equilibrium of nematic phases between
  cylinders,'' {\em Archive for Rational Mechanics and Analysis}, vol.~118,
  no.~2, pp.~149--168, 1992.

\bibitem{biscari1997local}
P.~Biscari and E.~G. Virga, ``Local stability of biaxial nematic phases between
  two cylinders,'' {\em International Journal of Non-linear Mechanics},
  vol.~32, no.~2, pp.~337--351, 1997.

\bibitem{canevari2013biaxiality}
G.~Canevari, ``Biaxiality in the asymptotic analysis of a 2-{D} {L}andau-de
  {G}ennes model for liquid crystals,'' {\em arXiv preprint arXiv:1307.8065},
  2013.

\bibitem{fatkullin2009vortices}
I.~Fatkullin and V.~Slastikov, ``Vortices in two-dimensional nematics,'' {\em
  Communications in Mathematical Sciences}, vol.~7, no.~4, pp.~917--938, 2009.

\bibitem{golovaty2014minimizers}
D.~Golovaty and J.~A. Montero, ``On minimizers of a {L}andau--de {G}ennes
  energy functional on planar domains,'' {\em Archive for Rational Mechanics
  and Analysis}, vol.~213, no.~2, pp.~447--490, 2014.

\bibitem{henao2012symmetry}
D.~Henao and A.~Majumdar, ``Symmetry of uniaxial global {L}andau--de {G}ennes
  minimizers in the theory of nematic liquid crystals,'' {\em SIAM Journal on
  Mathematical Analysis}, vol.~44, no.~5, pp.~3217--3241, 2012.

\bibitem{kralj1999biaxial}
S.~Kralj, E.~G. Virga, and S.~{\v{Z}}umer, ``Biaxial torus around nematic point
  defects,'' {\em Physical Review E}, vol.~60, no.~2, p.~1858, 1999.

\bibitem{mkaddem2000fine}
S.~Mkaddem and E.~Gartland~Jr, ``Fine structure of defects in radial nematic
  droplets,'' {\em Physical Review E}, vol.~62, no.~5, p.~6694, 2000.

\bibitem{frank1958liquid}
F.~C. Frank, ``I. {L}iquid crystals. {O}n the theory of liquid crystals,'' {\em
  Discussions of the Faraday Society}, vol.~25, pp.~19--28, 1958.

\bibitem{virga1995variational}
E.~G. Virga, {\em Variational theories for liquid crystals}, vol.~8.
\newblock CRC Press, 1995.

\bibitem{biscari2003expulsion}
P.~Biscari and T.~J. Sluckin, ``Expulsion of disclinations in nematic liquid
  crystals,'' {\em European Journal of Applied Mathematics}, vol.~14, no.~01,
  pp.~39--59, 2003.

\bibitem{biscari2005field}
P.~Biscari and T.~J. Sluckin, ``Field-induced motion of nematic
  disclinations,'' {\em SIAM Journal on Applied Mathematics}, vol.~65, no.~6,
  pp.~2141--2157, 2005.

\bibitem{sonnet1997dynamics}
A.~M. Sonnet and E.~G. Virga, ``Dynamics of nematic loop disclinations,'' {\em
  Physical Review E}, vol.~56, no.~6, p.~6834, 1997.

\bibitem{gartland2002elastic}
E.~C. Gartland, Jr., A.~M. Sonnet, and E.~G. Virga, ``Elastic forces on nematic
  point defects,'' {\em Continuum Mechanics and Thermodynamics}, vol.~14,
  no.~3, pp.~307--319, 2002.

\bibitem{hardt1988stable}
R.~Hardt, D.~Kinderlehrer, and F.-H. Lin, ``Stable defects of minimizers of
  constrained variational principles,'' vol.~5, pp.~297--322, 1988.

\bibitem{berlyand2005homogenization}
L.~Berlyand, D.~Cioranescu, and D.~Golovaty, ``Homogenization of a
  {G}inzburg--{L}andau model for a nematic liquid crystal with inclusions,''
  {\em Journal de Math{\'e}matiques Pures et Appliqu{\'e}es}, vol.~84, no.~1,
  pp.~97--136, 2005.

\bibitem{lin1995nonparabolic}
F.-H. Lin and C.~Liu, ``Nonparabolic dissipative systems modeling the flow of
  liquid crystals,'' {\em Communications on Pure and Applied Mathematics},
  vol.~48, no.~5, pp.~501--537, 1995.

\bibitem{lin2000existence}
F.-H. Lin and C.~Liu, ``Existence of solutions for the {E}ricksen-{L}eslie
  system,'' {\em Archive for Rational Mechanics and Analysis}, vol.~154, no.~2,
  pp.~135--156, 2000.

\bibitem{MR2804649}
N.~J. Walkington, ``Numerical approximation of nematic liquid crystal flows
  governed by the {E}ricksen-{L}eslie equations,'' {\em ESAIM Math. Model.
  Numer. Anal.}, vol.~45, no.~3, pp.~523--540, 2011.

\bibitem{ball2014discontinuous}
J.~M. Ball and S.~Bedford, ``Discontinuous order parameters in liquid crystal
  theories,'' {\em Molecular Crystals and Liquid Crystals}, vol.~612, no.~1,
  pp.~467--489, 2015.

\bibitem{gartland2015scalings}
E.~C. Gartland~Jr, ``Scalings and limits of the {L}andau-de {G}ennes model for
  liquid crystals: A comment on some recent analytical papers,'' {\em arXiv
  preprint arXiv:1512.08164}, 2015.

\bibitem{oseen1933theory}
C.~Oseen, ``The theory of liquid crystals,'' {\em Transactions of the Faraday
  Society}, vol.~29, no.~140, pp.~883--899, 1933.

\bibitem{ericksen1995remarks}
J.~L. Ericksen, ``Remarks concerning forces on line defects,'' in {\em
  Theoretical, Experimental, and Numerical Contributions to the Mechanics of
  Fluids and Solids}, pp.~247--271, Springer, 1995.

\bibitem{ericksen1991liquid}
J.~L. Ericksen, ``Liquid crystals with variable degree of orientation,'' {\em
  Archive for Rational Mechanics and Analysis}, vol.~113, no.~2, pp.~97--120,
  1991.

\bibitem{acharya2014dislocation}
A.~Acharya and X.~Zhang, ``From dislocation motion to an additive velocity
  gradient decomposition, and some simple models of dislocation dynamics,''
  {\em Chinese Annals of Mathematics, Series B}, vol.~36B, no.~5, pp.~645--658,
  2015.

\bibitem{zhang2015single}
X.~Zhang, A.~Acharya, N.~J. Walkington, and J.~Bielak, ``A single theory for
  some quasi-static, supersonic, atomic, and tectonic scale applications of
  dislocations,'' {\em Journal of the Mechanics and Physics of Solids},
  vol.~84, pp.~145--195, 2015.

\bibitem{eshelby1980force}
J.~Eshelby, ``The force on a disclination in a liquid crystal,'' {\em
  Philosophical Magazine A}, vol.~42, no.~3, pp.~359--367, 1980.

\bibitem{kohn2006energy}
R.~V. Kohn, ``Energy-driven pattern formation,'' in {\em Proceedings of the
  International Congress of Mathematicians} (J.~V. M.~Sanz-Sole, J.~Soria and
  J.~Verdera, eds.), vol.~1, pp.~359--384, 2006.

\bibitem{jerrard1998dynamics}
R.~L. Jerrard and H.~M. Soner, ``Dynamics of {G}inzburg-{L}andau vortices,''
  {\em Archive for Rational Mechanics and Analysis}, vol.~142, no.~2,
  pp.~99--125, 1998.

\bibitem{alicandro2014metastability}
R.~Alicandro, L.~De~Luca, A.~Garroni, and M.~Ponsiglione, ``Metastability and
  dynamics of discrete topological singularities in two dimensions: a
  {$\Gamma$}-convergence approach,'' {\em Archive for Rational Mechanics and
  Analysis}, vol.~214, no.~1, pp.~269--330, 2014.

\bibitem{ercolani2009variational}
N.~M. Ercolani and S.~C. Venkataramani, ``A variational theory for point
  defects in patterns,'' {\em Journal of nonlinear science}, vol.~19, no.~3,
  pp.~267--300, 2009.

\bibitem{newell2012pattern}
A.~C. Newell, ``Pattern quarks and leptons,'' {\em Applicable Analysis},
  vol.~91, no.~2, pp.~213--223, 2012.

\bibitem{das2013can}
A.~Das, A.~Acharya, J.~Zimmer, and K.~Matthies, ``Can equations of equilibrium
  predict all physical equilibria? {A} case study from field dislocation
  mechanics,'' {\em Mathematics and Mechanics of Solids}, vol.~18, no.~8,
  pp.~803--822, 2013.

\bibitem{vitek1968intrinsic}
V.~Vitek, ``Intrinsic stacking faults in body-centred cubic crystals,'' {\em
  Philosophical Magazine}, vol.~18, no.~154, pp.~773--786, 1968.

\end{thebibliography}
\bibliographystyle{ieeetr}
\end{document}